\documentclass[12pt]{elsarticle}                                          
\usepackage{axodraw4j}
\usepackage{hyperref}
\usepackage{graphicx}                                                         
\usepackage{a41}                                                                
\usepackage{xcolor}                     
\usepackage[rflt]{floatflt}
\usepackage{float}
\usepackage{lscape}
\usepackage{hyperref}
\hypersetup{colorlinks=true, 
linkcolor=blue, filecolor=blue, urlcolor=mygreen}
\usepackage{breakurl} 
\usepackage{times}
\usepackage{array}
\usepackage[english]{babel}
\usepackage[latin1]{inputenc}
\usepackage[T1]{fontenc}
\usepackage{ae}
\usepackage{url}
\usepackage{amsmath, amsthm, amssymb}
\usepackage{slashed}

\usepackage{rotating}
\usepackage{graphicx}
\usepackage{comment}
\newcounter{mmacnt}
\def\restartmma{\setcounter{mmacnt}{0}}
\restartmma \catcode`|=\active
\def|#1|{\mathrm{#1}}
\catcode`|=12
\newenvironment{mma}{
\par\smallskip
\catcode`|=\active
\parskip=0pt\parindent=0pt 
\small
\def\In##1\\{%
\def\linebreak{\hfill\break\null\qquad}%
\refstepcounter{mmacnt}
\hangindent=2.5em\hangafter=0
\leavevmode
\llap{\tiny\sffamily In[\arabic{mmacnt}]:=\kern.5em}%
\mathversion{bold}\footnotesize$
\displaystyle##1$\normalsize
\mathversion{normal}\par
 }%
\def\Print##1\\{%
\def\linebreak{\hfill\break}%
\hangindent=2.5em\hangafter=0
\leavevmode ##1\par}%
\def\Out##1\\{%
\def\linebreak{$\hfill\break\null\hfill$}%
\kern\abovedisplayskip\par
\hangindent=2.5em\hangafter=0
\leavevmode
\llap{\tiny\sffamily Out[\arabic{mmacnt}]=\kern.5em}
\footnotesize$\displaystyle##1$
\normalsize\hfill\null\par
\kern\belowdisplayskip
}%
\def\Warning##1##2\\{%
\def\linebreak{\hfill\break}%
\hangindent=2.5em\hangafter=0
\leavevmode
{\scriptsize##1 : ##2}\par}%
}{%
\par\smallskip
}

\usepackage{color}
\newenvironment{fshaded}{%
\MakeFramed {\FrameRestore}
}%
{\endMakeFramed}

\makeatletter
\def\ps@pprintTitle{%
\let\@oddhead\@empty
\let\@evenhead\@empty
\def\@oddfoot{\reset@font\hfil\thepage\hfil}
\let\@evenfoot\@oddfoot
}
\makeatother
\begin{document}   
\begin{frontmatter}
\title{\Large
\textbf{One-loop analytical
expressions for $\gamma \gamma 
\rightarrow \phi_i\phi_j$ in Higgs 
Extensions of the Standard Models
and its applications
}}
\author[1,2]{Khiem Hong Phan}
\ead{phanhongkhiem@duytan.edu.vn}
\author[1,2]{Dzung Tri Tran}
\author[3]{Thanh Huy Nguyen}
\address[1]{\it Institute of Fundamental
and Applied Sciences, Duy Tan University,
Ho Chi Minh City $70000$, Vietnam}
\address[2]{Faculty of Natural Sciences,
Duy Tan University, Da Nang City $50000$,
Vietnam}
\address[3]
{\it VNUHCM-University of Science, $227$ 
Nguyen Van Cu, District $5$, Ho Chi Minh 
City $70000$, Vietnam}
\pagestyle{myheadings}
\markright{}
\begin{abstract} 
General one-loop formulas 
for loop-induced processes
$\gamma \gamma
\rightarrow \phi_i\phi_j$ with 
$\phi_i\phi_j = hh,~hH,~HH$ are 
presented in the paper. Analytic 
expressions evaluated in this work 
are valid for a class of Higgs 
Extensions of the Standard Models, 
e.g. Inert Doublet Higgs Models, Two Higgs 
Doublet Models, Zee-Babu Models 
as well as Triplet Higgs Models, 
etc. Analytic expressions for one-loop
form factors are written in terms of 
the basic scalar one-loop two-, three- 
and four-point functions following 
the output format of both
the packages~{\tt LoopTools} and 
{\tt Collier}. Physical results can be 
hence evaluated numerically by using 
one of the mentioned packages. 
Analytic results are tested
by several checks such as
the ultraviolet finiteness,
infrared finiteness of the one-loop 
amplitudes. Furthermore, the amplitudes
also obey the ward identity
due to the on-shell initial photons.
This identity is also verified
numerically in this works. 
In the applications, we present the
phenomenological results for Zee-Babu model
as a typical example in this report.
Production cross-section for the processes
$\gamma \gamma\rightarrow hh$ are scanned 
over the parameter space of 
the Zee-Babu Models. 
\end{abstract}
\end{frontmatter}
\section{Introduction}
Discovering the structure of 
scalar potential and consequently 
answering for the nature of 
the electroweak symmetry breaking 
(EWSB) are the most important
purposes at future colliders, 
e.g. the High-luminosity Large Hadron 
Collider (HL-LHC)~\cite{Liss:2013hbb,
CMS:2013xfa}, future lepton colliders 
(LC)~\cite{ILC:2013jhg, Shiltsev:2019rfl}. 
For the purposes, probing for 
multi-scalar boson 
productions are great of interest at 
the future colliders. Because the accuracy  
production cross-sections 
provide us not only
a crucial information for 
extracting triple Higgs 
and quadruple Higgs 
self-couplings but also 
direct searches for new scalar particles.
In the former case, indirect searches 
for new physics contributions can be performed
through the corrected measurements 
for Higgs self-couplings.
Recently, Standard Model-like (SM-like) 
Higgs boson pair productions have 
been probed at the LHC, e.g. via 
the events of two bottom quarks 
associated with two photons, 
four bottom quarks events, etc,
as shown in~\cite{ATLAS:2021ifb,
ATLAS:2014pjm, ATLAS:2015zug,
ATLAS:2015sxd,CMS:2017rpp,ATLAS:2018dpp,
CMS:2018tla,CMS:2020tkr,CMS:2022cpr,
ATLAS:2022xzm,ATLAS:2023qzf,ATLAS:2024ish}. 
We know that the measurements for
the Higgs pair productions are 
rather challenging at the LHC. 
It is not only because that 
the production cross-sections 
are rather small but also we have to 
deal with the huge SM's background. 
The future lepton colliders are then
proposed for complementary
to the physics at 
the LHC, for examples, 
the LC can significantly improve accuracy
the LHC measurements on many observables~\cite{LHCLCStudyGroup:2004iyd}.
More important, photon-photon collision 
is considered as an option of the 
LC~\cite{ILC:2013jhg,Shiltsev:2019rfl} which
the scalar Higgs pair productions ($\phi_i\phi_j$)
can be measured through scattering processes
$\ell \bar{\ell} \rightarrow \ell \bar{\ell}
\gamma^*\gamma^* \rightarrow \ell \bar{\ell}
\phi_i\phi_j$ for $\ell \equiv e, \mu$.
In the perspective, the LC also open 
a good opportunity for discovering many of 
physics beyond the SM (BSM) via multi-scalar Higgs productions.

In order to match the higher-precision data 
at the future colliders, theoretical 
evaluations for one-loop contributing to  
the di-Higgs boson productions are mandatory. 
One-loop contributing to the processes
at the LHC in the frameworks 
of the SM, of the Higgs Extensions of 
the SM (HESM) and other BSMs have performed 
in many papers, e.g. refering  
typical works as in 
Refs.~\cite{Arhrib:2009hc,Grigo:2013rya,
Shao:2013bz, Ellwanger:2013ova,
Han:2013sga,Barr:2013tda, deFlorian:2013jea,
Haba:2013xla,Cao:2014kya, Enkhbat:2013oba,
Li:2013flc, Frederix:2014hta, Baglio:2014nea,
FerreiradeLima:2014qkf, Hespel:2014sla,
Barger:2014taa,Grigo:2014jma, Maltoni:2014eza, 
Goertz:2014qta, Azatov:2015oxa, 
Papaefstathiou:2015iba, Grober:2015cwa, 
deFlorian:2015moa, He:2015spf,
Grigo:2015dia, Zhang:2015mnh, 
Agostini:2016vze, Grober:2016wmf,
Degrassi:2016vss, Kanemura:2016tan, 
deFlorian:2016uhr,Borowka:2016ypz,
Bishara:2016kjn, Cao:2016zob, Nakamura:2017irk, 
Grober:2017gut, Heinrich:2017kxx, Jones:2017giv, 
Davies:2018ood, Goncalves:2018qas,
Chang:2018uwu, Buchalla:2018yce, 
Bonciani:2018omm, Banerjee:2018lfq,
AH:2018tzg, Baglio:2018lrj, 
Davies:2019xzc, Davies:2019dfy,Chen:2019lzz,
Chen:2019fhs, Baglio:2020ini, 
Wang:2020nnr, Abouabid:2021yvw,
Davies:2022ram, He:2022lhc, 
AH:2022elh, Iguro:2022fel, Alioli:2022dkj,
Davies:2023npk, 
Bagnaschi:2023rbx, 
Davies:2024znp,Brigljevic:2024vuv}.
In the high-energy photon-photon collisions, 
one-loop corrections to the di-Higgs boson 
productions within the SM and many of BSMs 
have computed in Refs.~\cite{Jikia:1992mt,
Sun:1995rd,Zhu:1997nz, Zhu:1998rh, Gounaris:2000ja,
Zhou:2003ss, Cornet:2008nq, Asakawa:2008se, Asakawa:2009ux,Takahashi:2009sf, Hodgkinson:2009uj, Arhrib:2009gg, Asakawa:2010xj, Hernandez-Sanchez:2011idv,Ma:2011zzb}. At the future linear lepton colliders, including future multi-TeV muon colliders, 
the equivalent calculations for the 
Higgs pair productions have considered in Refs.~\cite{Sola:2011ex, Heng:2013wia, Chiesa:2021qpr, Samarakoon:2023crt}.  Additionally, 
one-loop
corrections to the scattering processes
$\gamma\gamma \rightarrow A^0A^0$ ($A^0$ 
is CP-odd Higgs) in 
Two Higgs Doublet Model have shown 
in Ref.~\cite{Demirci:2019kop}.
In this paper, we present a general 
one-loop formulas 
for loop-induced processes
$\gamma \gamma\rightarrow
\phi_i\phi_j$ with $\phi_i\phi_j
= hh,~hH,~HH$
which are valid for a class of 
Higgs Extensions of the Standard Models,
e.g. Inert Doublet Higgs Models, Two
Higgs Doublet Models, Zee-Babu 
models as well as Triplet Higgs Models, 
etc. Analytic expressions for one-loop
form factors are written in terms of 
the basic scalar one-loop two-, three- 
and four-point functions with following 
the output format of the 
packages~{\tt LoopTools}~\cite{Hahn:1998yk} 
as well as 
{\tt Collier}~\cite{Denner:2016kdg}. 
Numerical investigation can be 
hence generated by using 
one of the mentioned packages. 
Analytic results are confirmed
by several checks such as such as
the ultraviolet finiteness,
infrared finiteness of the one-loop 
amplitudes. Furthermore, the amplitudes
also obey the ward identity
due to the on-shell initial photons.
This identity is also verified
numerically in the works. 
In the applications, the 
phenomenological results for Zee-Babu Model
are examined as a typical example in this article.
Production cross-section for the processes
$\gamma \gamma\rightarrow hh$ are scanned 
over the parameter space of 
the model under consideration. 

The paper is presented with 
the structure as follows. 
Detailed evaluations
for one-loop corrections to
$\gamma \gamma\rightarrow \phi_i\phi_j$
with CP-even Higgses $\phi_{i,j}\equiv h,~H_j$
in the HESMs are shown in section $2$.
We then discuss on the numerical checks 
for the calculation and show for the 
applications of this work in the section $3$. Conclusion 
and outlook are devoted in section $4$.
Analytic expressions for one-loop form factors
given in the appendix $B$. Deriving the 
additional couplings in the 
Zee-Babu models are presented in the 
appendix $C$. 
\section{One-loop contributions        
for processes $\gamma \gamma           
\rightarrow \phi_i\phi_j$ in the HESMs} 
In this section, detailed evaluations for
one-loop contributions for 
the scattering 
processes $\gamma \gamma
\rightarrow \phi_i\phi_j$ in the HESMs 
are shown in this section. 
Additional scalar bosons in the 
mentioned HESMs are included as CP-even 
Higgses $\phi_i$, CP-odd Higgses $A_j^{0}$ and 
singly (doubly) charged Higgses
$S\equiv S_k^Q$ with charged 
quantum number $Q$, 
for $i,j, k=1,2,\cdots$. In this work,
$S_k^Q$ can be singly charged Higgs
$H^{\pm}$ and 
doubly charged Higgs $K^{\pm\pm}$,
appropriately. Beyond the SM, 
the extra couplings relating to
the mentioned scalar particles in the HESMs
are parameterized as general form
$g_{\text{vertex}}$. Explicitly formulas 
for $g_{\text{vertex}}$ for each model
under investigation 
are presented in concrete,
seen Zee-Babu Model in the application 
of this work and our previous 
work~\cite{Phan:2024vfy} for examples.

\begin{figure}[H]
\centering
\begin{tabular}{c}
\includegraphics[width=15cm, height=10cm]
{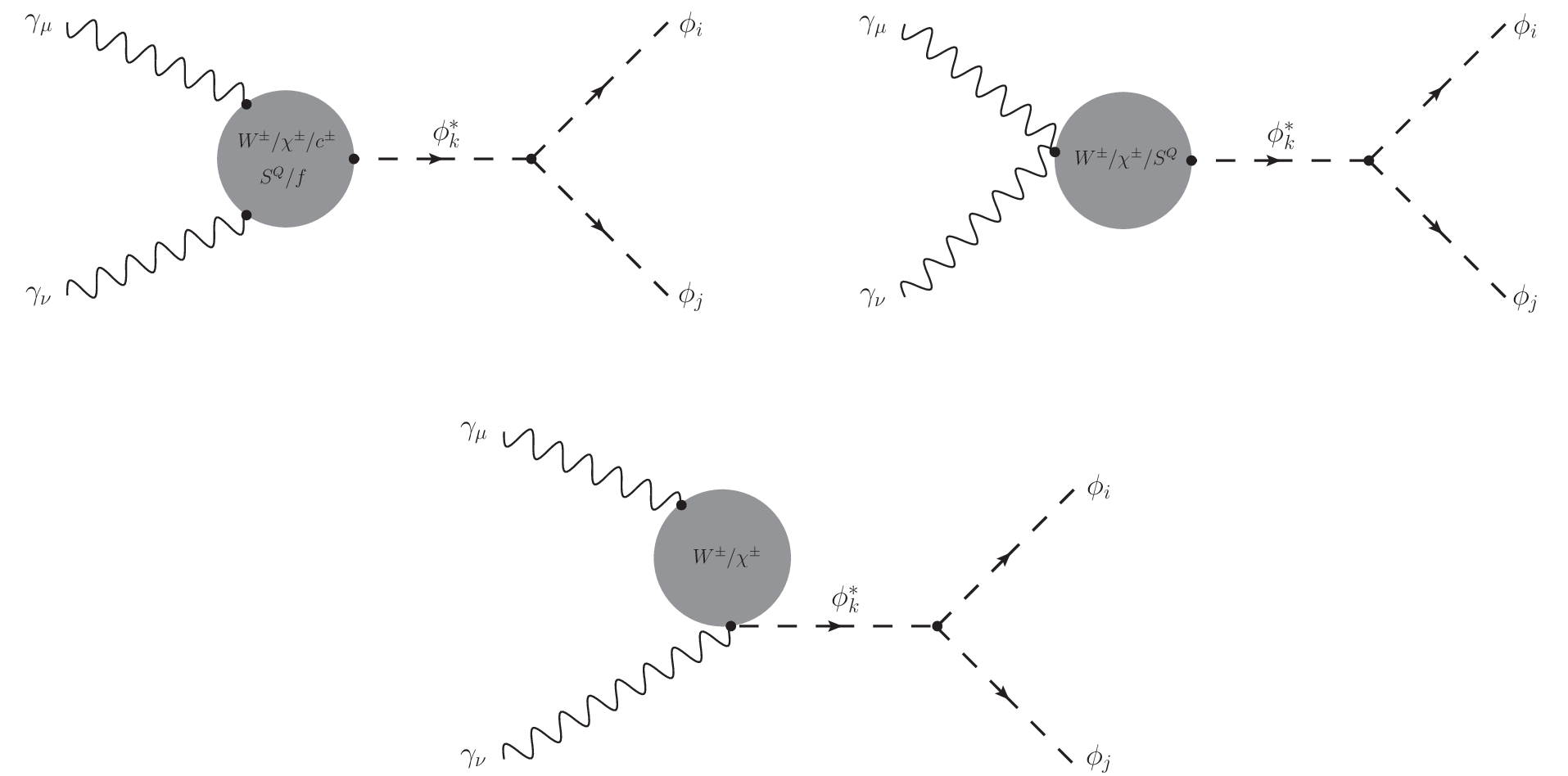}
\end{tabular}
\caption{\label{feynG1}
All one-loop diagrams
with fermions, W bosons 
(with 
charged Goldstone
$\chi^{\pm}$, Ghosht particles
$c^\pm$
) 
and charged Higges
exchanging in the loop
of $\phi_k^*$-poles, for
$\phi_k^*=h^*, H^*$. 
These kinds of
diagrams appear in this group
are included one-loop
contributions for off-shell
CP-even Higges decay 
like $\phi_k^* \rightarrow
\gamma\gamma$ 
in connecting with
the vertices 
$\phi_k^*\phi_i\phi_j$.
}
\end{figure}

\begin{figure}[H]
\centering
\begin{tabular}{c}
\includegraphics[width=15cm,
height=10cm]
{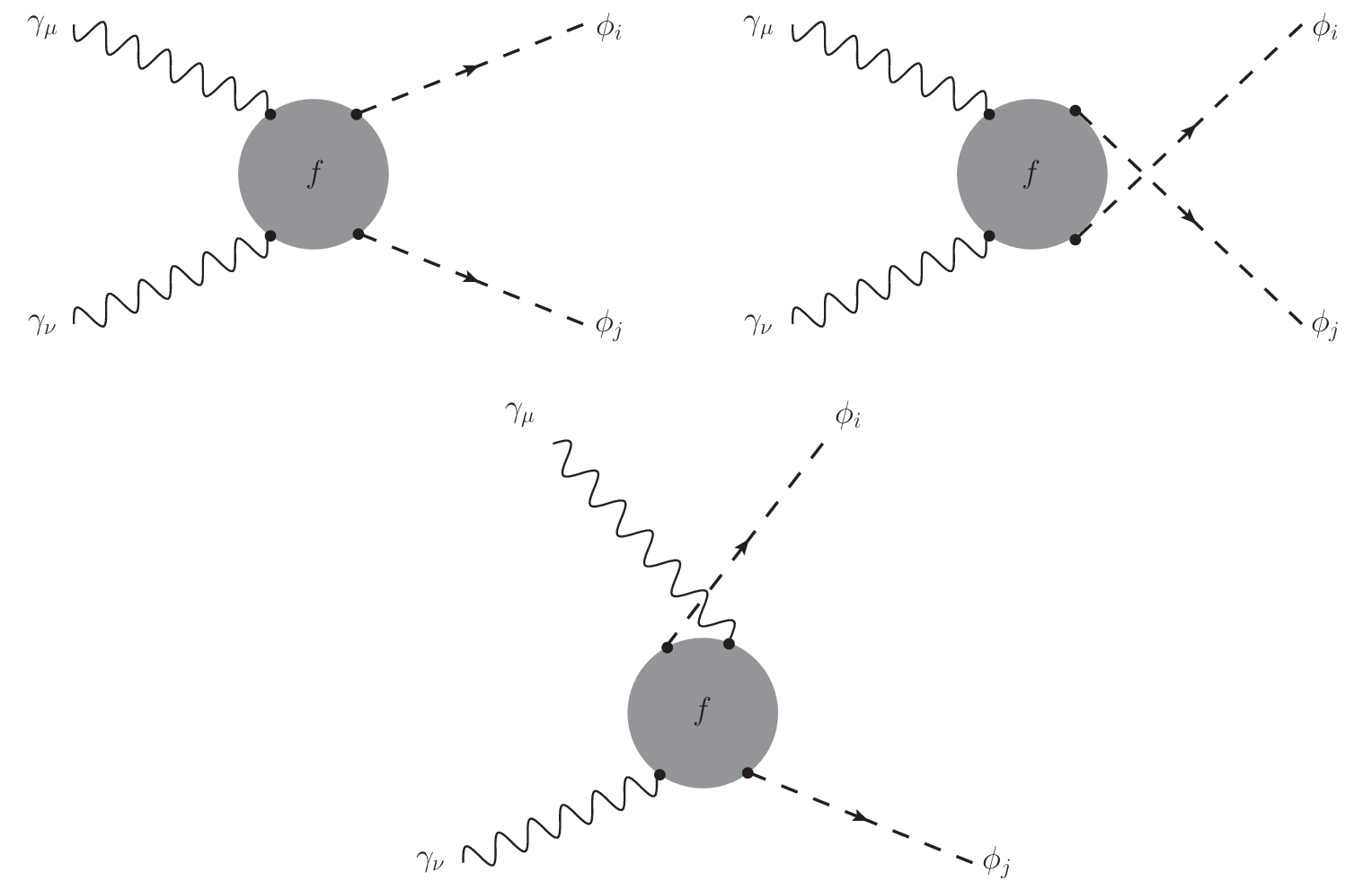}
\end{tabular}
\caption{\label{feynG4} 
One-loop four external legs with 
fermion internal lines contributing 
to the computed processes 
(noted as group $G_4$).}
\end{figure}

\begin{figure}[H]
\centering
\begin{tabular}{c}
\includegraphics[width=15cm,
height=10cm]
{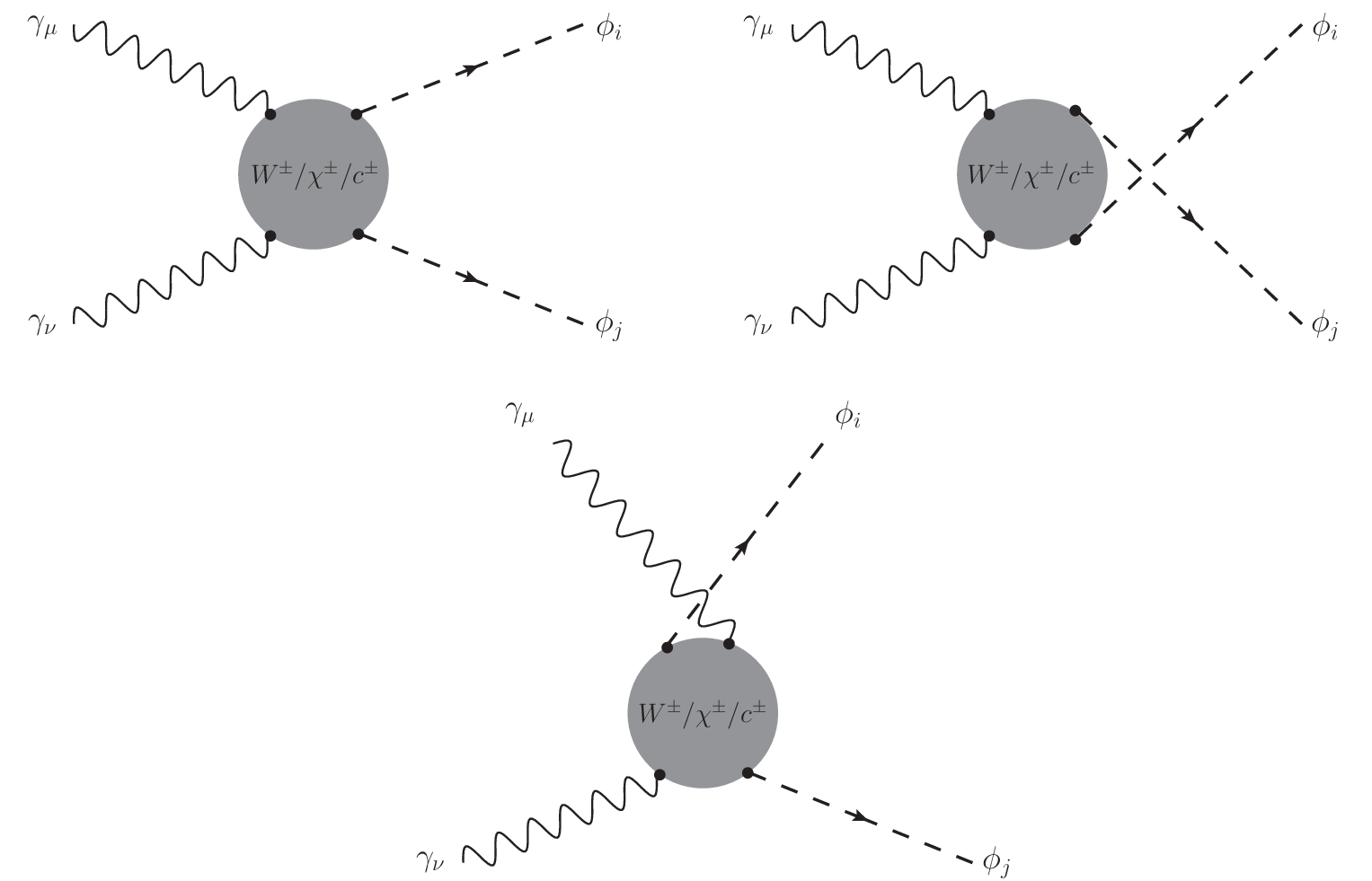}
\\
\includegraphics[width=15cm,
height=10cm]
{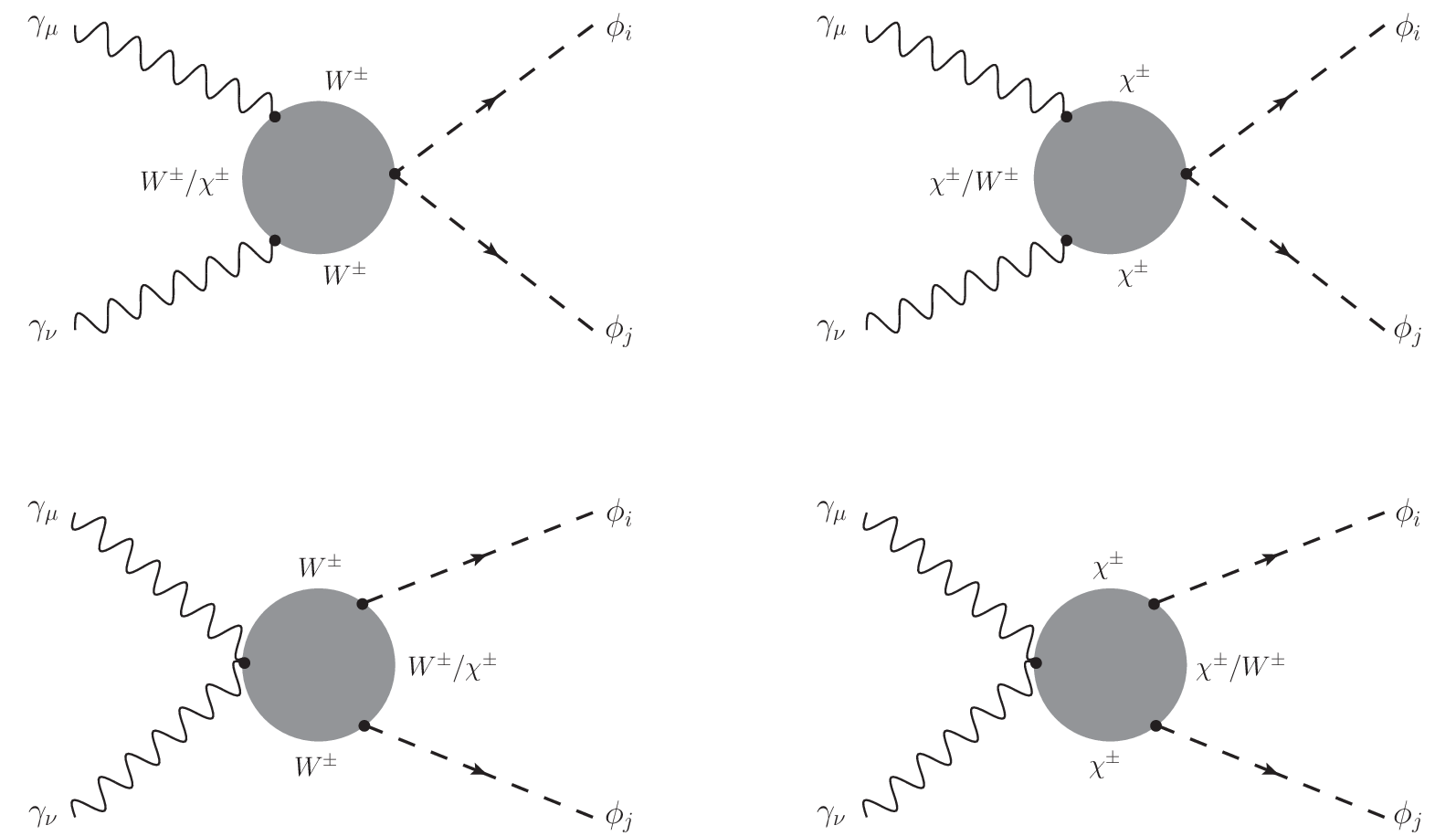}
\end{tabular}
\caption{\label{feynG5}
All one-loop
box diagrams contributing to
the processes with W-boson
exchanging in the loop
(putted into $G_5$). }
\end{figure}

\begin{figure}[H]
\centering
\begin{tabular}{c}
\includegraphics[width=15cm,
height=10cm]
{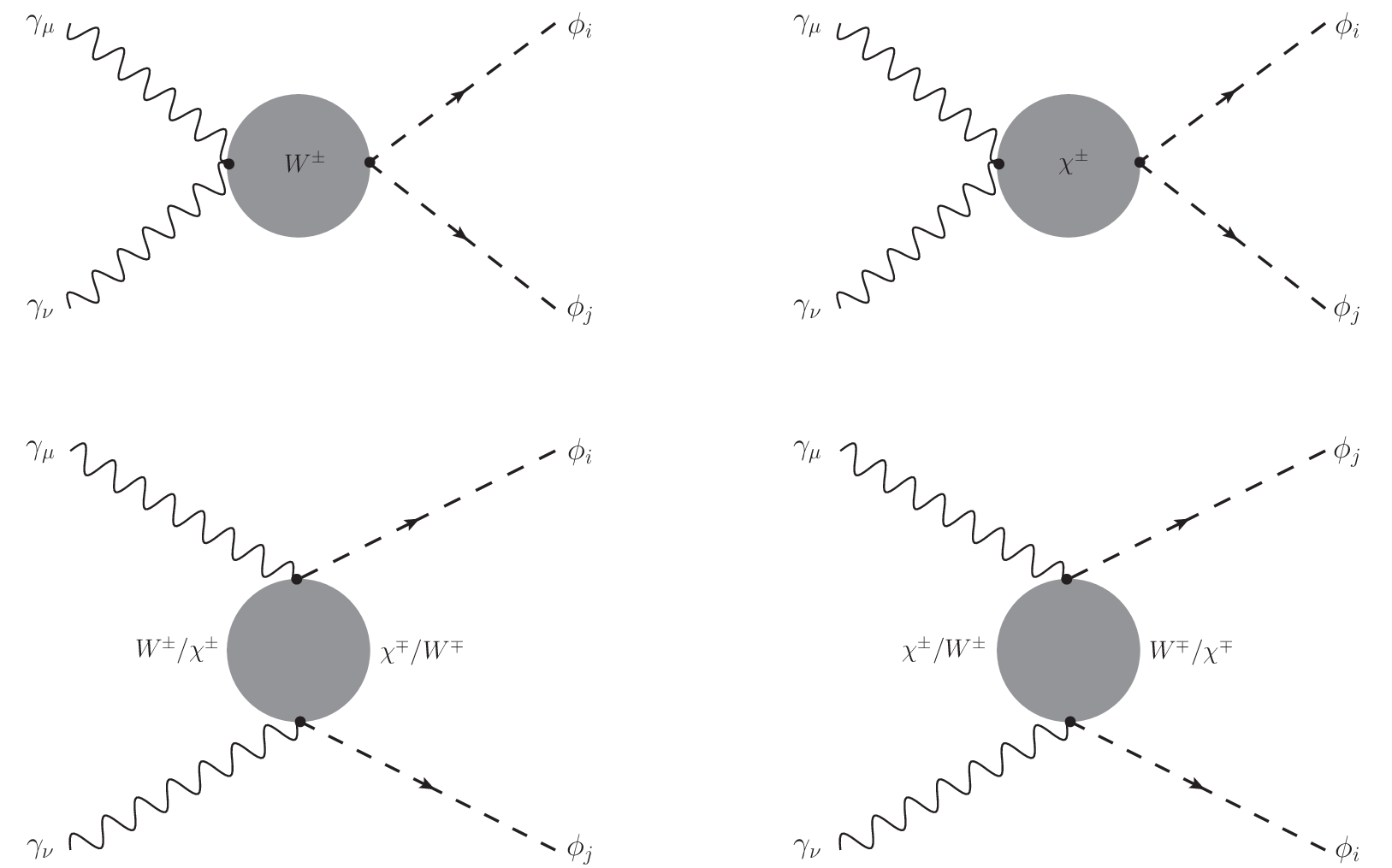}
\\
\includegraphics[width=15cm,
height=10cm]
{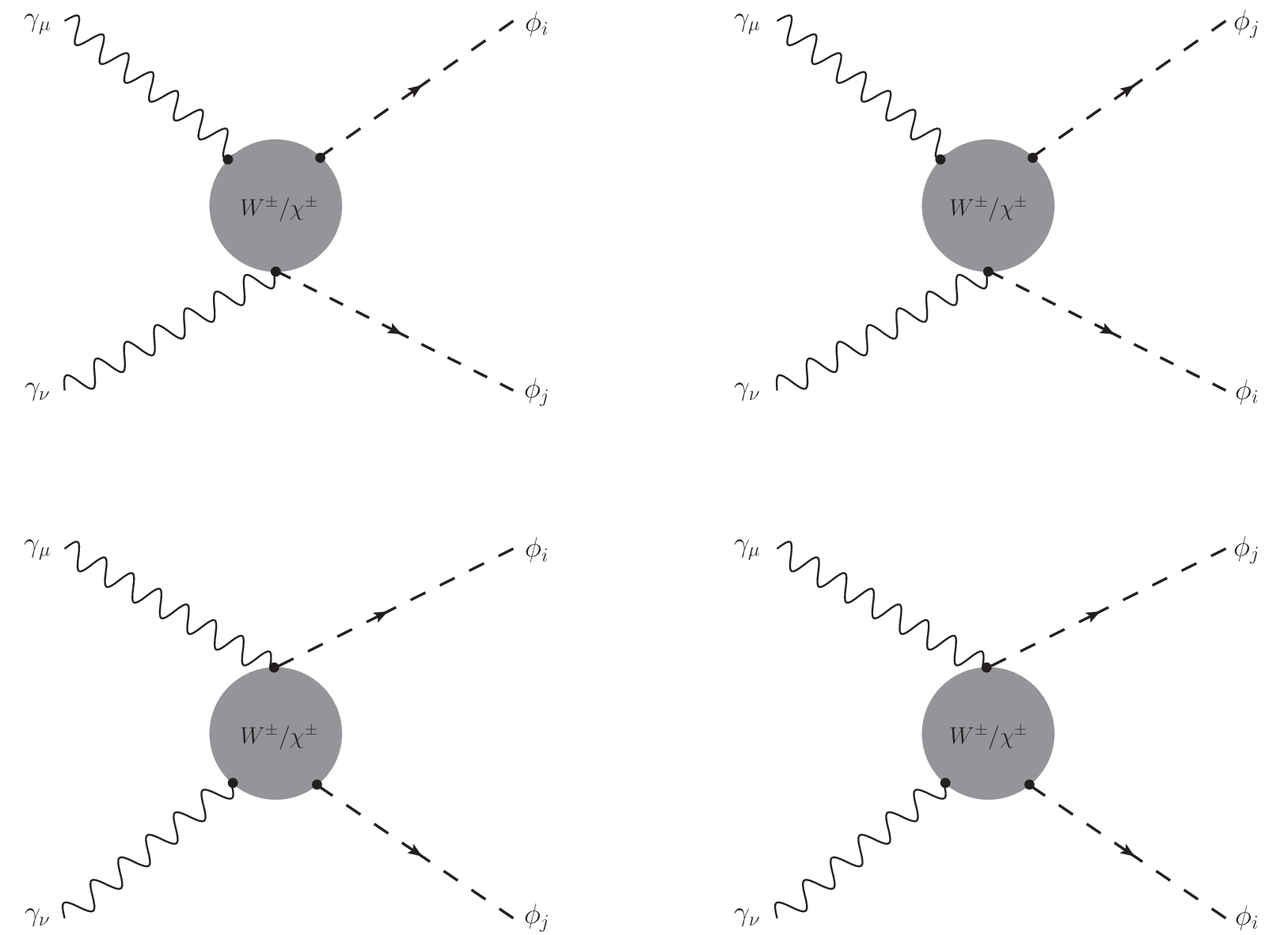}
\end{tabular}
\caption{\label{feynG5b}
All one-loop
box diagrams contributing to
the processes with W-boson
exchanging in the loop 
(putted into $G_5$).  }
\end{figure}

By employing the on-shell 
renormalization scheme developed 
in~\cite{Bohm:1986rj, Hollik:1988ii,Aoki:1982ed} 
for the fermion sector 
and gauge sector as well as 
the improved on-shell renormalization 
scheme for the scalar sector
following the method 
in~\cite{Kanemura:2017wtm}, 
one loop-induced Feynman diagrams
for the production processes
$\gamma \gamma \rightarrow \phi_i
\phi_j$ with CP-even Higgses
$\phi_{i,j} \equiv h, H$
in the HESMs are plotted 
in the following paragraphs. 
The calculations are handled in the
't Hooft-Feynman (HF) gauge which
one loop-induced Feynman diagrams
can be categorized into several
groups as explained in below.
We mention the first
classification Feynman diagrams 
as shown in Fig.~\ref{feynG1}. 
In this group, we list all 
one-loop diagrams with 
$\phi_k^*$-poles for $\phi_k^*
=h^*,~H^*$. These kinds of
diagrams appear in this group
are considered as one-loop
included contributions for 
off-shell CP-even Higges decay 
like $\phi_k^* \rightarrow
\gamma\gamma$ with fermions
(noted as $G_1$), W-boson, 
charged Goldstone $\chi^{\pm}$, 
Ghosht particles $c^\pm$ 
(as $G_2$)
and charged Higges $S^{Q}$ 
(as $G_3$) internal lines 
in connecting with
the vertices 
$\phi_k^*\phi_i\phi_j$.
The second classification of one-loop
Feynman diagrams is involed to the 
one-loop box diagrams. 
The first type of 
box diagrams contributing to the 
computed processes are plotted as 
in Fig.~\ref{feynG4}. 
In these topologies, 
all fermions internal lines
are taken into consideration 
(noted as group $G_4$).
Additionally, the second type of 
one-loop four-point Feynman diagrams 
with vector $W$-bosons, 
the charged Goldstone bosons 
$\chi^{\pm}$, Ghosht particles
$c^\pm$ internal lines are concerned
in the calculated processes.
These diagrams are grouped into 
$G_5$ as shown in
Figs.~\ref{feynG5},~\ref{feynG5b}.

\begin{figure}[H]
\centering
\begin{tabular}{c}
\includegraphics[width=15cm,
height=10cm]
{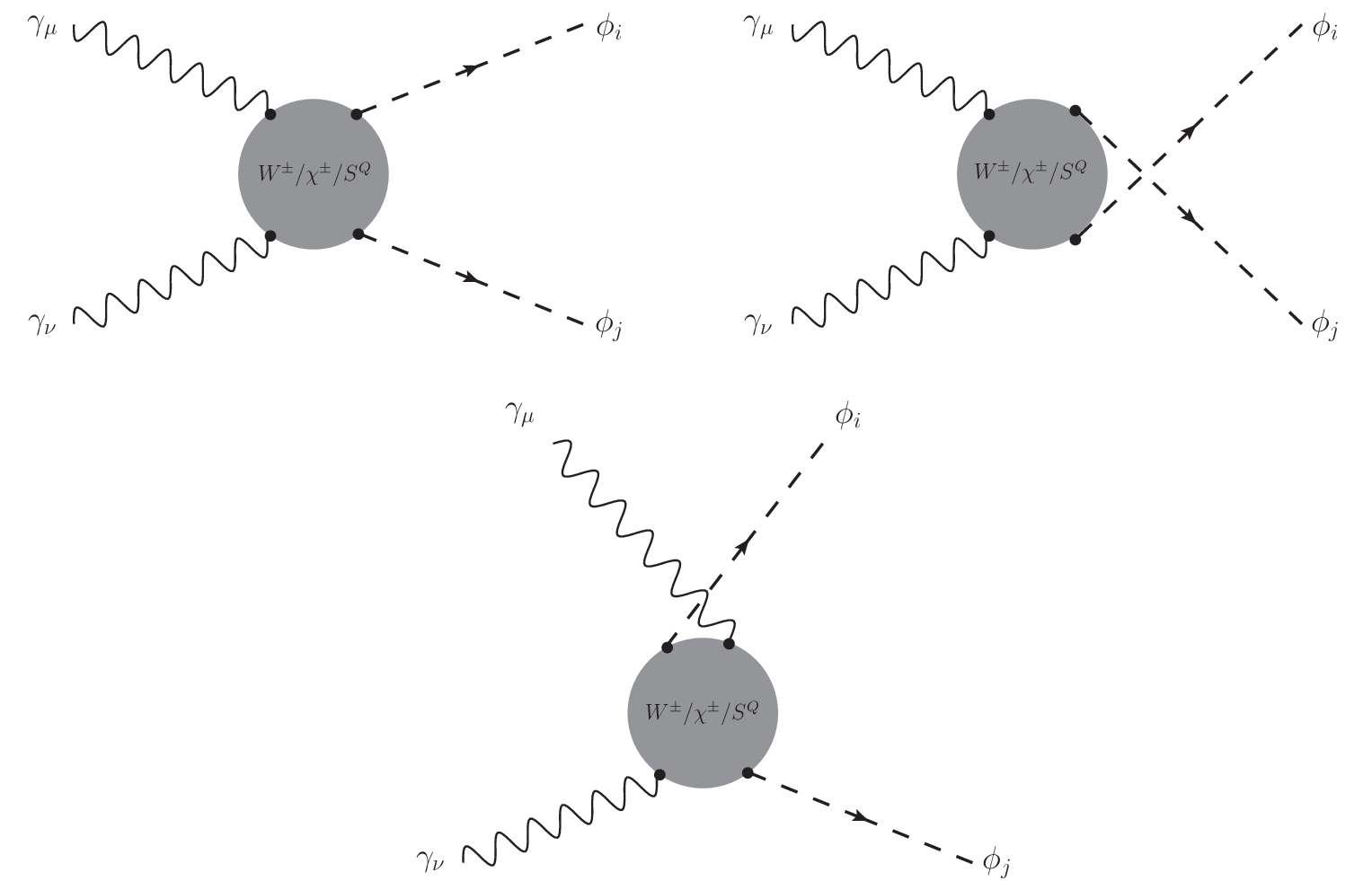}
\end{tabular}
\caption{\label{feynG6a} One-loop
box diagrams with both W-boson
and charged
Higgs propagating in the loop
(putted into $G_6$)
. }
\end{figure}

\begin{figure}[H]
\centering
\begin{tabular}{c}
\includegraphics[width=15cm,
height=10cm]
{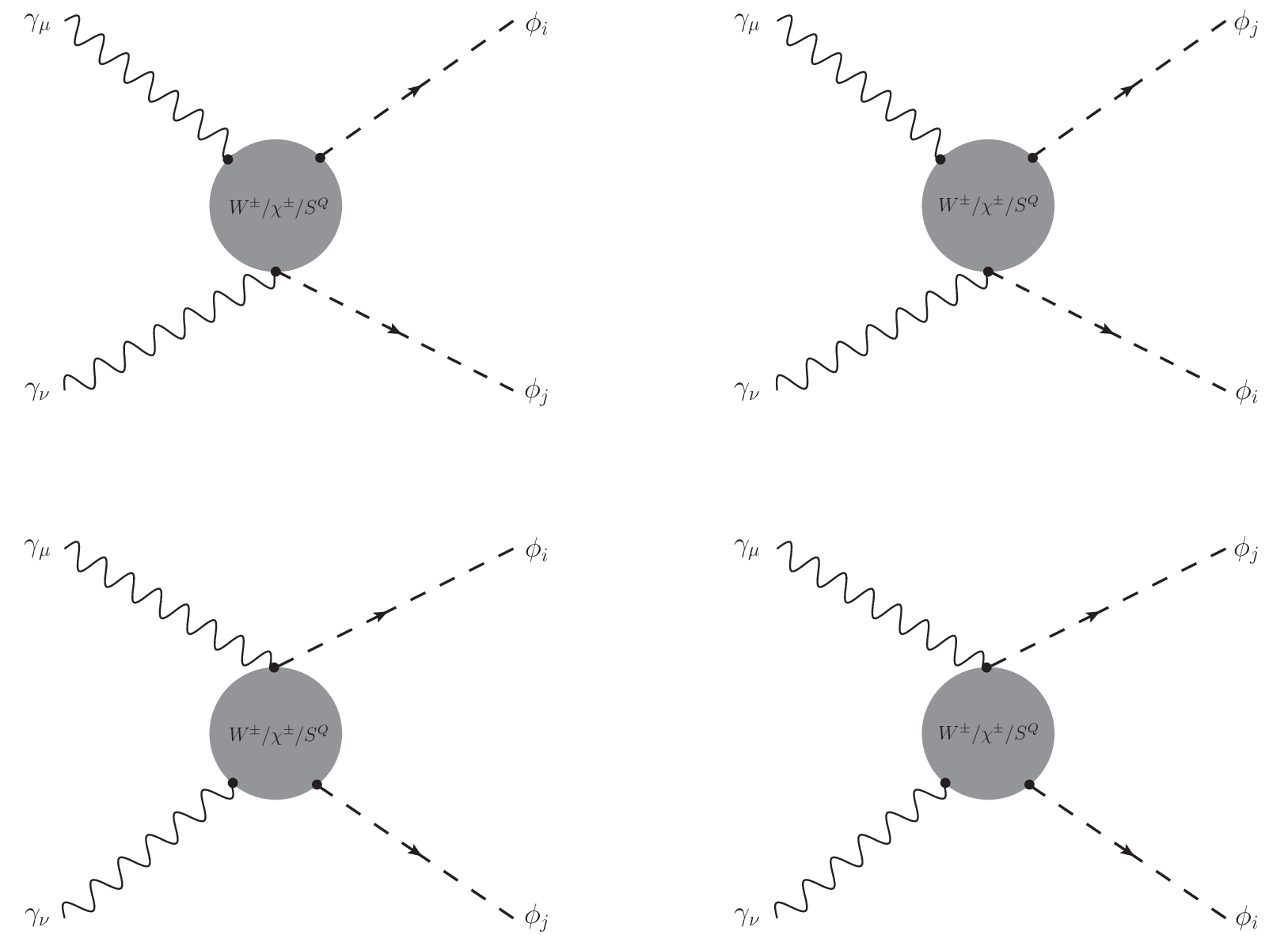}
\\
\includegraphics[width=15cm,
height=10cm]
{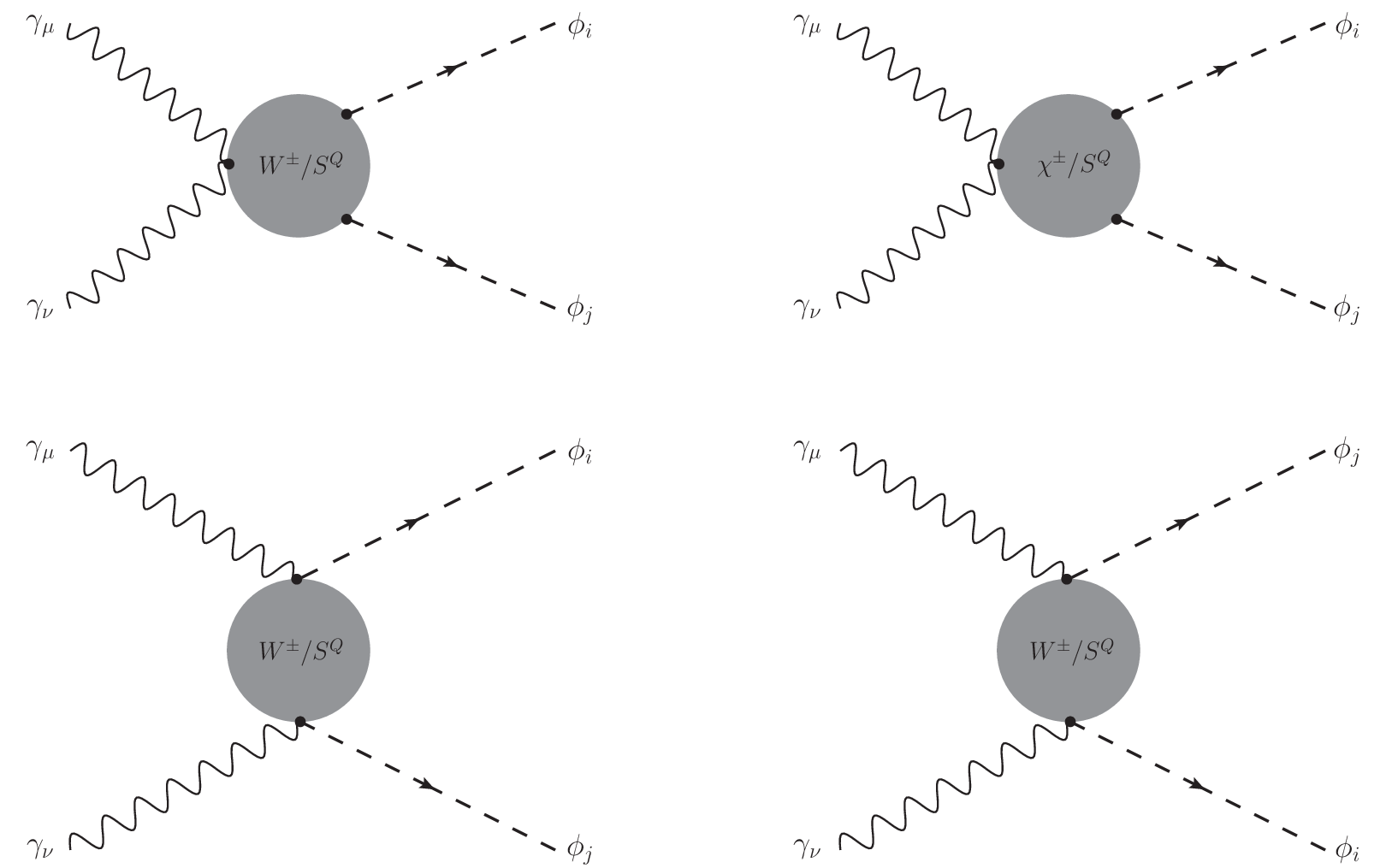}
\end{tabular}
\caption{
\label{feynG6b} Further
one-loop
box diagrams with both W-boson
and charged
Higgs propagating in the loop 
(putted into $G_6$).
}
\end{figure}

\begin{figure}[H]
\centering
\begin{tabular}{c}
\includegraphics[width=15cm,
height=10cm]
{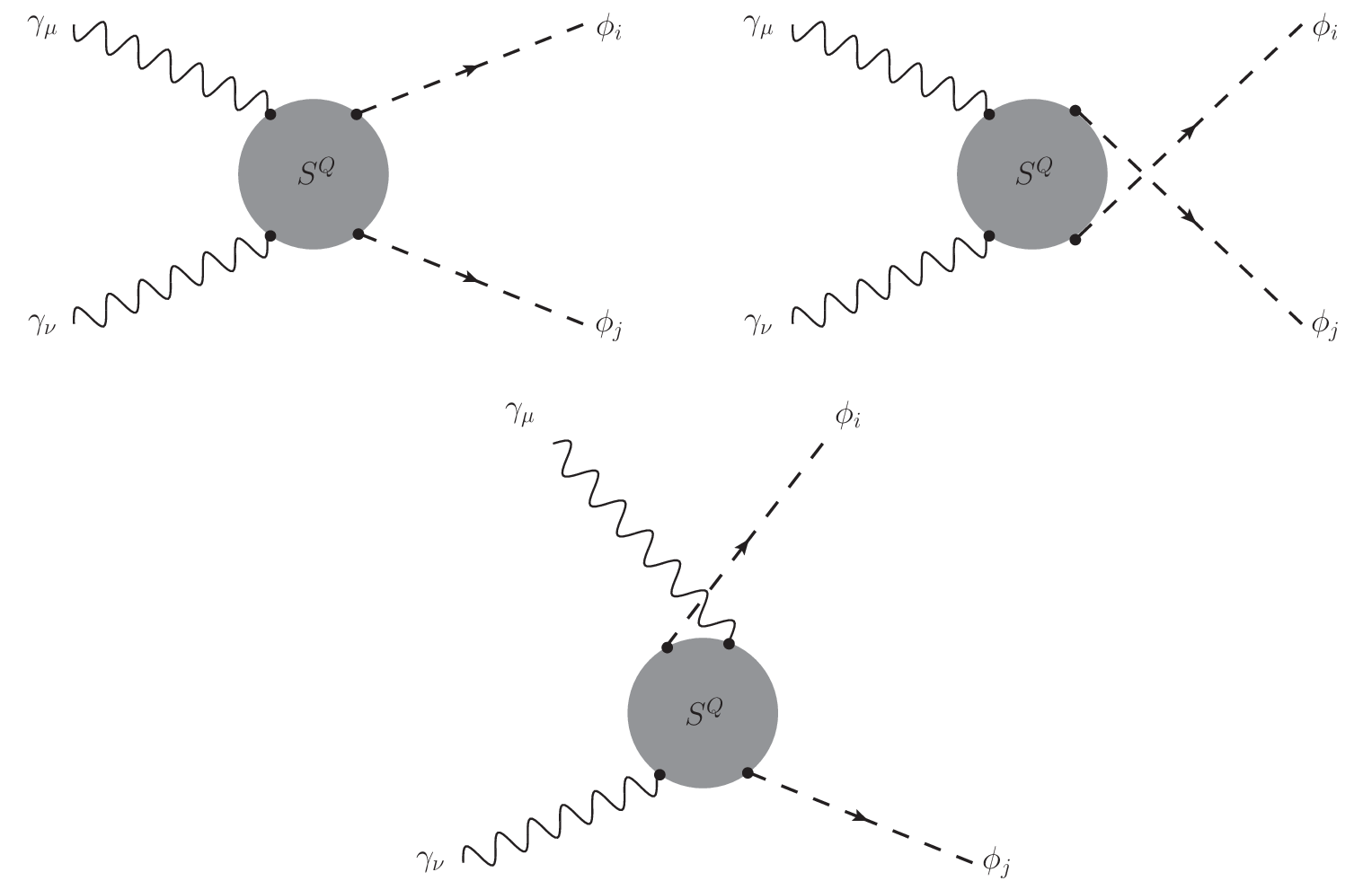}
\\
\includegraphics[width=15cm,
height=10cm]
{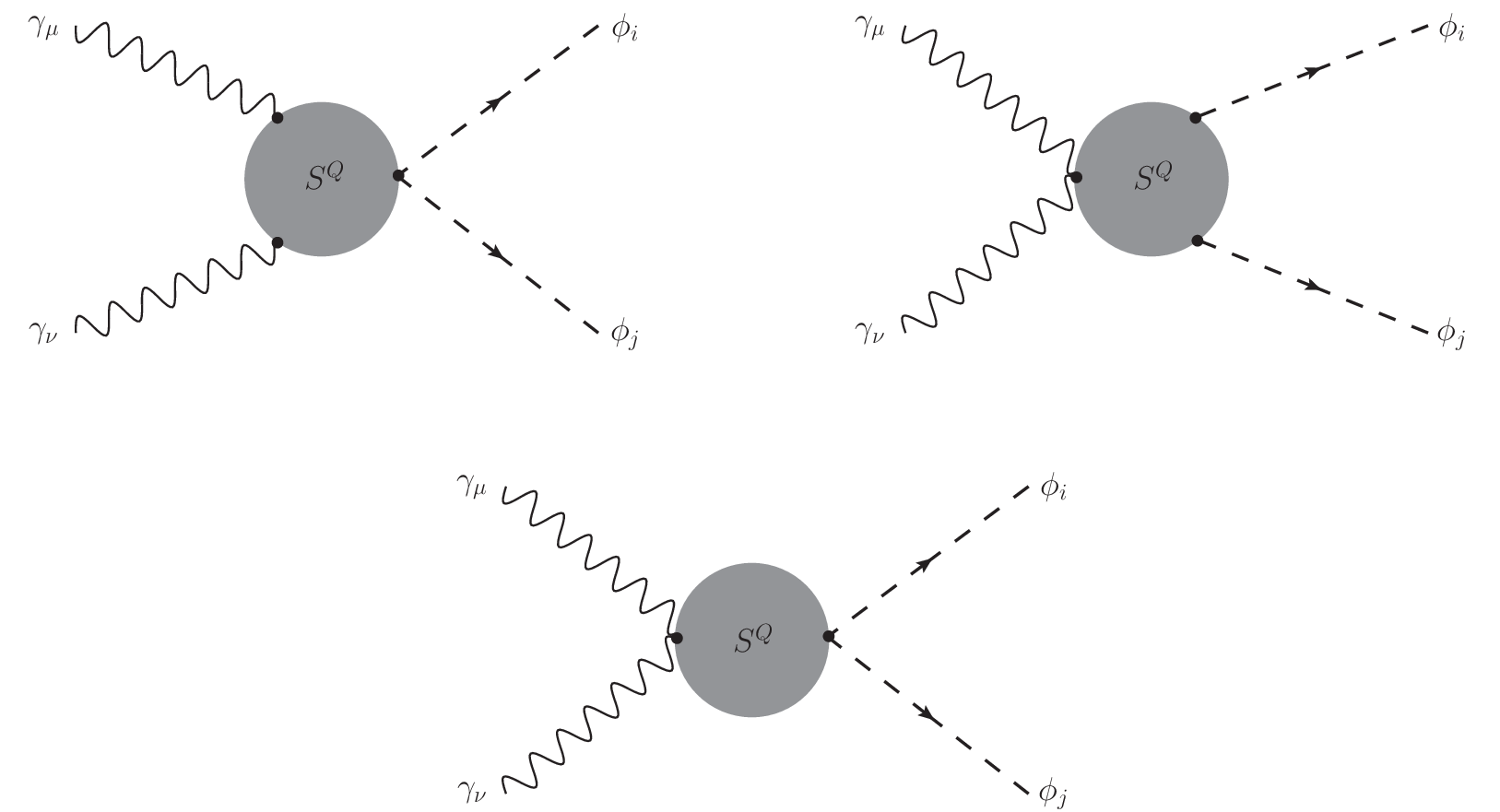}
\end{tabular}
\caption{\label{feynG7} All
one-loop
box diagrams with charged Higgs
in the loop (considered as $G_7$).
}
\end{figure}

In the scope of the HESMs concerned 
in this work, we have also another type pf 
one-loop box diagrams 
with both W-boson
and singly charged Higgs $S^{Q}
\equiv H^{\pm}$ 
propagating in the loop, 
seen Figs.~\ref{feynG6a},
~\ref{feynG6b}. We put 
these diagrams
into group $G_6$. 
Finally, we consider the last type of one-loop
box diagrams with charged Higgses $S^{Q}
\equiv H^{\pm},~K^{\pm\pm}$
in the loop, as shown in
Fig.~\ref{feynG7}. We then added 
these diagrams in to $G_7$. Both singly
charged and doubly charged Higges
are considered being internal line particles
in this case.

We turn out our attention to
discuss on analytic results.
In general, one-loop amplitude for
scattering processes
$\gamma_\mu (q_1) \, \gamma_\nu (q_2)
\rightarrow \phi_i (q_3) \, \phi_j (q_4)$
is presented in terms of
Lorentz structure as follows:
\begin{eqnarray}
\mathcal{A}_{\gamma \gamma
\rightarrow \phi_i \phi_j}
&=&
\Big[
F_{00}
\;
g^{\mu\nu}
+
\sum\limits_{i,j=1;
i\leq j}^{3}
F_{ij}
\;
q_i^{\nu} q_j^{\mu}
\Big]
\varepsilon_{\mu}(q_1)
\varepsilon_{\nu}(q_2).
\end{eqnarray}
In this formulas, the vector
$\varepsilon_{\mu}(q)$ is 
polarization vector of external 
photon with the $4$-dimension
momentum $q$. The scalar coefficients
$F_{ij}$ for $i,j=1,2,3$ are called as
one-loop form factors. They are written
as functions of the following
kinematic invariant
variables:
\begin{eqnarray}
\label{kine}
\hat{s} &=& (q_1+q_2)^2
= q_1^2 + 2 q_1 \cdot q_2 + q_2^2
= 2 q_1 \cdot q_2,
\\
\hat{t} &=& (q_1 - q_3)^2
= q_1^2 - 2 q_1 \cdot q_3 + q_3^2
= M_{\phi_i}^2 - 2 q_1 \cdot q_3,
\\
\hat{u} &=& (q_2 - q_3)^2
= q_2^2 - 2 q_2 \cdot q_3 + q_3^2
= M_{\phi_i}^2 - 2 q_2 \cdot q_3.
\end{eqnarray}
The kinematic invariant
masses for external legs 
are given as $q_1^2 = q_2^2 = 0$,
$q_3^2 = M_{\phi_i}^2$, 
$q_4^2 = M_{\phi_j}^2$. 
These variables obey the follow
identity as 
$\hat{s} + \hat{t} + \hat{u} =
M_{\phi_i}^2 + M_{\phi_j}^2$. 
Assuming two photons in 
initial states being 
on-shell particles, 
loop-induced
amplitude must satisfy 
the ward identity.
As a result, we derive
the following relations 
among the form factors as:
\begin{eqnarray}
\label{ward0}
F_{00}
&=&
\dfrac{ 
\hat{t} - M_{\phi_i}^2
}{2}
\,
F_{13}
-
\dfrac{
\hat{s}
}{2}
\,
F_{12},
\\
\label{ward1}
F_{00}
&=&
\dfrac{
\hat{u}
- M_{\phi_i}^2}{2}
\,
F_{23}
-
\dfrac{
\hat{s}
}{2}
\,
F_{12},
\\
\label{ward2}
F_{13}
&=&
\dfrac{
\hat{u} 
- M_{\phi_i}^2}
{
\hat{s} 
}
\,
F_{33},
\\
\label{ward3}
F_{23}
&=&
\dfrac{
\hat{t}
- M_{\phi_i}^2}
{
\hat{s}
}
\,
F_{33}.
\end{eqnarray}
Using the mentioned above
relations, one-loop amplitude
is expressed
via two independent one-loop
form factors, e.g. taking
$F_{12}$ and $F_{33}$ as an
example.
In detail, the amplitude
can be rewritten as follows:
\begin{eqnarray}
\label{ampred}
\mathcal{A}_{\gamma \gamma
\rightarrow \phi_i \phi_j}
&=&
\Big[
\mathcal{P}^{\mu\nu}
\cdot F_{12}
+
\mathcal{Q}^{\mu\nu}
\cdot 
F_{33}
\Big]
\,
\varepsilon_{\mu}(q_1)
\varepsilon_{\nu}(q_2).
\nonumber
\end{eqnarray}
Where two given tensors
are defined as
\begin{eqnarray}
\mathcal{P}^{\mu\nu}
&=&
q_2^{\mu}
q_1^{\nu}
-
\dfrac{
\hat{s}
}{2}
\cdot g^{\mu\nu},
\\
\mathcal{Q}^{\mu\nu}
&=&
\dfrac{(M_{\phi_i}^2 - \hat{t})
(M_{\phi_i}^2 - \hat{u})}{2 \hat{s}
}
\cdot
g^{\mu\nu}
+
q_3^{\mu} q_3^{\nu}
 +
\dfrac{(
\hat{t}
-M_{\phi_i}^2)}
{
\hat{s}
}
\cdot
q_2^{\mu} q_3^{\nu}.
\end{eqnarray}

Analytic results for one-loop form factors
$F_{12}$ and $F_{33}$ for the considered
processes in the HESMs are collected in terms 
of the basic scalar one-loop functions. 
The form factors $F_{ab}$ for $ab = {12, 33}$
are decomposed into triangle and box parts
which are corresponding to the contributions 
from one-loop triangle and one-loop box 
diagrams shown in above paragraphs. 
In detail, the form factor are 
expressed as follows:
\begin{eqnarray}
\label{form-factor12}
F_{12}
&=&
\sum\limits_{\phi_k^*=h^*, H^*}
\dfrac{g_{\phi_k^* \phi_i \phi_j}}
{\Big[
s - M_{\phi_k}^2 +i \Gamma_{\phi_k}
M_{\phi_k}
\Big]
}
\times
\nonumber\\
&& \times
\Bigg[
\sum\limits_{f}
g_{\phi_k^* ff}
\cdot
F_{12,f}^{\text{Trig}}
+
g_{\phi_k^* WW }
\cdot
F_{12,W}^{\text{Trig}}
+
\sum\limits_{S=H^{\pm}, 
K^{\pm\pm}}
g_{\phi_k^*SS }
\cdot
F_{12,S}^{\text{Trig}}
\Bigg]
\nonumber\\
&& +
\sum\limits_{f}
g_{\phi_i ff}
\cdot
g_{\phi_j ff}
\cdot
F_{12,f}^{\text{Box}}
\nonumber\\
&&
+
\Big[
g_{\phi_i W W}
\cdot
g_{\phi_j WW }
\cdot
F_{12,W}^{\text{Box, 1}}
+
g_{\phi_i \phi_j WW }
\cdot
F_{12,W}^{\text{Box, 2}}
+
g_{\phi_i \phi_j \chi\chi }
\cdot
F_{12,W}^{\text{Box, 3}}
\Big]
\nonumber\\
&&
+
\sum\limits_{S
=H^{\pm}, K^{\pm\pm}}
\Big[
g_{\phi_i SS}
\cdot
g_{\phi_j SS}
\cdot
F_{12,S}^{\text{Box},1}
+
g_{\phi_i \phi_j SS}
\cdot
F_{12,S}^{\text{Box},2}
\Big]
\nonumber\\
&&
+
g_{\phi_i H^{\pm} W^{\mp}}
\cdot
g_{\phi_j H^{\pm} W^{\mp}}
\cdot
F_{12,W, H^\pm}^{\text{Box}},
\\
&&
\nonumber
\\
\label{form-factor33}
F_{33}
&=&
\sum\limits_{f}
g_{\phi_i ff }
\cdot
g_{\phi_j ff }
\cdot
F_{33,f}^{\text{Box}}
+
g_{\phi_i W W}
\cdot
g_{\phi_j W W}
\cdot
F_{33,W}^{\text{Box}}
\nonumber\\
&&
+
\sum\limits_{S^Q=H^{\pm}, K^{\pm\pm}}
g_{\phi_i SS}
\cdot
g_{\phi_j SS}
\cdot
F_{33,S}^{\text{Box}}
+
g_{\phi_i H^{\pm} W^{\mp}}
\cdot
g_{\phi_j H^{\pm} W^{\mp}}
\cdot
F_{33,W, H^\pm}^{\text{Box}},
\end{eqnarray}
Analytic results for one-loop form
factors given in Eqs.~\ref{form-factor12},~\ref{form-factor33} are shown explicitly
in the appendix $B$. Having all the necessary form factors, the cross sections
are then evaluated as follows
\begin{eqnarray}
\hat{\sigma}_{\textrm{HESM}}
^{\gamma\gamma \rightarrow 
\phi_i \phi_j}(\hat{s})
&=&
\dfrac{1}{n!}
\dfrac{1}{16 \pi \hat{s}^2}
\int \limits_{t_\text{min}} ^{t_\text{max}}
d \hat{t}\;
\frac{1}{4}
\sum \limits_\text{unpol.}
\big|
\mathcal{A}_{\gamma \gamma
\rightarrow \phi_i \phi_j}
\big|^2
\end{eqnarray}
with $n = 2$ if the final particles
are identical such as $\gamma \gamma
\rightarrow hh, HH$, and 1 otherwise
like $\gamma \gamma \rightarrow h H$.
The integration limits are
\begin{eqnarray}
t_{\text{min} (\text{max})}
&=&
-\dfrac{\hat{s}}{2}
\Bigg\{
1
-
\dfrac{M_{\phi_i}^2 + M_{\phi_j}^2}{s}
\pm
\Bigg[
1
-
2
\Bigg(
\dfrac{M_{\phi_i}^2 + M_{\phi_j}^2}{s}
\Bigg)
+
\Bigg(
\dfrac{M_{\phi_i}^2 - M_{\phi_j}^2}{s}
\Bigg)^2
\Bigg]^{1/2}
\Bigg\}.
\end{eqnarray}
Total amplitude is given
\begin{eqnarray}
\frac{1}{4}
\sum \limits_\text{unpol.}
\big|
\mathcal{A}_{\gamma \gamma
\rightarrow \phi_i \phi_j}
\big|^2
&=&
\dfrac{
M_{\phi_i}^4
\; \hat{s}^2
+
(
M_{\phi_i}^2
M_{\phi_j}^2
-
\hat{t}\; \hat{u}
)^2
}
{8\; \hat{s}^2}
\,
\big|
F_{33}
\big|^2
\nonumber\\
&&
-
\dfrac{
M_{\phi_i}^2 \hat{s}
}
{4}
\,
\mathcal{R}e
\Big[
F_{33}
\cdot
\big(
F_{12}
\big)^*
\Big]
+
\dfrac{
\hat{s}^2
}
{8}
\,
\big|
F_{12}
\big|^2
.
\nonumber
\end{eqnarray}
\section{Applications}             
We are going to present
the phenomenological results for this work.
We consider two typical applications 
in this work which are the SM and 
Zee-Babu Model. We work in 
the $G_{\mu}$-scheme
and use the input parameters in the SM
as same as our previous papers
~\cite{Phan:2024jbx, Phan:2024zus}. 
In the phenomenological results for Zee-Babu 
model, the parameter spaces will be taken 
appropriately in the next subsections. 
\subsection{Numerical tests}
We first perform numerical tests for the calculations.
The form factors must be the ultraviolet finiteness
and the infrared finiteness. Furthermore,
one-loop amplitude must follow 
the ward identity due to the initial
photon. Numerical checks for
the ultraviolet finiteness,
infrared finiteness for one-loop
form factors are shown in 
Table~\ref{F1233UVIR}. For this test,
we set the couplings 
$\lambda_{H\Phi}=+2$, $\lambda_{K\Phi}=-1$ 
and the charged scalar masses as 
follows: $M_{H^\pm}=500$ GeV, 
$M_{K^{\pm\pm}}=1000$
GeV. Additional, one 
sets $\hat{s} =1500^2$ GeV, 
$\hat{t} = -200^2$ GeV$^2$.
By varying the parameters
$C_{UV},~\lambda^2$ in a wide range,
we find that the results
are good stability up to last digits
(over $15$ digits at the amplitude level).
\begin{table}[H]
\centering
\begin{tabular}
{|l|l|l|}
\hline\hline
$\big(
C_{UV},
\lambda^2
\big)$
&
$F_{12}$
&
$F_{33}$
\\
\hline \hline
$(0,1)$
&
$-4.432377929637615
\cdot 10^{-10}$
&
$-1.274104987282513
\cdot 10^{-8} $
\\
&
$+ 7.427714924247333
\cdot 10^{-10} \, i$
&
$+ 2.439535005865118
\cdot 10^{-7} \, i$
\\ \hline
$(10^2, 10^4)$
&
$-4.432377929637721
\cdot 10^{-10}$
&
$-1.274104987282583
\cdot 10^{-8}$
\\
&
$+ 7.427714924247333
\cdot 10^{-10} \, i$
&
$+ 2.439535005865118
\cdot 10^{-7} \, i$
\\ \hline
$(10^4, 10^8)$
&
$-4.432377929637489
\cdot 10^{-10}$
&
$-1.274104987282561
\cdot 10^{-8}$
\\
&
$+ 7.427714924247333
\cdot 10^{-10} \, i$
&
$+ 2.439535005865118
\cdot 10^{-7} \, i$
\\
\hline\hline
\end{tabular}
\caption{
\label{F1233UVIR}
Numerical checks for the UV- and IR- finiteness
for one-loop form factor $F_{12}$, $F_{33}$. 
For this test,
we set the couplings 
$\lambda_{H\Phi}=+2$, $\lambda_{K\Phi}=-1$ 
and the charged scalar Higgs masses as 
follows: $M_{H^\pm}=500$ GeV, $M_{K^{\pm\pm}}=1000$
GeV. Additional, one sets $\hat{s} =1500^2$ GeV$^2$, 
$\hat{t} = -200^2$ GeV$^2$.}
\end{table}
Due to the initial photons taking part 
in the scattering processes, 
one-loop amplitude must follow
the ward identity. The identity is
verified numerically in this work. This can
be done as follows. We collect analytic results for 
one-loop form factors $F_{00}$, $F_{12}$,
$F_{23}$ and $F_{33}$ independent way. 
All the relations in Eqs.~\ref{ward0},
\ref{ward1},
\ref{ward2},
\ref{ward3} are confirmed
numerically. For illustrations,
we show numerical check for 
the following relation
\begin{eqnarray}
\label{FTT} 
F_{TT}
&=&
\dfrac{\hat{u} - M_{\phi_i}^2}{2}
\cdot 
\dfrac{\hat{t} - M_{\phi_i}^2}{\hat{s} }
\cdot 
F_{33}
-
\dfrac{\hat{s} }{2}
\cdot 
F_{12}.
\end{eqnarray}
One then verifies numerically 
the equation that $F_{TT}=F_{00}$. 
In Table~\ref{wardMUP}, we show
numerical results for the test.
In this Table, we fix the value
of 
$(\hat{t},
\lambda_{H\Phi},
\lambda_{K\Phi})$ in the first column.
The results of
form factors $F_{TT}$ obtaining 
from $F_{12}$ and $F_{33}$ as shown in
Eq.~\ref{FTT} are shown in the second column. 
The last column is presented for the results
of $F_{00}$. The relation
$F_{00}=F_{TT}$ is confirmed
numerically in this Table. 
From the data, we find that
the results are good stability 
over $12$ digits. 
\begin{table}[H]
\centering
\begin{tabular}
{|l|l|l|}
\hline\hline
&
$F_{12}$
&
$-$
\\
$\big(
\hat{t},
\lambda_{H\Phi},
\lambda_{K\Phi}
\big)$
&
$F_{33}$
&
$-$
\\
&
$F_{TT}$ [given in Eq.~\ref{FTT}]
&
$F_{00}$
\\
\hline \hline
&
$-3.506673731227688
\cdot 10^{-9}$
&
$-$
\\
&
$- 9.48914729381836
\cdot 10^{-9} \, i$
&
\\
&
&
\\
$\big(
+300^2
,
+1.5
,
+0.5
\big)$
&
$-2.073875880451545
\cdot 10^{-7}$
&
$-$
\\
&
$+ 3.018956669634884
\cdot 10^{-8} \, i$
&
\\
&
&
\\
&
$-0.01566372086843976$
&
$-0.01566372086843974$
\\
&
$+ 0.01122720571181518 \, i$
&
$+ 0.01122720571181515 \, i$
\\  \hline
&
$-2.473869485576741
\cdot 10^{-9}$
&
$-$
\\
&
$- 1.389444047877029
\cdot 10^{-8} \, i$
&
\\
&
&
\\
$\big(
+300^2
,
-1.5
,
+0.5
\big)$
&
$-2.073875880451545
\cdot 10^{-7}$
&
$-$
\\
&
$+ 3.018956669634884
\cdot 10^{-8} \, i$
& \\
&
&
\\
&
$+0.005051569661633232$
&
$+0.005051569661633237$
\\
&
$+ 0.0118652133892367 \, i$
&
$+ 0.0118652133892366 \, i$
\\
\hline
&
$-3.691602634060829
\cdot 10^{-9}$
&
$-$
\\
&
$- 1.389444047877029
\cdot 10^{-8} \, i$
&
\\
&
&
\\
$\big(
+300^2
,
-1.5
,
-0.5
\big)$
&
$-2.073875880451545
\cdot 10^{-7}$
&
$-$
\\
&
$+ 3.018956669634884
\cdot 10^{-8} \, i$
&
\\
&
&
\\
&
$+0.03914990554641669$
&
$+0.03914990554641665$
\\
&
$+ 0.0118652133892367
\, i$
&
$+ 0.0118652133892364
\,i$
\\
\hline
&
$+1.601959722985257
\cdot 10^{-9}$
&
$-$
\\
&
$- 3.179364680100479
\cdot 10^{-11} \, i$
&
\\
&
&
\\
$\big(
-300^2
,
+1.5
,
+0.5
\big)$
&
$-1.286260135515678
\cdot 10^{-8}$
&
$-$
\\
&
$+ 1.785310771633653
\cdot 10^{-7} \, i$
&
\\
&
&
\\
&
$-0.03002288877158604$
&
$-0.03002288877158601$
\\
&
$+ 0.01073513714169602 \, i$
&
$+ 0.01073513714169605 \, i$
\\
\hline
&
$+1.417030820152116
\cdot 10^{-9}$
&
$-$
\\
&
$- 4.437086831752933
\cdot 10^{-9} \, i$
&
\\
&
&
\\
$\big(
-300^2
,
-1.5
,
-0.5
\big)$
&
$-1.286260135515678
\cdot 10^{-8}$
&
$-$
\\
&
$+ 1.785310771633653
\cdot 10^{-7} \, i$
&
\\
&
&
\\
&
$+0.02479073764327042$
&
$+0.02479073764327045$
\\
&
$+ 0.01137314481911753 \, i$
&
$+ 0.01137314481911754 \, i$
\\
\hline\hline
\end{tabular}
\caption{
\label{wardMUP}
The ward identity check, confirming the relation
$F_{TT} =F_{00}$, for the case of $M_{H^\pm}=500$ GeV, $M_{K^{\pm\pm}} =1000$ GeV, $\hat{s} = 1500^2$ GeV$^2$ and varying of $\hat{t},~\lambda_{H\Phi},~\lambda_{K\Phi}$.}
\end{table}
\subsection{Standard model}
In this case, we have no contributions of
charged Higgs as well as the 
mixing of charged Higgs with $W$
bosons in the loop. By replacing the 
general couplings to the SM's couplings 
respectively, we arrive at the analytical 
results for the process 
$\gamma \gamma \rightarrow hh$ in the SM.
Additionally, cross sections for
the process $\gamma \gamma \rightarrow hh$
have calculated in the SM in many previous
works. For example, we take 
Ref.~\cite{Chiesa:2021qpr}
in which the results have shown
in $\alpha$-scheme,
$\alpha = 1/137.035999 084(21)$,
cross-section for
$\gamma \gamma \rightarrow hh$ at
$\sqrt{\hat{s}_{\gamma\gamma}}= 470$ GeV
is about $\sim 0.28$ fb.
In our work, the result is 
corresponding
to $0.275$ fb. This value is good
agreement with the result in 
Ref.~\cite{Chiesa:2021qpr}.
We note that part of the 
results $\gamma\gamma \rightarrow hh$
in the SM are shown together with 
$\gamma\gamma \rightarrow hh,~hH,~HH$
in the Inert Higgs Doublet model, 
Two Higgs Doublet Models as 
in our previous paper~\cite{Phan:2024vfy}. 
In this paper,
we are not going to present 
the phenomenological
results for $\gamma\gamma \rightarrow hh$
in the SM in further. 
\subsection{Zee-Babu model}  
The Zee-Babu model is cosnidered as 
another typical application. We first 
review briefly the Zee-Babu model based on 
the papers~\cite{Zee:1985id, Babu:1988ki}. 
The model is added to two complex scalars which 
are a singly charged scalar $H^{\pm}$ 
and a doubly charged scalar $K^{\pm\pm}$
with the quantum numbers 
as $H^{\pm}\sim(1,1,\pm1)$,
$K^{\pm\pm}\sim(1,1,\pm2)$, respectively. 
The Lagrangian of the Zee-Babu 
model constructed as follows:
\begin{eqnarray}
\mathcal{L}_{ZB}&=&\mathcal{L}_{SM}
+\mathcal{L}_{K}^{ZB}
- \mathcal{V}_{ZB}
+\mathcal{L}_{Y}^{ZB}.
\end{eqnarray}
In the Lagrangian, 
the kinetic term for the scalar fields 
$K$ and $H$ is expressed explicitly
by
\begin{eqnarray}
\mathcal{L}_{K}^{ZB}&=&(D_{\mu}H)^{\dagger}
(D^{\mu}H)+(D_{\mu}K)^{\dagger}(D^{\mu}K)
\end{eqnarray}
with the following covariant derivatives given
$D_{\mu}=\partial_\mu+ig_{H/K}Y_{H/K} B_{\mu}$.
The electromagnetic charge operator is 
given $Q{H/K}=T_L^3+Y$
in which the hypercharge is taken
$Y_{H^\pm}=\pm{1}$ $(Y_{K^{\pm\pm}} 
=\pm{2})$, respectively. Two additional 
scalars don't carry the color or weak isopin. 
As a result, additional scalar particles 
only interact with the $U(1)_Y$ group.

The Zee-Babu scalar potential is taken 
the form of
\begin{eqnarray}
\mathcal{V}_{ZB}&=& \mu_1^2H^\dagger{H}
+\mu^2_2K^\dagger{K}+\lambda_H(H^\dagger{H})^2
+\lambda_K(K^\dagger{K})^2
+\lambda_{HK}(H^\dagger{H})(K^\dagger{K})
\nonumber\\
&&
+(\mu_L{HHK^\dagger}
+\mu^{\dagger}_L{H^{\dagger}H^{\dagger}K}) 
+\lambda_{K\Phi}(K^{\dagger}K)(\Phi^\dagger\Phi)
+\lambda_{H\Phi}(H^{\dagger}H)(\Phi^\dagger\Phi).
\end{eqnarray}
For the EWSB, the Higgs doublet field of SM 
$\Phi$ is parameterized as follows:
\begin{eqnarray}
\Phi&=&
\left(\begin{array}{c}
\chi^{\pm}  \\
\dfrac{v+h+i\chi_0 }{\sqrt{2}}
\end{array}\right)
\end{eqnarray}
with $v\sim 246.$ GeV 
for coinciding with the SM case.
The particles $h,~\chi_0$ 
and $\chi^{\pm}$ are corresponding to 
Higgs bosons of SM and neutral
Goldstone bosons, charged Goldstone
bosons. From the Zee-Babu 
scalar potential, the masses of 
$H^\pm$ and $K^{\pm\pm}$ 
can be collected as 
\begin{eqnarray}
M_{K^{\pm\pm}}^2=\mu_2^2+\frac{\lambda_{K\Phi}}{2}v^2, 
\quad  M_{H^{\pm}}^2=\mu_1^2+\frac{\lambda_{H\Phi}}{2}v^2.
\end{eqnarray}

The Yukawa lagragian $\mathcal{L}_{ZB}$ part which 
describes the interactions between the SM leptons to 
the additional scalar fields $K$ and $H$ are 
given by
\begin{eqnarray}
\mathcal{L}_{Y}^{ZB}
&=&f_{ij}\overline{\tilde{L^i}}L^{j}H^\dagger
+g_{ij}\overline{(e_R^c)^i}e_R^jK^\dagger
+\overline{f}_{ij}(\overline{\tilde{L^i}}L^{j})^{\dagger}H
+\overline{g}_{ij}(\overline{(e_R^c)^i}e_R^j
)^{\dagger}K.
\end{eqnarray}
In the Yukawa sector $\mathcal{L}_{Y}^{ZB}$,
we denote that 
$L^i=(\nu_L^i,e_L^i)$ and 
$\tilde{L}^i=i\sigma_2(L^{*})^i$ 
with generation index $i=1,2,3$. 
We have also noted as 
$L\sim(1,2,{\mp}1/2)\sim(\nu_L,\ell_L)^T$ 
and $\ell_R \sim(1,1,{\mp}1)$. 
The $3\times{3}$ Yukawa coupling matrices 
$f_{ij}$ and $g_{ij}$ are anti-symmetric 
$(f_{ij}=-f_{ji})$ and 
symmetric $(g_{ij}=g_{ji})$, 
respectively. 

All the additional couplings from the Zee-Babu
model are listed in Table~\ref{ZBcouplings}.
Several couplings in this Table are taken part
in to the processes under investigation. 
\begin{table}[H]
 \centering
\begin{tabular}{|l|l|l|}
\hline\hline
Vertices 
& Notations 
& Couplings
\\
\hline\hline
$Z_{\mu}H^{\pm}(K^{\pm\pm})H^{\mp}(K^{\mp\mp})$ 
& $g_{ZHH(KK)}\times$
& $ie\frac{s_W}{c_W}Q_{H(K)}\times$ \\
&
$\times \Big(
p_{H^\pm(K^{\pm\pm})}
-p_{H^\mp(K^{\mp\mp})}
\Big)_{\mu}$ 
&
$\times
\Big(p_{H^\pm(K^{\pm\pm})}-p_{H^\mp(K^{\mp\mp})}
\Big)_{\mu}$ 
\\ \hline
$A_{\mu}H^{\pm}
(K^{\pm\pm})H^{\mp}(K^{\mp\mp})$ 
& $
g_{AH^{\pm}H^{\mp}
(K^{\pm\pm}K^{\mp\mp})}\times$ 
& $-ieQ_{H(K)} \times$\\
&
$\times
\Big(
p_{H^\pm(K^{\pm\pm})}
-p_{H^\mp(K^{\mp\mp})}
\Big)_{\mu}$
&
$\times
\Big(
p_{H^\pm(K^{\pm\pm})}
-
p_{H^\mp(K^{\mp\mp})}
\Big)_{\mu}$ 
\\
\hline
$A_{\mu}A_{\nu}H^{\pm}H^{\mp}
(K^{\pm\pm}K^{\mp\mp})$ 
& $g_{AAH^{\pm}H^{\mp}
(K^{\pm\pm}K^{\mp\mp})}
\cdot g_{\mu\nu}$ 
& $ie^2Q_{H(K)}^2
\cdot g_{\mu\nu}$
\\ 
\hline
$Z_{\mu}Z_{\nu}
H^{\pm}H^{\mp}(K^{\pm\pm}K^{\mp\mp})$ 
& $g_{ZZH^{\pm}H^{\mp}
(K^{\pm\pm}K^{\mp\mp})}
\cdot g_{\mu\nu}$ 
& $ie^2
\left(
\dfrac{s_W^2}{c_W^2}Q_{H(K)}^2
\right)
\cdot g_{\mu\nu}$ 
\\
\hline 
$A_{\mu}Z_{\nu}H^{\pm}
H^{\mp}
(K^{\pm\pm} 
K^{\mp\mp})$ 
& $g_{AZH^{\pm}H^{\mp}
(K^{\pm\pm}K^{\mp\mp})}
\cdot g_{\mu\nu}$ 
& $-ie^2
\left(
\dfrac{s_{2W}}
{c_W^2}Q_{H(K)}^2
\right)
\cdot g_{\mu\nu}$
\\
\hline
$H^{\pm}H^{\pm}K^{\mp\mp}$ 
& $g_{H^{\pm}H^{\pm}K^{\mp\mp}}$ 
& $-i\mu_L$ 
\\
\hline
$hH^{\mp}H^{\pm}$ 
& $g_{hH^{\mp}H^{\pm}}$ 
& $-iv\lambda_{H\Phi}
=i\dfrac{2(\mu_1^2-M_{H^\pm}^2)}{v}$
\\
\hline
$hK^{\mp\mp}K^{\pm\pm}$ 
& $g_{hK^{\mp\mp}K^{\pm\pm}}$ 
& $-iv\lambda_{K\Phi}
=i\dfrac{2(\mu_2^2-M_{K^{\pm\pm}}^2)}{v}$
\\ 
\hline
$hhH^{\mp}H^{\pm}$ 
& $g_{hhH^{\mp}H^{\pm}}$ 
& $-i\lambda_{H\Phi}
=i\dfrac{2(\mu_1^2-M_{H^\pm}^2) }{v^2}
$
\\
\hline
$hhK^{\mp\mp}K^{\pm\pm}$ 
& $g_{hhK^{\mp\mp}K^{\pm\pm}}$ 
& $-i\lambda_{K\Phi}
=i\dfrac{2(\mu_2^2-M_{K^{\pm\pm} }^2)}{v^2}
$
\\ 
\hline
$H^{\pm}H^{\mp}K^{\mp\mp}K^{\pm\pm}$ 
& $g_{H^{\pm}H^{\mp}K^{\mp\mp}K^{\pm\pm}}$ 
& $-i\lambda_{HK}$ 
\\
\hline
$H^{\pm}H^{\mp}\chi^{\mp}\chi^{\pm}$ 
& $g_{H^{\pm}H^{\mp}\chi^{\mp}\chi^{\pm}}$ 
& $-i\lambda_{H\Phi}
=i\dfrac{2(\mu_1^2-M_{H^\pm}^2)}{v^2}$ 
\\
\hline
$K^{\pm\pm}K^{\mp\mp}\chi^{\mp}\chi^{\pm}$
& $g_{K^{\pm\pm}K^{\mp\mp}\chi^{\mp}\chi^{\pm}}$ 
& $-i\lambda_{K\Phi}
=i\dfrac{2(\mu_2^2-M_{K^{\pm\pm}}^2)}{v^2}$ 
\\
\hline\hline
\end{tabular}
\caption{\label{ZBcouplings} 
The additional 
couplings from the ZB 
model are listed in this Table.
Some of these ones are taken part
in to the processes under investigation.
}
\end{table}
It is stressed that the Yukawa 
Lagragian $\mathcal{L}_{ZB}$ part
presented in above is not related 
to the computed processes. As a result, 
the parameter space of the Zee-Babu model
for our next phenomenological studies 
is included as 
$\mathcal{P}_{\text{ZB}}
=\{M_{H^\pm}^2,M_{K^{\pm\pm}},
\lambda_{K\Phi}, 
\lambda_{H\Phi} \}$. 
For the updated parameter space in the Zee-Babu,
we refer the papers~\cite{Nebot:2007bc,Herrero-Garcia:2014hfa,
AristizabalSierra:2006gb,Alcaide:2017dcx,Ruiz:2022sct,
Jueid:2023qcf}.

We pay our attention to 
phenomenological
results for the Zee-Babu model.
To our knowledge, we emphasize that all phenomenological results presented
in the following paragraphs for 
the Zee-Babu model can be considered 
to be first results from this study.
First, 
cross-sections are
presented as functions
of center-of-mass (CoM) energies
($\sqrt{\hat{s}}$). For this plot,
we fix $M_{H^\pm}^2=400$ GeV, 
$M_{K^{\pm\pm}} =800$ GeV and 
$\lambda_{K\Phi}=\pm 0.7, 
\lambda_{H\Phi}=\pm 2$. 
In the following plot,
The CoM energies are vsaried from 
$500$ GeV to $2000$ GeV. The red
line shows for cross sections with
$\lambda_{K\Phi}=+0.7, \lambda_{H\Phi}=2$,
the blue line presents for the data with
$\lambda_{K\Phi}=+0.7, \lambda_{H\Phi}=-2$
and the green line is corresponding to 
the results for 
$\lambda_{K\Phi}=-0.7, \lambda_{H\Phi}=-2$
(and pink color line for the data at 
$\lambda_{K\Phi}=-0.7, \lambda_{H\Phi}=+2$). 
The black line is for the data of 
the process in the SM.
In general, we observe two peaks of 
cross sections at $\sqrt{\hat{s}}\sim 
2M_{H^{\pm} } = 800$ GeV and 
$\sqrt{\hat{s}}\sim 
2M_{K^{\pm\pm} }= 1600$ GeV.  
If we consider the same input 
configurations for both singly charged 
Higgs and doubly charged Higgs
in the loop, e.g. the same values 
for the masses and the couplings. 
Approximately, one-loop amplitude 
with doubly charged Higgs internal 
lines may be estimated as $4$ 
times of the ones with singly 
charged Higgs in the loop.
It is because 
the couplings of $g_{A K^{\pm\pm}
K^{\mp\mp} } = 2g_{A H^{\pm}H^{\mp} } $. 
The contributions
of doubly chagred Higgs are domimant in 
comparison with 
the corresponding ones from singly 
charged Higgs. Examining the artributions 
from doubly chagred Higgs in the loop 
in further concrete, we find that 
the couplings $g_{h K^{\pm\pm}
K^{\mp\mp}}^2$ appear in the one-loop
box diagrams. However the couplings 
$g_{h K^{\pm\pm}K^{\mp\mp}}$ are only 
taken into account in one-loop 
triangle diagrams. As a result, 
the squared amplitudes 
of one-loop box diagrams 
may cancel with the ones 
from mixing of one-loop 
triangle diagrams and box diagrams when 
the couplings $g_{h K^{\pm\pm}K^{\mp\mp}}$
being negative values. This explains 
that cross-sections in the cases of 
$\lambda_{K\Phi}=+0.7, \lambda_{H\Phi}=2$
($\lambda_{K\Phi}=-0.7, \lambda_{H\Phi}=2$
as same reason)
tend to the SM case. Other cases, one finds
the large contributions from chagred scalars
in the loop around the peaks. 
\begin{figure}[H]
\centering
\begin{tabular}{c}
\hspace{-11cm}
$\hat{\sigma}_{hh}\times 10^{-4}$
[pb]
\\
\includegraphics[width=16cm, height=8cm]
{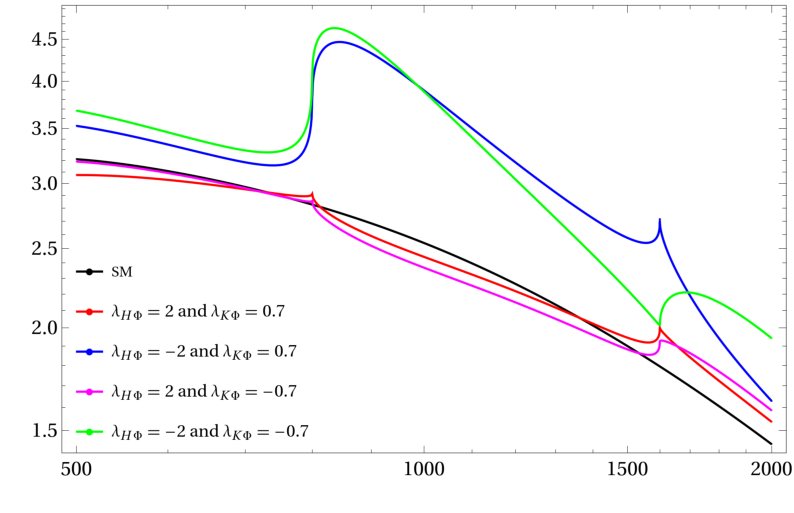}\\
\hspace{14cm} $\sqrt{\hat{s}}$ [GeV]
\end{tabular}
\caption{\label{ZBdata0}
Cross section as functions of 
C.o.M. 
$M_{H^\pm}^2=400$ GeV, $M_{K^{\pm\pm}} =800$ GeV
and $\lambda_{K\Phi}=\pm 0.7,~\lambda_{H\Phi}=\pm 2$. 
In the following plots,
we vary $\sqrt{\hat{s}} 
=500$ GeV to $\sqrt{\hat{s}} 
=2$ TeV.
}
\end{figure}
We next study enhancement factor
which are given by 
\begin{eqnarray}
 \mu_{hh}^{\textrm{ZB}}
 =
 \dfrac{\hat{\sigma}_{hh}^{\textrm{ZB}}
 }{\hat{\sigma}_{hh}^{\textrm{SM}} }
 (\sqrt{\hat{s}}, \mathcal{P}_{\text{ZB}})
\end{eqnarray}
over the parameters of the Zee-Babu model.

\begin{figure}[H]
\centering
\begin{tabular}{cc}
\hspace{-6cm}
$\mu_{hh}^{\textrm{ZB}}$
& 
\hspace{-6cm}
$\mu_{hh}^{\textrm{ZB}}$
\\
\includegraphics[width=8cm, height=6cm]
{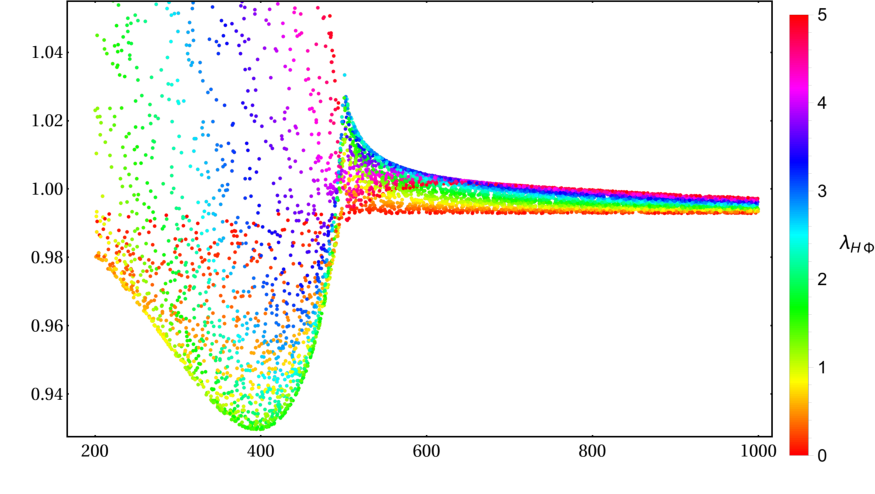}
&
\includegraphics[width=8cm, height=6cm]
{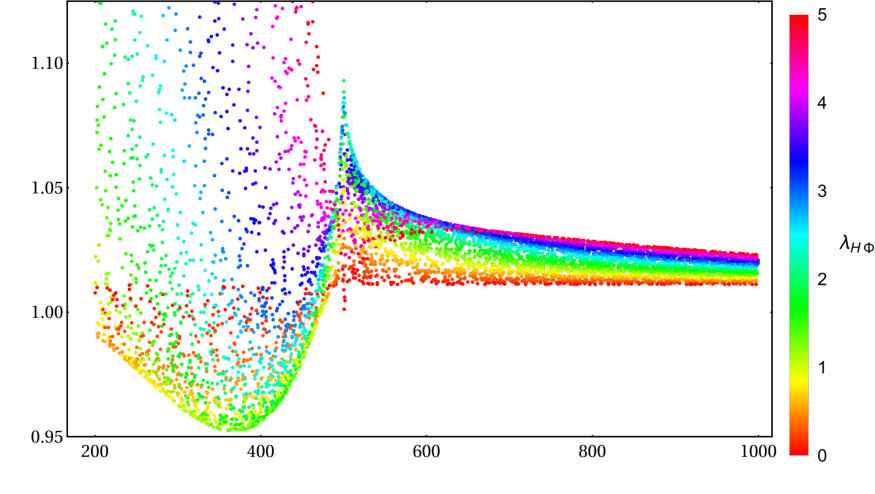}
\\
\hspace{5cm}
$M_{H^{\pm} }$ [GeV]
& 
\hspace{5cm}
$M_{H^{\pm} }$ [GeV]
\\
&
\\
\includegraphics[width=8cm, height=6cm]
{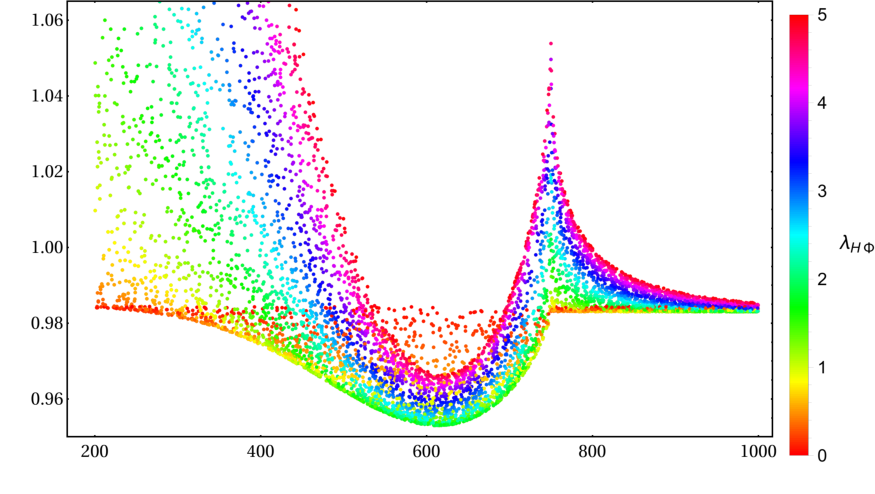}
&
\includegraphics[width=8cm, height=6cm]
{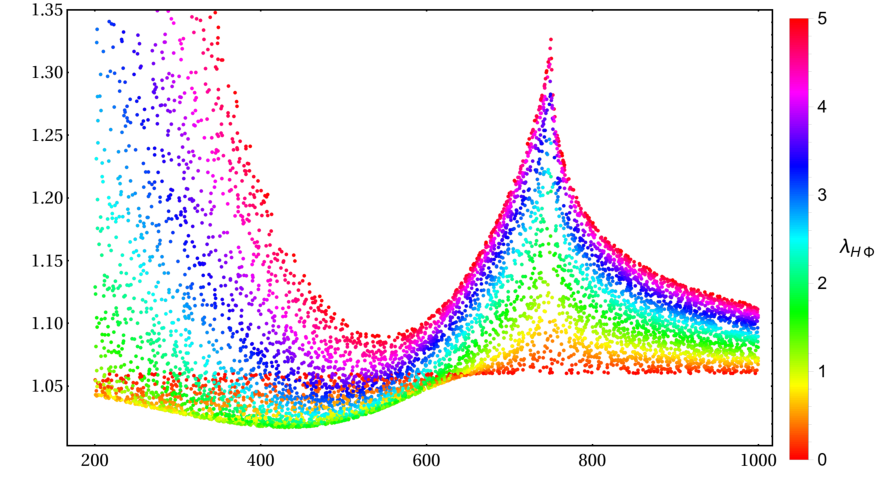}
\\
\hspace{5cm}
$M_{H^{\pm} }$ [GeV]
& 
\hspace{5cm}
$M_{H^{\pm} }$ [GeV]
\\
\end{tabular}
\caption{\label{ZBdata1}
The scatter plots as functions of 
$(M_{H^\pm}^2, \lambda_{H\Phi})$. 
In these plots, we vary $200$ GeV $\leq M_{H^\pm}
\leq 1000$ GeV and $0\leq \lambda_{H\Phi}\leq  5$. 
}
\end{figure}

In Fig.~\ref{ZBdata1}, the factors 
are scanned over the singly 
charged Higgs masses 
$M_{H^\pm}$ and $\lambda_{H\Phi}$.
In the scatter plots, we fix 
$M_{K^{\pm\pm}} =800$ GeV
and $\lambda_{K\Phi}=-0.7$
(left panel plots), 
$\lambda_{K\Phi}=+0.7$
(right panel plots). 
In the following plots
we set $\sqrt{\hat{s}} =1000$ GeV
(for all above plots) 
and $\sqrt{\hat{s}} =1500$ GeV  
(for below plots), respectively. 
We vary $200$ GeV 
$\leq M_{H^\pm} \leq 1000$ GeV and 
$0\leq \lambda_{H\Phi} \leq 5$. 
We find that the factors tend to $1$
(tend to the SM case)
when $\lambda_{H\Phi}\rightarrow 0$.
In this limit, the contributions of 
singly charged Higges are 
going to zero and because of the 
small contributions
of doubly charged Higgs due to 
the large value of 
$M_{K^{\pm\pm}}$
and the small 
value of the couplings
$\lambda_{K\Phi}$.
At $\sqrt{\hat{s}}=1$ TeV, 
we observe 
a narrove peak of producing 
two charged Higgses 
at $500$ GeV. The factors are sensitive 
with $\lambda_{H\Phi}$ 
in the $M_{H^\pm}$ regions 
of the below the peak. Above the peak 
region of $M_{H^{\pm}}$, 
the factor depends slightly on
$\lambda_{H\Phi}$ and its value
goes to $1$. It shows that the 
contributions of singly and 
doubly charged Higges
being small in the 
concerned regions. We note that 
for the case $\lambda_{K\Phi}=+0.7$
the factors are bigger than the ones in 
case of $\lambda_{K\Phi}=-0.7$. 
This can be explained
the same reasons as in Fig.~\ref{ZBdata0}.
We obtain the same behavor for the enhancement
factors at $\sqrt{\hat{s}}=1.5$ TeV. 

\begin{figure}[H]
\centering
\begin{tabular}{cc}
$\mu_{hh}^{\textrm{ZB}}$
$\quad\quad
(\lambda_{H\Phi}=- 0.7$, $\lambda_{K\Phi}=- 0.7)$
& 
$\mu_{hh}^{\textrm{ZB}}$
$\quad \quad
(\lambda_{H\Phi}=- 0.7$, $\lambda_{K\Phi}=+ 0.7)$
\\
\includegraphics[width=8cm, height=6cm]
{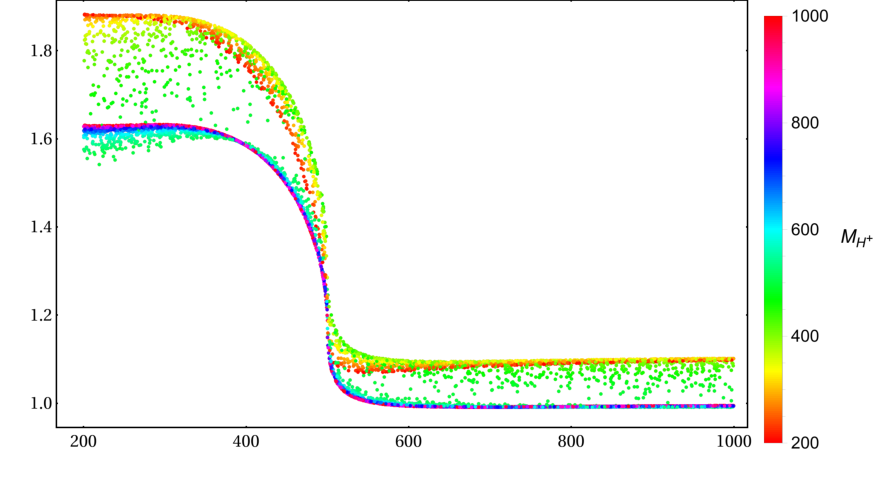}
&
\includegraphics[width=8cm, height=6cm]
{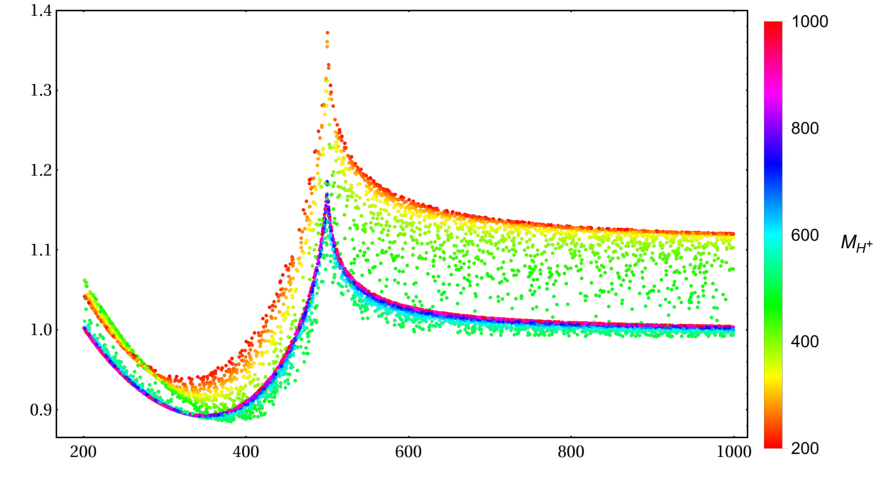}
\\
\hspace{5cm}
$M_{K^{\pm\pm} }$ [GeV]
& 
\hspace{5cm}
$M_{K^{\pm\pm} }$ [GeV]
\\
&
\\
$\mu_{hh}^{\textrm{ZB}}$
$\quad\quad
(\lambda_{H\Phi}=+ 0.7$, $\lambda_{K\Phi}=- 0.7)$
& 
$\mu_{hh}^{\textrm{ZB}}$
$\quad\quad
(\lambda_{H\Phi}=+ 0.7$, $\lambda_{K\Phi}=+ 0.7)$
\\
\includegraphics[width=8cm, height=6cm]
{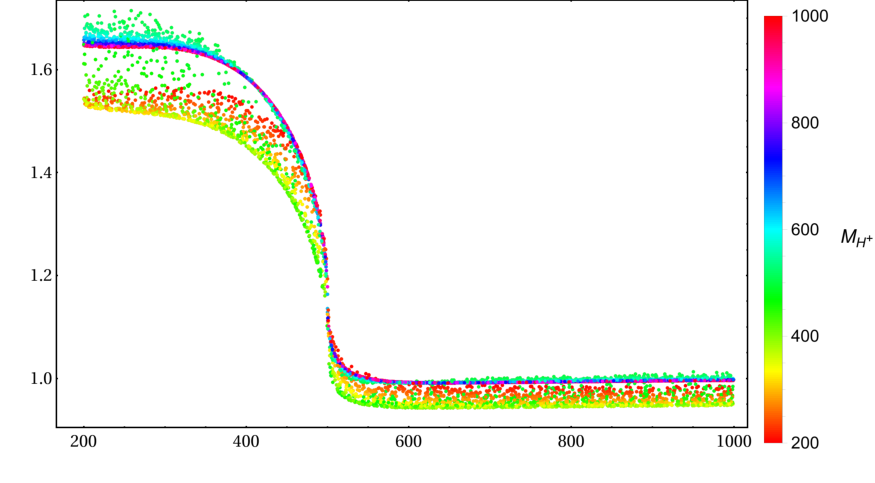}
&
\includegraphics[width=8.cm, height=6cm]
{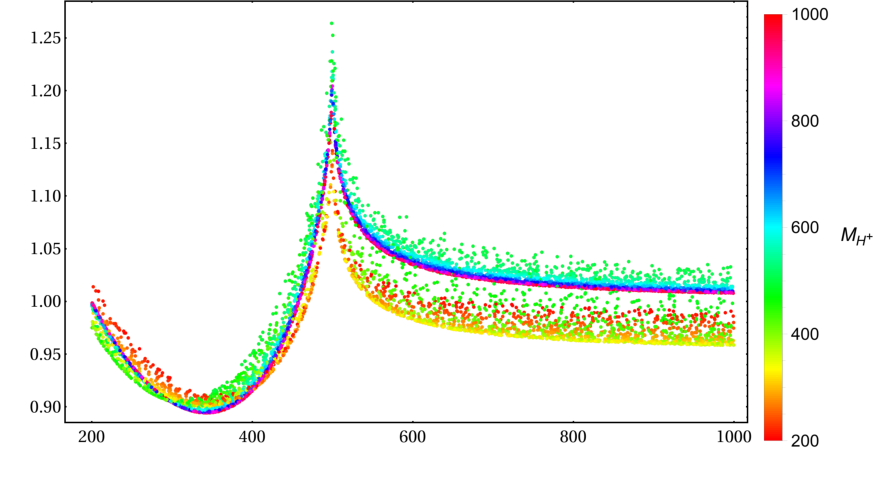}
\\
\hspace{5cm}
$M_{K^{\pm\pm} }$ [GeV]
& 
\hspace{5cm}
$M_{K^{\pm\pm} }$ [GeV]
\end{tabular}
\caption{\label{ZBdata2}
The scatter plots as functions of 
$(M_{H^\pm},~M_{K^{\pm\pm}})$
at $1$ TeV of CoM. 
In these plots, we vary  
$200$ GeV 
$\leq M_{H^\pm},~M_{K^{\pm\pm}} 
\leq 1000$ GeV. 
}
\end{figure}

\begin{figure}[H]
\centering
\begin{tabular}{cc}
$\mu_{hh}^{\textrm{ZB}}$
$\quad\quad
(\lambda_{H\Phi}=- 0.7$, $\lambda_{K\Phi}=- 0.7)$
& 
$\mu_{hh}^{\textrm{ZB}}$
$\quad \quad
(\lambda_{H\Phi}=- 0.7$, $\lambda_{K\Phi}=+ 0.7)$
\\
\includegraphics[width=8cm, height=6cm]
{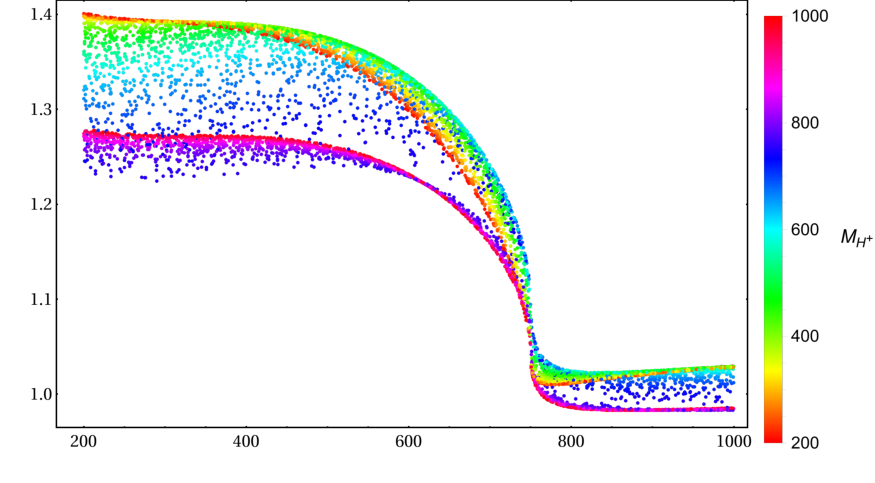}
&
\includegraphics[width=8cm, height=6cm]
{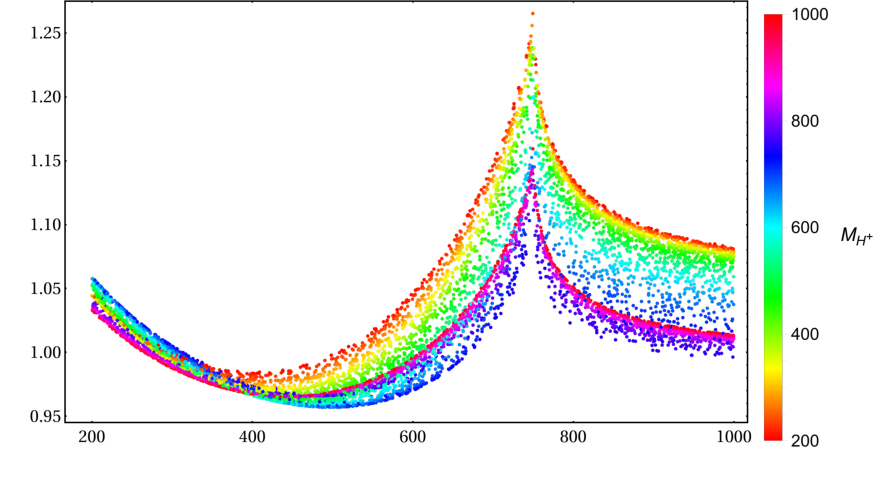}
\\
\hspace{5cm}
$M_{K^{\pm\pm} }$ [GeV]
& 
\hspace{5cm}
$M_{K^{\pm\pm} }$ [GeV]
\\
&
\\
$\mu_{hh}^{\textrm{ZB}}$
$\quad\quad
(\lambda_{H\Phi}=+ 0.7$, $\lambda_{K\Phi}=- 0.7)$
& 
$\mu_{hh}^{\textrm{ZB}}$
$\quad\quad
(\lambda_{H\Phi}=+ 0.7$, $\lambda_{K\Phi}=+ 0.7)$
\\
\includegraphics[width=8cm, height=6cm]
{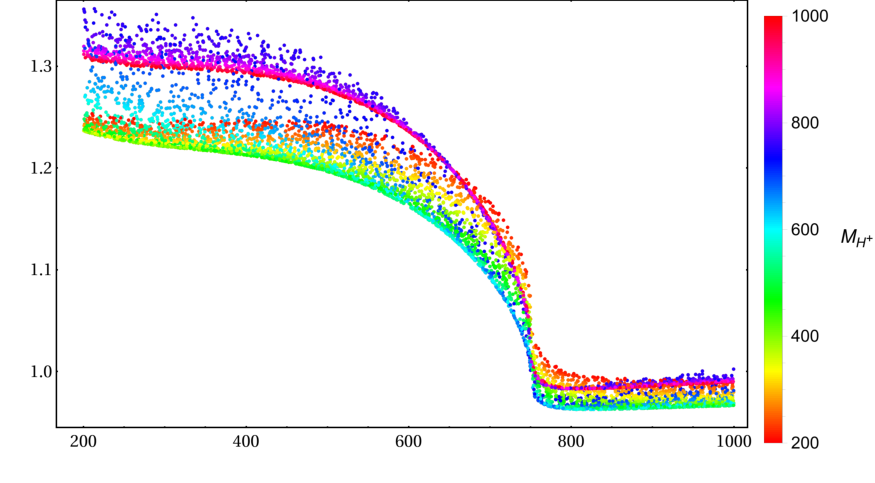}
&
\includegraphics[width=8cm, height=6cm]
{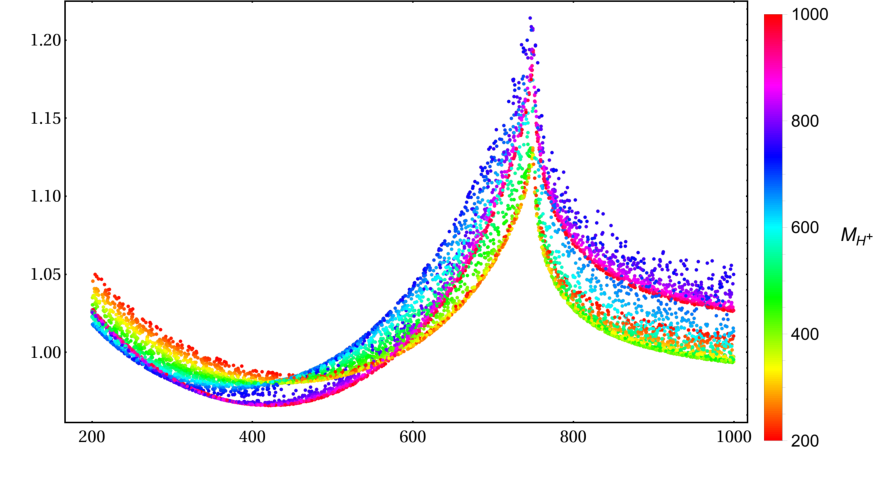}
\\
\hspace{5cm}
$M_{K^{\pm\pm} }$ [GeV]
& 
\hspace{5cm}
$M_{K^{\pm\pm} }$ [GeV]
\end{tabular}
\caption{\label{ZBdata3}
The scatter plots as functions of 
$(M_{H^\pm},~M_{K^{\pm\pm}})$
at $1.5$ TeV of CoM.
In these plots, we vary  
$200$ GeV 
$\leq M_{H^\pm},~M_{K^{\pm\pm}} 
\leq 1000$ GeV. 
}
\end{figure}

The investigations for 
the enhancement factors in the parameter 
space of 
$(M_{H^\pm},~M_{K^{\pm\pm}})$ are next 
considered. For this study, we take 
$\lambda_{K\Phi}=\lambda_{H\Phi}=\pm 0.7$. 
In the following plots, we vary $200$ GeV 
$\leq M_{H^\pm},~M_{K^{\pm\pm}} 
\leq 1000$ GeV at fixing 
$\sqrt{\hat{s}}=1$ TeV (for 
all plots~\ref{ZBdata2}) and 
$\sqrt{\hat{s}}=1.5$ TeV (for 
all plots~\ref{ZBdata3}). 
In general, the factors are inversly 
propotional to $M_{H^\pm},~M_{K^{\pm\pm}}$. 
For the cases of 
$\lambda_{H\Phi}= \mp 
0.7$, $\lambda_{K\Phi}=- 0.7$ 
(for left pannel Figures)
the enhancement are domimant at low mass regions
of $M_{H^\pm},~M_{K^{\pm\pm}}$. We have no peak 
of the factor around $500$ GeV in this case. 
The factors tend to $1$ beyond 
the regions of $M_{K^{\pm\pm}}> 500$ GeV. 
We next coments on the cases of 
$\lambda_{H\Phi}=\mp 0.7$,
$\lambda_{K\Phi}=+ 0.7$
(for right pannel Figures), 
the factors are suppressed
in the low mass regions of $M_{H^\pm},~M_{K^{\pm\pm}}$.
They develop to the peak 
$2M_{K^{\pm\pm}}=2M_{H^\pm}=500$ GeV. The factors 
are in the range of $[\sim 1.0, \sim 1.15]$
beyond the peak regions.
The enhancement factors are examined
in the parameter space of 
$(M_{H^\pm},~M_{K^{\pm\pm}})$
at $1.5$ TeV of CoM. 
We observe the same 
behavor of the factors
as previous $1$ TeV of CoM. 
In both CoM energies mentioned, 
the factors in the case of 
$(\lambda_{H\Phi}=+ 0.7$, 
$\lambda_{K\Phi}=+ 0.7)$ are smallest 
in comparison with other cases. 
It is explained as
the data in Fig.~\ref{ZBdata0}.
\section{Conclusions} 
A general one-loop formulas 
for loop-induced processes
$\gamma \gamma\rightarrow
\phi_i\phi_j$ with CP-even Higges 
$\phi_i,\phi_j
= h,~H_j$
which are valid for a class of 
Higgs Extensions of the Standard Models,
e.g. Inert Doublet Higgs Models, Two
Higgs Doublet Models, Zee-Babu 
models as well as Triplet Higgs Models, 
etc, are presented in the paper. 
Analytic expressions for one-loop
form factors are written in terms of 
the basic scalar one-loop 
functions. The scalar functions are output
in the packages~{\tt LoopTools} and 
{\tt Collier}. Physical results are 
hence evaluated numerically by using 
one of the mentioned packages. 
Analytic results are tested
by several checks such as
the ultraviolet finiteness,
infrared finiteness of the one-loop 
amplitudes. 
Furthermore, the amplitudes
also obey the so-called ward identity
due to the on-shell initial photons.
This identity is also verified
numerically in the works.
Additionally, both the packages
{\tt LoopTools} and {\tt Collier}
are used for cross-checking for
the final results before generating physical
results.  In the applications, we show 
phenomenological results for Zee-Babu model
as a typical example in this reference.
Production cross-section for the processes
$\gamma \gamma\rightarrow hh$ scanned 
over the masses of singly charged Higgs, 
doubly chagred Higges as well as
their couplings to 
SM-like Higgs are studied. 
\\

\noindent
{\bf Acknowledgment:}~
This research is funded by Vietnam
National Foundation for Science and
Technology Development (NAFOSTED) under
the grant number $103.01$-$2023.16$.
Khiem Hong Phan and Dzung Tri Tran
express their gratitude to all the
valuable support from Duy Tan
University, for the 30th anniversary
of establishment (Nov. 11, 1994 -
Nov. 11, 2024) towards "Integral,
Sustainable and Stable Development".

\section*{Appendix $A$: Tensor reduction for
one-loop integrals }          
We apply tensor reduction
method developed in Ref.~\cite{Denner:2005nn}
for this computation. The method is described 
briefly in the appendix. Following the 
technique, tensor one-loop integrals rank $P$
with $N$-external legs can be decomposed into
the basic scalar one-loop 
functions with $N\leq 4$ (they are labeled 
as $A_0$, $B_0$, $C_0$, $D_0$). Definition
of tensor integrals with rank $P$ (taking  
$N\leq 4$ external legs for examples)
is
\begin{eqnarray}
\{A; B; C; D\}^{\mu_1\mu_2\cdots \mu_P}= (\mu^2)^{2-d/2}
\int \frac{d^dk}{(2\pi)^d}
\dfrac{k^{\mu_1}k^{\mu_2}\cdots 
k^{\mu_P}}{\{D_1;~D_1 D_2;~D_1D_2D_3;
~D_1D_2D_3D_4\}}.
\end{eqnarray}
In this formula,
$D_j$ ($j=1,\cdots, 4$) 
are the inverse Feynman
propagators
\begin{eqnarray}
D_j &=&
(k+ q_j)^2 -m_j^2 +i\rho,
\end{eqnarray}
$q_j = \sum\limits_{i=1}^j p_i$,~$p_i$ 
are the external momenta, 
$m_j$ are internal masses in the loops.
One-loop integrals are 
handled in the space-time dimension
$d=4-2\varepsilon$. The parameter
$\mu^2$ plays a key role of a 
renormalization scale. 
Explicit reduction formulas 
for one-loop one-, two-, three- 
and four-point tensor integrals 
up to rank $P=3$~\cite{Denner:2005nn}
are presented as follows. For 
one-loop one-, two-, three-point tensor integrals, 
one has
\begin{eqnarray}
A^{\mu}        &=& 0, \\
A^{\mu\nu}     &=& g^{\mu\nu} A_{00}, \\
A^{\mu\nu\rho} &=& 0,\\
B^{\mu}        &=& q^{\mu} B_1,\\
B^{\mu\nu}     &=& g^{\mu\nu} B_{00} + q^{\mu}q^{\nu} B_{11}, \\
B^{\mu\nu\rho} &=& \{g, q\}^{\mu\nu\rho} B_{001}
+ q^{\mu}q^{\nu}q^{\rho} B_{111}, 
\end{eqnarray}
and 
\begin{eqnarray}
C^{\mu}        &=& q_1^{\mu} C_1 + q_2^{\mu} C_2
 = \sum\limits_{i=1}^2q_i^{\mu} C_i,
\\
C^{\mu\nu}    &=& g^{\mu\nu} C_{00}
 + \sum\limits_{i,j=1}^2q_i^{\mu}q_j^{\nu} C_{ij},
\\
C^{\mu\nu\rho} &=&
\sum_{i=1}^2 \{g,q_i\}^{\mu\nu\rho} C_{00i}+
\sum_{i,j,k=1}^2 q^{\mu}_i q^{\nu}_j q^{\rho}_k C_{ijk}.
\end{eqnarray}
For one-loop four-point  tensor 
functions, 
the reduction formulas are given 
\begin{eqnarray}
D^{\mu}        &=& q_1^{\mu} D_1 + q_2^{\mu} D_2 + q_3^{\mu}D_3
 = \sum\limits_{i=1}^3q_i^{\mu} D_i, \\
 D^{\mu\nu}    &=& g^{\mu\nu} D_{00}
 + \sum\limits_{i,j=1}^3q_i^{\mu}q_j^{\nu} D_{ij},
\\
D^{\mu\nu\rho} &=&
	\sum_{i=1}^3 \{g,q_i\}^{\mu\nu\rho} D_{00i}+
	\sum_{i,j,k=1}^3 q^{\mu}_i q^{\nu}_j q^{\rho}_k D_{ijk}.
\end{eqnarray}
The tensor $\{g, q_i\}^{\mu\nu\rho}$~\cite{Denner:2005nn} 
is given by
$\{g, q_i\}^{\mu\nu\rho} = g^{\mu\nu} q^{\rho}_i
+ g^{\nu\rho} q^{\mu}_i + g^{\mu\rho} q^{\nu}_i$.
The scalar Passarino-Veltman 
functions (PV-functions)~\cite{Denner:2005nn}
are $A_{00}, B_1, \cdots, D_{333}$
in the right hand sides. The PV-functions 
are calculated in terms of the basic 
scalar one-loop functions with $N\leq 4$, 
e.g. $A_0$-, $B_0$-, $C_0$- and 
$D_0$- scalar functions which are
implemented into~{\tt LoopTools}
~\cite{Hahn:1998yk} and
{\tt Collier}~\cite{Denner:2016kdg}
for numerical computations.
\section*{Appendix $B$: Analytic results
for one-loop form factors}
In this appendix,
we show analytic results for one-loop 
form factors given in the equations
~\ref{form-factor12},~\ref{form-factor33}.
In the analytic expressions, we use 
the following kinematic variables
as 
\begin{eqnarray}
 x_{t(u)} &=& 
 \dfrac{\hat{t}(\hat{u})}{\hat{s} },
 \quad 
 x_{\phi_{i,j,k}} =
 \dfrac{M_{\phi_{i,j,k}}^2 }
 {\hat{s} },
 \\
 x_{f} &=&
 \dfrac{m_f^2}{\hat{s}},
 \quad 
 x_{W} = \dfrac{M_W^2} {\hat{s} },
 \quad 
 x_{S} = \dfrac{M_S^2}{\hat{s} }
\end{eqnarray}
for $S\equiv S^Q \equiv H^{\pm}$, 
$K^{\pm\pm}$ in the below results.
We first arrive at
the factors $F_{12,f/W/S}^{\text{Trig}}$
calculated from one-loop triangle 
with connecting to $\phi_k^*$-poles. 
The factors are written in terms of 
scalar one-loop three-point 
functions $C_0$. We first take into account
all fermions propagating in the loop, 
as ploted in Fig.~\ref{feynG1} ($G_1$),
the factors are given
explicitly
\begin{eqnarray}
F_{12,f}^{\text{Trig}}
&=&
\dfrac{
e^2 Q_f^2
}{4 \pi ^2}
\; N_C^f
\Big[
4 x_f
+
2 m_f^2
\big(
4 x_f-1
\big) 
C_0(0,\hat{s},0;
m_f^2,m_f^2,m_f^2)
\Big].
\end{eqnarray}
Where $Q_f~(N_C^f)$ is denoted for 
charged (color) quantum number 
of the corresponding fermion $f$. 

One next considers one-loop diagrams with
$W$ boson exchanged in the loop in connecting
with $\phi_k^*$-poles ($G_2$), as shown in
Fig.~\ref{feynG1}. The corresponding factors
are given by
\begin{eqnarray}
F_{12,W}^{\text{Trig}}
&=&
\dfrac{e^2}{8\pi^2\; M_W^2}
\Big[
x_{\phi_k}
+
6 x_W
+
2M_W^2
\big(
x_{\phi_k}
+
6 x_W
-
4
\big) 
C_0(0,\hat{s},0;M_W^2,M_W^2,M_W^2)
\Big].
\end{eqnarray}
Furthermore, considering singly 
(as well as doubly)
charged Higgses in the loop, 
as plotted in Fig.~\ref{feynG1} ($G_3$), 
the respective factors are 
collected as
\begin{eqnarray}
F_{12,S}^{\text{Trig}}
&=&
-
\dfrac{
e^2 Q_S^2
}{4 \pi ^2
\; \hat{s}
}
\Big[
1
+
2 M_S^2
\,
C_0(0,\hat{s},0; M_S^2,M_S^2,M_S^2)
\Big].
\end{eqnarray}
Here we note $Q_S$ for charged 
quantum numbers
for charged scalars ($S$).

We then arrive at the factors
contributing from the one-loop
box diagrams with 
fermions $f$, $W$-bosons and
singly (doubly) charged Higgses
$S$ internal lines, 
noted as $F_{ab, P}^{\text{Box}}$ 
with $ab \equiv 12, 33$ and 
$P=f,~W,~S$. For all fermions
propagating in the loop, 
the factors
are casted into the form of
\begin{eqnarray}
F_{ab,f}^{\text{Box}}
&=&
\dfrac{
e^2 Q_f^2
}{4 \pi ^2}
\; N_C^f
\;
\Bigg[
\delta_{ab}^f
+
\eta_{ab,f}^{(0)}
\cdot
C_0(0,\hat{s},0;
m_f^2,m_f^2,m_f^2)
\\
&& 
\hspace{2cm}
+
\eta_{ab,f}^{(1)}
\cdot
C_0(M_{\phi_i}^2,M_{\phi_j}^2,
\hat{s};m_f^2,m_f^2,m_f^2)
\nonumber \\
&&
\hspace{2cm}
+
\eta_{ab,f}^{(2)}
\cdot
C_0(\hat{t},M_{\phi_i}^2,0
;m_f^2,m_f^2,m_f^2)
\nonumber
\\
&& 
\hspace{2cm}
+
\eta_{ab,f}^{(3)}
\cdot
C_0(M_{\phi_i}^2,0,
\hat{u};m_f^2,m_f^2,m_f^2)
\nonumber \\
&&
\hspace{2cm}
+
\eta_{ab,f}^{(4)}
\cdot
C_0(0,M_{\phi_j}^2,\hat{t}
;m_f^2,m_f^2,m_f^2)
\nonumber
\\
&& 
\hspace{2cm}
+
\eta_{ab,f}^{(5)}
\cdot
C_0(\hat{u},M_{\phi_j}^2,0
;m_f^2,m_f^2,m_f^2)
\nonumber \\
&&
\hspace{2cm}
+
\zeta_{ab,f}^{(0)}
\cdot
D_0(0,M_{\phi_j}^2,M_{\phi_i}^2,0
;\hat{t},\hat{s};
m_f^2,m_f^2,m_f^2,m_f^2)
\nonumber \\
&&
\hspace{2cm}
+
\zeta_{ab,f}^{(1)}
\cdot
D_0(0,M_{\phi_i}^2,M_{\phi_j}^2,0
;\hat{u},\hat{s};
m_f^2,m_f^2,m_f^2,m_f^2)
\nonumber \\
&&
\hspace{2cm}
+
\zeta_{ab,f}^{(2)}
\cdot
D_0(M_{\phi_i}^2,0,M_{\phi_j}^2,0
;\hat{u},\hat{t};
m_f^2,m_f^2,m_f^2,m_f^2)
\Bigg].
\nonumber
\end{eqnarray}
In this formulas, we have used
the notations:
$\delta_{12}^f = 4 x_f$, 
$\delta_{33}^f = 0$. 
All presented coefficients in
the above-mentioned
factors are given by
\begin{eqnarray}
\eta_{12,f}^{(0)}
&=&
\dfrac{
m_f^2 
}{
\Big[
(x_{\phi_i}-x_t)
(x_{\phi_j}-x_t)
+x_t
\Big]^2
}
\times 
\\
&&
\times 
\Bigg\{
2 x_{\phi_i}^2
\Big[
4 x_f
[
1 + (x_{\phi_j}-x_t)^2
]
-2 x_{\phi_j}+x_t+1
\Big]
+
8 x_t^2 x_f
(
1-x_{\phi_j}+x_t
)^2
-
x_{\phi_i}^3
\nonumber\\
&&
+x_{\phi_i}
\Big[
2 x_t^2
[
8 x_f
(2 x_{\phi_j}
-
x_t
-
1)
-
1
]
+
(1 - x_{\phi_j})
\big(
16 x_{\phi_j} x_t x_f
-
8 x_f
-
2 x_t
+
x_{\phi_j}
-
1
\big)
\Big]
\Bigg\},
\nonumber
\end{eqnarray}

\begin{eqnarray}
\eta_{12,f}^{(1)}
&=&
\dfrac{
m_f^2
\;
x_{\phi_i}
\big(
x_{\phi_i}+x_{\phi_j}- 8 x_f-1
\big)
\Big[
x_{\phi_i}^2
+
x_{\phi_j}^2
+
1
+
2 (x_t + 1)
\big(
x_t
-
x_{\phi_i}
-
x_{\phi_j}
\big)
\Big]
}{\Big[
(x_{\phi_i}-x_t) (x_{\phi_j}-x_t)
+
x_t
\Big]^2},
\end{eqnarray}

\begin{eqnarray}
\eta_{12,f}^{(2)}
&=&
\dfrac{
m_f^2
\;
(x_{\phi_i}-x_t)^2
}{\Big[
(x_{\phi_i}-x_t) (x_{\phi_j}-x_t)
+
x_t
\Big]^2}
\times
\nonumber\\
&&\times
\Bigg\{
(8 x_f-x_{\phi_i}-x_{\phi_j})
\Big[
x_t^2
(x_{\phi_i}+2 x_{\phi_j} - x_t - 2)
+
x_{\phi_i} x_{\phi_j}^2
\Big]
\nonumber \\
&&
+x_t
\Big[
(2 x_{\phi_i}+x_{\phi_j}-2)
\big[
x_{\phi_j}
(x_{\phi_i}+x_{\phi_j})
-
8 x_f
x_{\phi_j}
\big]
-
8 x_f
+
x_{\phi_j}
\Big]
+
x_{\phi_i} x_{\phi_j}
\Bigg\}.
\end{eqnarray}
We also have the following
coefficients:
\begin{eqnarray}
\zeta_{12,f}^{(0)}
&=&
\dfrac{
\hat{s}^2 \; x_f }
{
\Big[
(x_{\phi_i}-x_t)
(x_{\phi_j}-x_t)
+ x_t
\Big]^2}
\Bigg\{
16 x_f^2
\Big[
(x_{\phi_i}-x_t)
(x_{\phi_j}-x_t)
+x_t
\Big]
\times 
\nonumber\\
&&
\hspace{5.5cm}
\times
\Big[
x_{\phi_i} (x_{\phi_j}
+
1)
-
x_t
(x_{\phi_i}+x_{\phi_j}
-
x_t
-
1)
\Big]
\nonumber\\
&&
+2 x_f
\Big\{
-x_{\phi_i}^2 x_{\phi_j}
\Big[
x_{\phi_j}
(x_{\phi_i}+x_{\phi_j}+2)
+
x_{\phi_i}
-
1
\Big]
\nonumber \\
&&
\hspace{1.3cm}
+
x_t^3
(x_{\phi_i}+x_{\phi_j}+1)
\Big[
-
x_t
+
2 \big( x_{\phi_i} 
+x_{\phi_j}-1 \big)
\Big]
\nonumber \\
&&
\hspace{1.3cm}
-x_t^2
\Big[
x_{\phi_i}
\big[
x_{\phi_j}
\big(
5 x_{\phi_i}
+
5 x_{\phi_j}
+
1
\big)
+
x_{\phi_i}^2
+
2
\big]
+(x_{\phi_j}-1)^2 (x_{\phi_j}+1)
\Big]
\nonumber \\
&&
\hspace{1.3cm}
+
x_{\phi_i} x_t 
(x_{\phi_i}+x_{\phi_j}-1)
\Big[
x_{\phi_i}
(2 x_{\phi_j} + 1)
+
x_{\phi_j} (2 x_{\phi_j}+3)
-
1
\Big]
\Big\}
\nonumber \\
&&
+
x_{\phi_i} x_t
\big(
x_{\phi_i} x_{\phi_j}+x_t^2
\big)
\Bigg\},   
\end{eqnarray}

\begin{eqnarray}
\zeta_{12,f}^{(2)}
&=&
\dfrac{
\hat{s}^2 \, x_f
}{\Big[
(x_{\phi_i}-x_t) (x_{\phi_j}-x_t)
+
x_t
\Big]}
\Bigg\{
16 x_f^2
\Big[
-x_t (x_{\phi_i}+x_{\phi_j}-x_t-1)
+
x_{\phi_i} (x_{\phi_j}+1)
\Big]
\nonumber\\
&&
-2 x_f
\Big\{
x_t^2
\Big[
4 \big( x_{\phi_i}^2+x_{\phi_j}^2 \big)
- 7 \big( x_{\phi_i}+x_{\phi_j} \big)
+ 16 x_{\phi_i} x_{\phi_j}
+ 5
\Big]
\nonumber\\
&&
\hspace{1.0cm}
+
x_{\phi_i}
\Big[
x_{\phi_i}
\big(
4 x_{\phi_j}^2+x_{\phi_j}+1
\big)
+
x_{\phi_j} (x_{\phi_j}+2)-1
\Big]
\nonumber \\
&&
\hspace{1.0cm}
-x_t (x_{\phi_i}+x_{\phi_j}-1)
\Big[
x_{\phi_i}
\big(
8 x_{\phi_j}
+
1
\big)
+
x_{\phi_j}
+
1
\Big]
+
4 x_t^3
\big[
x_t
-
2 (x_{\phi_i}
+
x_{\phi_j}
-1
)
\big]
\Big\}
\nonumber\\
&&
+
(x_{\phi_i}+x_{\phi_j})
\Big[
(x_{\phi_i}-x_t) (x_{\phi_j}-x_t)
+
x_t
\Big]^2
\Bigg\}.
\end{eqnarray}
The corresponding
coefficients for the factors
$F_{33,f}^{\text{Box}}$ are shown 
as follows:
\begin{eqnarray}
\eta_{33,f}^{(0)}
&=&
\dfrac{
m_f^2
}{\Big[
(x_{\phi_i}-x_t) (x_{\phi_j}-x_t)
+
x_t
\Big]^2}
\Bigg\{
8 x_f (x_{\phi_i}+x_{\phi_j}-1)
\\
&&
+
x_{\phi_i} \big(
2 x_t -4 x_{\phi_j}-x_{\phi_i} + 2
\big)
+
2 x_t (x_{\phi_j}-x_t-1)
-
(x_{\phi_j}-1)^2
\Bigg\},
\nonumber
\\
\eta_{33,f}^{(1)}
&=&
\dfrac{1}{x_{\phi_i}}
\times
\eta_{12,f}^{(1)},
\end{eqnarray}

\begin{eqnarray}
\eta_{33,f}^{(2)}
&=&
\dfrac{
m_f^2
\;
\big(
x_{\phi_i}-x_t \big)
\Big[
x_t \big(x_t-8 x_f \big)
+
x_{\phi_i} x_{\phi_j}
\Big]
}{\Big[
(x_{\phi_i}-x_t) (x_{\phi_j}-x_t)
+
x_t
\Big]^2},
\\
\zeta_{33,f}^{(0)}
&=&
\dfrac{
s^2 \, x_f
}{\Big[
(x_{\phi_i}-x_t) (x_{\phi_j}-x_t)
+
x_t
\Big]^2}
\times
\\
&&\times
\Bigg\{
16 x_f^2 \Big[
(x_{\phi_i}-x_t) (x_{\phi_j}-x_t)
+
x_t
\Big]
+
x_t
\big(
x_{\phi_i} x_{\phi_j}
+
x_t^2
\big)
\nonumber\\
&&
+2 x_f
\Big[
-x_t^2 (x_{\phi_i}+x_{\phi_j}+3)
+
\big[
x_t
(x_{\phi_i}+x_{\phi_j}-1)
-
x_{\phi_i} x_{\phi_j}
\big]
(x_{\phi_i}+x_{\phi_j}-1)
\Big]
\Bigg\}, \nonumber
\end{eqnarray}
and also have
\begin{eqnarray}
\zeta_{33,f}^{(2)}
&=&
\dfrac{
2m_f^4
\;
\big(
8 x_f-x_{\phi_i}-x_{\phi_j}+1
\big)
}{\Big[
(x_{\phi_i}-x_t) (x_{\phi_j}-x_t)
+
x_t
\Big]}.
\end{eqnarray}
We next consider one-loop box diagrams with
vector boson $W$ in the loop. The factors
are then presented as follows:
\begin{eqnarray}
F_{ab,W}^{\text{Box},1}
&=&
\dfrac{e^2}{(4 \pi)^2}
\dfrac{1}{
M_W^2
}
\Bigg\{
\delta_{ab}^W
+
\varepsilon_{ab}^W
\Big[
B_0(\hat{s};
M_W^2,M_W^2)
-
B_0(0;
M_W^2,M_W^2)
\Big]
\nonumber \\
&&
\hspace{2cm}
+
\eta_{ab,W}^{(0)}
\cdot
C_0(0,\hat{s},0;
M_W^2,M_W^2,M_W^2)
\nonumber \\
&&
\hspace{2cm}
+
\eta_{ab,W}^{(1)}
\cdot
C_0(M_{\phi_i}^2,
M_{\phi_j}^2,\hat{s};
M_W^2,M_W^2,M_W^2)
\nonumber \\
&&
\hspace{2cm}
+
\eta_{ab,W}^{(2)}
\cdot
C_0(\hat{t},M_{\phi_i}^2,0;
M_W^2,M_W^2,M_W^2)
\nonumber \\
&&
\hspace{2cm}
+
\eta_{ab,W}^{(3)}
\cdot
C_0(M_{\phi_i}^2,0,\hat{u};
M_W^2,M_W^2,M_W^2)
\nonumber \\
&&
\hspace{2cm}
+
\eta_{ab,W}^{(4)}
\cdot
C_0(0,M_{\phi_j}^2,\hat{t};
M_W^2,M_W^2,M_W^2)
\nonumber \\
&&
\hspace{2cm}
+
\eta_{ab,W}^{(5)}
\cdot
C_0(\hat{u},M_{\phi_j}^2,0;
M_W^2,M_W^2,M_W^2)
\nonumber \\
&&
\hspace{2cm}
+
\zeta_{ab,W}^{(0)}
\cdot
D_0(0,M_{\phi_j}^2,M_{\phi_i}^2,0
;\hat{t},\hat{s};
M_W^2,M_W^2,M_W^2,M_W^2)
\nonumber \\
&&
\hspace{2cm}
+
\zeta_{ab,W}^{(1)}
\cdot
D_0(0,M_{\phi_i}^2,M_{\phi_j}^2,0
;\hat{u},\hat{s};
M_W^2,M_W^2,M_W^2,M_W^2)
\nonumber \\
&&
\hspace{2cm}
+
\zeta_{ab,W}^{(2)}
\cdot
D_0(M_{\phi_i}^2,0,M_{\phi_j}^2,0
;\hat{u},\hat{t};
M_W^2,M_W^2,M_W^2,M_W^2)
\Bigg\},
\\
F_{12,W}^{\text{Box, 2}}
&=&
\dfrac{e^2}{(4\pi)^2}
\dfrac{2}{\hat{s} }
\Bigg\{
5
+2
\Big[
B_0(\hat{s},M_W^2,M_W^2)
-
B_0(0,M_W^2,M_W^2)
\Big]
\nonumber
\\
&&\hspace{4cm} +
2 \hat{s}
\big(
5 x_W
-
2
\big)
C_0(0,\hat{s},0,M_W^2,M_W^2,M_W^2)
\Bigg\},
\\
F_{12,W}^{\text{Box, 3}}
&=&
-\dfrac{e^2}{(4\pi)^2}
\dfrac{4}{\hat{s}}
\Bigg[
1
+
2 x_W 
\;
\hat{s}
\;
C_0(0,\hat{s},0;
M_W^2,M_W^2,M_W^2)
\Bigg].
\end{eqnarray}
Where we definded the following
functions as: 
$\varepsilon_{12}^W = \dfrac{2}{\hat{s} }$,
$\delta_{12}^W = - \dfrac{1}{\hat{s} }$,
$\varepsilon_{33}^W = 0$, 
$\delta_{33}^W = 0$. 
All coefficients involed with one-loop
form factors calculated from
the $W$-boson box diagrams
contributions. The coefficients
are given explicitly as follows:
\begin{eqnarray}
\eta_{12,W}^{(0)}
&=&
\dfrac{1}{x_W \Big[
(x_{\phi_i}-x_t) (x_{\phi_j}-x_t)
+
x_t
\Big]^2}
\times
\nonumber\\
&& \times 
\Bigg\{
-
2 x_t^2 x_W \big(x_W+2\big)
\big(1 -x_{\phi_j}+x_t \big)^2
\\
&&
\hspace{0.5cm}
+
x_{\phi_i}^2
\Big\{
x_{\phi_j}
- x_{\phi_j}^2
\Big[
2 x_W
\big(x_W+2 \big)
+
1
\Big]
+
4 x_{\phi_j} x_W
\Big[
x_t
\big(x_W+2 \big)
+
3
\Big]
\nonumber \\
&&
\hspace{6cm}
-
2 x_W
\Big[
x_t
\big(
2 x_t
+4
\big)
+
\big(
x_t^2
+
6
\big)
x_W
+
3
\Big]
\Big\}
\nonumber\\
&&\hspace{0.5cm}
+
2 x_{\phi_i} x_W
\Big\{
x_{\phi_j}^2
\Big[
2 x_t \big(x_W+2 \big)
+
1
\Big]
\nonumber \\
&&
\hspace{2.5cm}
-
x_{\phi_j}
\Big[
2 x_t
\big(
2 x_t x_W + 4 x_t+x_W+4
\big)
+6 x_W+3
\Big]
\nonumber \\
&&
\hspace{4cm}
+
2 x_t \big(x_t+1\big)
\Big[
x_t \big(x_W+2\big) +2
\Big]
+6 x_W+2
\Big\}
\nonumber \\
&&
\hspace{0.5cm}
-
x_{\phi_i}^3
\big( x_{\phi_j} - 2 x_W \big)
\Bigg\},
\nonumber
\end{eqnarray}

\begin{eqnarray}
\eta_{12,W}^{(1)}
&=&
\dfrac{
x_{\phi_i}
}
{
x_W
\Big[
(x_{\phi_i}-x_t)
(x_{\phi_j}-x_t)
+
x_t
\Big]^2
}
\times
\Big[
x_{\phi_i} x_{\phi_j}
-
2 x_W
(x_{\phi_i}+x_{\phi_j}-6 x_W-2)
\Big]
\nonumber\\
&&\times
\Big[
x_{\phi_i}^2
+
x_{\phi_j}^2
+
1
+
2 \big(
x_t
-
x_{\phi_i}
-
x_{\phi_j}
\big)
\big( x_t+1 \big)
\Big],
\end{eqnarray}

\begin{eqnarray}
\eta_{12,W}^{(2)}
&=&
- \dfrac{
(x_{\phi_i}-x_t)^2
}{
x_W \Big[
(x_{\phi_i}-x_t)
(x_{\phi_j}-x_t)
+
x_t
\Big]^2
}
\times 
\\
&& \times 
\Bigg\{
x_{\phi_i}
\big( x_{\phi_j}-x_t \big)^2
\Big[
x_{\phi_i}
\big( x_{\phi_j}-2 x_W \big)
+
12 x_W^2
\Big]
\nonumber\\
&& 
\hspace{0.7cm}
-2 x_{\phi_i}  x_W
\Big[
x_{\phi_j}
\big(x_{\phi_j}^2 + 2 x_t -2 \big)
-
x_t^2 \big(
x_t
-
3 x_{\phi_j}
+
2
\big)
+
x_t
\big(
1
-
3 x_{\phi_j}^2
\big)
\Big]
\nonumber \\
&&\hspace{4.5cm}
+
x_t
\big( 1-x_{\phi_j}+x_t \big)^2
\Big[
2 x_W
\big( x_{\phi_j}-6 x_W \big)
-
x_{\phi_i}  x_{\phi_j}
\Big]
\Bigg\}.
\nonumber
\end{eqnarray}
All coefficients
for scalar one-loop four-point
functions in the above-equations
are given by
\begin{eqnarray}
\zeta_{12,W}^{(0)}
&=&
\dfrac{ \hat{s} }{x_W \Big[
(x_{\phi_i}-x_t) (x_{\phi_j}-x_t)
+
x_t
\Big]^2}
\times
\\
&&\times
\Bigg\{
2 x_W x_{\phi_i}^3
\big( x_t - x_{\phi_j} \big)
\Big[
x_{\phi_j}^2
-
x_{\phi_j}
\big(x_t+2 x_W+1 \big)
+
2 x_W \big(x_t-1 \big)
+
2 x_t
\Big]
\nonumber \\
&&
+
x_{\phi_i}^2
\Big\{
2 x_t x_W
\Big[
x_{\phi_j}^2 \big(2 x_{\phi_j}-5 \big)
+
2 x_t^2 \big(x_{\phi_j}-2 \big)
+
x_{\phi_j} x_t \big(9-4 x_{\phi_j} \big)
+
x_{\phi_j}-3 x_t
\Big]
\nonumber \\
&&
\hspace{1cm}
+
4 x_W^2
\Big[
x_t^2 \big(5 x_{\phi_j}+1 \big)
+
x_t \big( -4 x_{\phi_j}^2
-4 x_{\phi_j} + 3 \big)
\nonumber \\
&&
\hspace{5cm}
+
x_{\phi_j}
\big( x_{\phi_j}^2
+ 3 x_{\phi_j} - 2 \big)
-
2 x_t^3
\Big]
\nonumber \\
&&
\hspace{1cm}
-
24 x_W^3 \big(x_{\phi_j}-x_t \big)
\big(x_{\phi_j}-x_t+1 \big)
+
x_{\phi_j} x_t^2
\Big\}
\nonumber \\
&&
+
2 x_{\phi_i} x_t x_W
\Big\{
x_t
\Big[
2 x_{\phi_j}^2 \big(x_t+2 \big)
-
x_{\phi_j}
\big(x_t^2 + 6x_t + 4 \big)
+
2 \big(x_t^2+x_t+1\big)
-
x_{\phi_j}^3
\Big]
\nonumber \\
&&
\hspace{2cm}
+
2 x_W
\Big[
x_t^3
-
2 x_t
-
2
-
2 x_t^2 \big(2 x_{\phi_j}+1 \big)
+
5 x_{\phi_j} x_t \big(x_{\phi_j}+1 \big)
\nonumber \\
&&
\hspace{9cm}
+
x_{\phi_j}
\big(7- 2 x_{\phi_j}^2
- 3 x_{\phi_j} \big)
\Big]
\nonumber \\
&&
\hspace{2cm}
+
12 x_W^2
\big(x_{\phi_j}-x_t-1 \big)
\big(2 x_{\phi_j}
-2 x_t+1 \big)
\Big\}
\nonumber\\
&&
\hspace{0cm}
+
4 x_t^2 x_W^2
\big(x_{\phi_j}-6 x_W+2 \big)
\big(1-x_{\phi_j}+x_t \big)^2
\Bigg\},
\nonumber
\end{eqnarray}

\begin{eqnarray}
\zeta_{12,W}^{(2)}
&=&
\dfrac{\hat{s} }
{x_W \Big[
(x_{\phi_i}-x_t) (x_{\phi_j}-x_t)
+
x_t
\Big]}
\Bigg\{
x_{\phi_i}^3
\big(x_{\phi_j}-2 x_W \big)
\big(x_{\phi_j}-x_t \big)^2
\\
&&
\hspace{0.5cm}
+
2 x_{\phi_i}^2
\Big\{
2 x_W^2
\Big[
x_{\phi_j}
\big( 3 x_{\phi_j}
- 6 x_t + 1 \big)
+
x_t \big(3 x_t-1 \big)
+
1
\Big]
\nonumber \\
&&
\hspace{1.5cm}
-
x_W
\Big[
x_{\phi_j}^2
\big(x_{\phi_j}-4 x_t+1 \big)
+
x_{\phi_j}
\big(5 x_t^2+x_t-1\big)
-
2 x_t
\big(x_t^2+x_t-1\big)
\Big]
\nonumber \\
&&
\hspace{1.5cm}
-
x_{\phi_j} x_t
\big(x_{\phi_j}-x_t \big)
\big(x_{\phi_j}-x_t-1 \big)
\Big\}
\nonumber \\
&&
\hspace{0.5cm}
+
x_{\phi_i}
\Big\{
-4 x_W^2
\Big[
x_{\phi_j}^2
\big(6 x_t-1 \big)
-x_{\phi_j}
\big( 12 x_t^2 + 4 x_t + 3 \big)
+
x_t
\big(6 x_t^2 + 5 x_t + 1 \big)
+
2
\Big]
\nonumber \\
&&
\hspace{2cm}
+
2 x_t x_W
\big(x_{\phi_j}-x_t-1 \big)
\Big[
x_t
\big(
x_t
+
1
\big)
-
2
+
x_{\phi_j}
\big(
2 x_{\phi_j}
-
3 x_t
+
1
\big)
\Big]
\nonumber \\
&&
\hspace{2cm}
+
x_{\phi_j} x_t^2
\big(-x_{\phi_j}+x_t+1 \big)^2
-
24 x_W^3
\big(x_{\phi_j}-x_t+1 \big)
\Big\}
\nonumber \\
&&
\hspace{0.5cm}
-
2 x_t x_W
\big(x_{\phi_j}-x_t-1 \big)
\times
\nonumber \\
&&
\hspace{3.5cm}
\times
\Big[
x_t
\big(x_{\phi_j}-6 x_W \big)
\big(
x_{\phi_j}
-
x_t
-
1
\big)
+
2 x_W
\big(x_{\phi_j}-6 x_W + 2 \big)
\Big]
\Bigg\}.
\nonumber
\end{eqnarray}
Furthermore, one gives
the coefficients
appear in the factors
$F_{33,W}^{\text{Box}}$
\begin{eqnarray}
\eta_{33,W}^{(0)}
&=&
\dfrac{1}{x_W \Big[
(x_{\phi_i}-x_t) 
(x_{\phi_j}-x_t)
+
x_t
\Big]^2}
\Bigg\{
\big(
12 x_W^2
+
x_{\phi_i} x_{\phi_j}
\big)
\big(
1-x_{\phi_i}-x_{\phi_j}
\big)
\\
&&
+
2 x_W \Big[
x_{\phi_i} \big(
x_{\phi_i}
+6 x_{\phi_j}-4 x_t-3
\big)
+
\big(x_{\phi_j}-2 x_t \big)^2
-
3 x_{\phi_j}
+
4 x_t
+
2
\Big]
\Bigg\},
\nonumber
\end{eqnarray}

\begin{eqnarray}
\eta_{33,W}^{(1)}
&=&
\dfrac{1}{x_{\phi_i}}
\times
\eta_{12,W}^{(1)},
\\
\eta_{33,W}^{(2)}
&=&
\dfrac{\big(x_{\phi_i}-x_t \big)
\Big[
x_{\phi_i} x_{\phi_j}
\big(x_t-4 x_W \big)
+
2 x_t x_W
\big(x_{\phi_j} 
+ x_{\phi_i} 
- 2 x_t+6 x_W \big)
\Big]
}{
x_W \Big[
(x_{\phi_i}-x_t) (x_{\phi_j}-x_t)
+
x_t
\Big]^2}.
\end{eqnarray}
Furthermore, one gives
\begin{eqnarray}
\zeta_{33,W}^{(0)}
&=&
\dfrac{\hat{s} }{x_W \Big[
(x_{\phi_i}-x_t) (x_{\phi_j}-x_t)
+x_t
\Big]^2}
\times  \\
&&\times
\Bigg\{
2 x_{\phi_i}^2 x_W
\big(x_{\phi_j}-x_t \big)
\big(x_{\phi_j}-2 x_t+2 x_W \big)
+
8 x_W^2 x_{\phi_i} x_t
\nonumber \\
&&
+
x_{\phi_i}
\Big\{
4 x_W^2
\Big[
x_{\phi_j}^2 +
\big(x_t - 2 x_{\phi_j}\big)
\big(x_t+1\big)
\Big]
\nonumber\\
&& \hspace{1cm}
+
2 x_t x_W
\Big[
x_{\phi_j}
\big(
- 3 x_{\phi_j}
+ 7 x_t  + 1\big)
\nonumber\\
&& \hspace{1cm}
-
x_t \big(4 x_t+3\big)
\Big]
+
24 x_W^3
\big(x_t-x_{\phi_j}\big)
+
x_{\phi_j} x_t^2
\Big\}
\nonumber\\
&& \hspace{0cm}
+2 x_t x_W
\Big\{
x_t
\Big[
x_{\phi_j}
\big(2 x_{\phi_j}
- 4 x_t - 3\big)
+
2 \big(x_t^2+x_t+1\big)
\Big]
\nonumber\\
&& \hspace{1.5cm}
+12 x_W^2
\big(x_{\phi_j}-x_t-1 \big)
\nonumber\\
&& \hspace{1.5cm}
+2 x_W
\Big[
x_{\phi_j}
\big(-x_{\phi_j}+x_t+3 \big)
+
x_t-2
\Big]
\Big\}
\;
\Bigg\},
\nonumber
\\
\zeta_{33,W}^{(2)}
&=&
\dfrac{2 \hat{s}
\Big[
x_{\phi_i}
\big( x_{\phi_j}
- 2 x_t+2 x_W \big)
+ 2 x_W
\big(x_{\phi_j}-6 x_W-2 \big)
+
2 x_t
\big(
x_t
-
x_{\phi_j}
+
1
\big)
\Big]
}{\Big[
(x_{\phi_i}-x_t)
(x_{\phi_j}-x_t)
+
x_t
\Big]}.
\end{eqnarray}
Besides, the remaining
factors with a shorted
abbreviation like
$P \equiv f, W, S$
are directly expressed
by the following
relations as shown,
\begin{eqnarray}
\eta_{ab,P}^{(3)}
\equiv
\eta_{ab,P}^{(2)}
\,
\big(x_t
\leftrightarrow x_u \big),
\hspace{2.5cm}
\eta_{ab,P}^{(4)}
=
\dfrac{x_{\phi_j}
- x_t}{x_{\phi_i} - x_t}
\times
\eta_{ab,P}^{(2)},
\\
\nonumber \\
\eta_{ab,P}^{(5)}
=
\dfrac{x_{\phi_j}
- x_u}{x_{\phi_i} - x_u}
\times
\eta_{ab,P}^{(3)},
\hspace{2.5cm}
\zeta_{ab,P}^{(1)}
\equiv
\zeta_{ab,P}^{(0)}
\,
\big(x_t \leftrightarrow x_u \big).
\end{eqnarray}
We pay attention to
the contributions
of singly (doubly) charged Higges
exchanging in the loop.
The factors are given by
\begin{eqnarray}
F_{ab,S}^{\text{Box}, 1}
&=&
\dfrac{
e^2 Q_S^2
}{4 \pi ^2}
\Bigg\{
\eta_{ab,S}^{(0)}
\cdot
C_0(0,\hat{s},0
;M_S^2,M_S^2,M_S^2)
\\
&&
\hspace{1cm}
+
\eta_{ab,S}^{(1)}
\cdot
C_0(M_{\phi_i}^2,M_{\phi_j}^2
,\hat{s} ;M_S^2,M_S^2,M_S^2)
\nonumber \\
&&
\hspace{1cm}
+
\eta_{ab,S}^{(2)}
\cdot
C_0(\hat{t},M_{\phi_i}^2,0
;M_S^2,M_S^2,M_S^2)
\nonumber \\
&&
\hspace{1cm}
+
\eta_{ab,S}^{(3)}
\cdot
C_0(M_{\phi_i}^2,0,\hat{u}
;M_S^2,M_S^2,M_S^2)
\nonumber \\
&&
\hspace{1cm}
+
\eta_{ab,S}^{(4)}
\cdot
C_0(0,M_{\phi_j}^2,\hat{t}
;M_S^2,M_S^2,M_S^2)
\nonumber \\
&&
\hspace{1cm}
+
\eta_{ab,S}^{(5)}
\cdot
C_0(\hat{u},M_{\phi_j}^2,0
;M_S^2,M_S^2,M_S^2)
\nonumber \\
&&
\hspace{1cm}
+
\zeta_{ab,S}^{(0)}
\cdot
D_0(0,M_{\phi_j}^2,M_{\phi_i}^2,0
;\hat{t},\hat{s};M_S^2,M_S^2,M_S^2,M_S^2)
\nonumber \\
&&
\hspace{1cm}
+
\zeta_{ab,S}^{(1)}
\cdot
D_0(0,M_{\phi_i}^2,M_{\phi_j}^2,0
;\hat{u},\hat{s};M_S^2,M_S^2,M_S^2,M_S^2)
\nonumber \\
&&
\hspace{1cm}
+
\zeta_{ab,S}^{(2)}
\cdot
D_0(M_{\phi_i}^2,0,M_{\phi_j}^2,0
;\hat{u},\hat{t};M_S^2,M_S^2,M_S^2,M_S^2)
\Bigg\},
\nonumber 
\\
F_{12,S}^{\text{Box},2}
&=&
\dfrac{
e^2 Q_S^2
}{4 \pi ^2}
\Big[
-\dfrac{1}{\hat{s} }
- 2 x_{S}
\,
C_0(0,\hat{s},0,M_S^2,M_S^2,M_S^2)
\Big].
\end{eqnarray}
All coefficients related to
the above-formulas are given explicitly
\begin{eqnarray}
\eta_{12,S}^{(0)}
&=&
-\dfrac{x_{\phi_i} 
\big(
x_{\phi_i}+x_{\phi_j}-1
\big)}{s 
\Big[\big(x_{\phi_i}-x_t\big)
\big(x_{\phi_j}-x_t\big)
+
x_t
\Big]^2},
\\
\eta_{12,S}^{(1)}
&=&
\dfrac{x_{\phi_i}
\Big[
x_{\phi_i}^2
+
x_{\phi_j}^2
+ 2
\big(
x_t -
x_{\phi_i} -
x_{\phi_j}
\big) 
\big(
x_t+1 
\big)
+
1
\Big]
}{s 
\Big[\big(x_{\phi_i}-x_t\big)
\big(x_{\phi_j}-x_t\big)
+
x_t
\Big]^2},
\\
\eta_{12,S}^{(2)}
&=&
-\dfrac{(x_{\phi_i}-x_t)^2
\Big[
x_{\phi_i} 
\big(
x_{\phi_j}-x_t
\big)^2
-
x_t 
\big(
-x_{\phi_j}+x_t+1
\big)^2
\Big]
}{s 
\Big[
\big(x_{\phi_i}-x_t\big)
\big(x_{\phi_j}-x_t\big)
+
x_t
\Big]^2}.
\end{eqnarray}
The coefficients of
scalar one-loop four-point
integrals
in the above equations are
\begin{eqnarray}
\zeta_{12,S}^{(0)}
&=&
\dfrac{1}{ 
\Big[
\big(x_{\phi_i}-x_t\big) 
\big(x_{\phi_j}-x_t\big)
+
x_t
\Big]^2}
\times
\\
&&
\times
\Bigg\{
x_{\phi_i} x_t^2
-
2 x_{S} 
\Big[
\big(
x_{\phi_i}-x_t
\big) 
\big(
x_{\phi_j}-x_t
\big)
+
x_t
\Big] 
\times
\nonumber\\
&&\times
\Big[
-x_t 
\big(
x_{\phi_i}+x_{\phi_j} - 1
\big)
+x_{\phi_i} 
\big( x_{\phi_j}+1 \big)
+
x_t^2
\Big]
\Bigg\},
\nonumber \\
\zeta_{12,S}^{(2)}
&=&
\dfrac{1}{ 
\Big[
\big(x_{\phi_i}-x_t\big)
\big(x_{\phi_j}-x_t\big)
+
x_t
\Big]}
\times
\\
&&
\times
\Bigg\{
\Big[
\big(x_{\phi_i}-x_t\big) 
\big(x_{\phi_j}-x_t\big)
+
x_t
\Big]^2
\nonumber\\
&&
-
2 x_{S} 
\Big[
-x_t \big(x_{\phi_i}+x_{\phi_j} - 1\big)
+
x_{\phi_i} \big(x_{\phi_j}+1\big)
+
x_t^2
\Big]
\Bigg\}.
\nonumber 
\end{eqnarray}
Other coefficients are calculated
as
\begin{eqnarray}
\eta_{33,S}^{(0)}
&=&
\dfrac{1}{x_{\phi_i}}
\times
\eta_{12,S}^{(0)},
\\
\eta_{33,S}^{(1)}
&=&
\dfrac{1}{x_{\phi_i}}
\times
\eta_{12,S}^{(1)},
\\
\eta_{33,S}^{(2)}
&=&
\dfrac{x_t
\big(x_{\phi_i}-x_t\big)}{s \Big[
\big(x_{\phi_i}-x_t\big) 
\big(x_{\phi_j}-x_t\big)
+
x_t
\Big]^2},
\\
\zeta_{33,S}^{(0)}
&=&
\dfrac{x_t^2-2 x_{S} 
\Big[
\big(x_{\phi_i}-x_t\big) 
\big(x_{\phi_j}-x_t\big)
+
x_t
\Big]}{ 
\Big[\big(x_{\phi_i}-x_t\big) 
\big(x_{\phi_j}-x_t\big)
+
x_t
\Big]^2},
\\
\zeta_{33,S}^{(2)}
&=&
-\dfrac{2 x_{S}}{ 
\Big[\big(x_{\phi_i}-x_t\big) 
\big(x_{\phi_j}-x_t\big)
+
x_t
\Big]}.
\end{eqnarray}
We arrive at contributions
of mixing charged Higgs $H^\pm$
and vector boson $W^\pm$
boxes diagrams as follows:
\begin{eqnarray}
F_{ab,W, H^\pm}^{\text{Box}}
&=&
\dfrac{e^2}{4 \pi^2}
\Bigg\{
\delta_{ab}^{W, H^\pm}
+
\varepsilon_{ab}^{W, H^\pm}
\Big[
B_0(\hat{s},M_W^2,M_W^2)
-
B_0(0,M_W^2,M_W^2)
\Big]
\nonumber \\
&&
\hspace{1cm}
+
\eta_{ab,W, H^\pm}^{(0)}
\cdot
C_0(0,\hat{s},0
;M_W^2,M_W^2,M_W^2)
\nonumber \\
&&
\hspace{1cm}
+
\eta_{ab,W, H^\pm}^{(1)}
\cdot
C_0(0,\hat{s},0;
M_{H^\pm}^2,M_{H^\pm}^2,
M_{H^\pm}^2)
\nonumber \\
&&
\hspace{1cm}
+
\eta_{ab,W, H^\pm}^{(2)}
\cdot
\Big[
C_0(M_{\phi_i}^2,\hat{s},M_{\phi_j}^2
;M_{H^\pm}^2,M_W^2,M_W^2)
\nonumber \\
&&
\hspace{3.5cm}
+
C_0(\hat{s},M_{\phi_i}^2,M_{\phi_j}^2
;M_{H^\pm}^2,M_{H^\pm}^2,M_W^2)
\Big]
\nonumber \\
&&
\hspace{1cm}
+
\eta_{ab,W, H^\pm}^{(3)}
\cdot
\Big[
C_0(\hat{t},0,M_{\phi_i}^2
;M_{H^\pm}^2,M_W^2,M_W^2)
\nonumber \\
&&
\hspace{3.5cm}
+
C_0(0,\hat{t},M_{\phi_i}^2
;M_{H^\pm}^2,M_{H^\pm}^2,M_W^2)
\Big]
\nonumber \\
&&
\hspace{1cm}
+
\eta_{ab,W, H^\pm}^{(4)}
\cdot
\Big[
C_0(\hat{t},0,M_{\phi_j}^2
;M_{H^\pm}^2,M_W^2,M_W^2)
\nonumber \\
&&
\hspace{3.5cm}
+
C_0(0,\hat{t},M_{\phi_j}^2
;M_{H^\pm}^2,M_{H^\pm}^2,M_W^2)
\Big]
\nonumber \\
&&
\hspace{1cm}
+
\eta_{ab,W, H^\pm}^{(5)}
\cdot
\Big[
C_0(\hat{u},0,M_{\phi_i}^2
;M_{H^\pm}^2,M_W^2,M_W^2)
\nonumber \\
&&
\hspace{3.5cm}
+
C_0(0,\hat{u},M_{\phi_i}^2
;M_{H^\pm}^2,M_{H^\pm}^2,M_W^2)
\Big]
\nonumber \\
&&
\hspace{1cm}
+
\eta_{ab,W, H^\pm}^{(6)}
\cdot
\Big[
C_0(\hat{u},0,M_{\phi_j}^2;
M_{H^\pm}^2,M_W^2,M_W^2)
\nonumber \\
&&
\hspace{3.5cm}
+
C_0(0,\hat{u},M_{\phi_j}^2;
M_{H^\pm}^2,M_{H^\pm}^2,M_W^2)
\Big]
\nonumber \\
&&
\hspace{1cm}
+
\zeta_{ab,W, H^\pm}^{(0)}
\cdot
D_0(\hat{t},0,\hat{s},M_{\phi_i}^2
;M_{\phi_j}^2,0;
M_{H^\pm}^2,M_W^2,M_W^2,M_W^2)
\nonumber \\
&&
\hspace{1cm}
+
\zeta_{ab,W, H^\pm}^{(1)}
\cdot
D_0(\hat{u},0,\hat{s},M_{\phi_j}^2
;M_{\phi_i}^2,0;
M_{H^\pm}^2,M_W^2,M_W^2,M_W^2)
\nonumber \\
&&
\hspace{1cm}
+
\zeta_{ab,W, H^\pm}^{(2)}
\cdot
D_0(0,\hat{s},M_{\phi_i}^2,\hat{t}
;0,M_{\phi_j}^2;
M_{H^\pm}^2,M_{H^\pm}^2,
M_{H^\pm}^2,M_W^2)
\nonumber \\
&&
\hspace{1cm}
+
\zeta_{ab,W, H^\pm}^{(3)}
\cdot
D_0(0,\hat{s},M_{\phi_j}^2,\hat{u};
0,M_{\phi_i}^2;
M_{H^\pm}^2,M_{H^\pm}^2,
M_{H^\pm}^2,M_W^2)
\nonumber \\
&&
\hspace{1cm}
+
\zeta_{ab,W, H^\pm}^{(4)}
\cdot
D_0(0,\hat{u},0,\hat{t}
;M_{\phi_i}^2,M_{\phi_j}^2;
M_{H^\pm}^2,M_{H^\pm}^2,M_W^2,M_W^2)
\Bigg\}.
\end{eqnarray}
In the above-equations, we have used
the following functions as:
$\varepsilon_{12}^{W, H^\pm} =
2/\hat{s}$, 
$\delta_{12}^{W, H^\pm} =
- 1/\hat{s},$ 
$\varepsilon_{33}^{W, H^\pm}
= 0$ and 
$\delta_{33}^{W, H^\pm} =0.$
Remaining coefficients
relating to the formulas
are given
\begin{eqnarray}
\eta_{12,W, H^\pm}^{(0)}
&=&
\dfrac{1}{x_W \Big[
(x_{\phi_i}-x_t) (x_{\phi_j}-x_t)
+
x_t
\Big]^2}
\times 
\\
&&  
\times 
\Bigg\{
x_{\phi_i} 
x_W 
\Big[
2 x_{H^\pm} 
\big(
x_{\phi_i}
+x_{\phi_j}-3 x_{H^\pm}+1
\big)
\nonumber
\\
&&
\hspace{2cm}
+
4 x_{\phi_i} x_{\phi_j}
\big( 2 - x_{\phi_j} \big)
+
x_{\phi_i} 
\big( x_{\phi_i} - 3 \big)
+
x_{\phi_j}
\big( x_{\phi_j} - 3 \big)
+
2
\Big]
\nonumber
\\
&&
\hspace{0.5cm}
-
4 x_t^2
x_W 
\Big[
x_{\phi_i}^2
+
x_t^2
+
\big(
4 x_{\phi_i} 
+
x_{\phi_j}
-
1
\big)
\big( x_{\phi_j}-1 \big)
\Big]
\nonumber
\\
&&
\hspace{0.5cm}
+
8 x_t x_W 
\big(
x_{\phi_i}+x_{\phi_j}-1
\big)
\Big[
x_t^2
+
x_{\phi_i} 
\big( x_{\phi_j}-1 \big) 
\Big]
\nonumber \\
&&
\hspace{0.5cm}
+
x_{\phi_i} 
x_W^2 
\Big[
4 x_t^2 
\big(
2 x_{\phi_j}-x_t-1
\big)
+
\big(
1 - 4 x_{\phi_j} x_t
\big) 
\big( x_{\phi_j}-1 \big)
+
6 x_{H^\pm}
-
2
\Big]
\nonumber
\\
&&
\hspace{0.5cm}
+
x_{\phi_i}^2 x_W^2
\Big[
1
+
2 \big( x_{\phi_j}-x_t \big)^2
\Big]
+
2 x_t^2 x_W^2
\big( x_{\phi_j}-x_t-1 \big)^2
\nonumber
\\
&&
\hspace{0.5cm}
+
x_{\phi_i} 
\Big[
\big( x_{H^\pm}-x_{\phi_i} \big) 
\big( x_{H^\pm}-x_{\phi_j} \big) 
\big(
2 x_{H^\pm}-x_{\phi_i}-x_{\phi_j}+1
\big)
-
2 x_W^3
\Big]
\Bigg\},
\nonumber
\end{eqnarray}

\begin{eqnarray}
\eta_{12,W, H^\pm}^{(1)}
&=&
\dfrac{1}{x_W \Big[
(x_{\phi_i}-x_t) (x_{\phi_j}-x_t)
+
x_t
\Big]^2}
\times 
\\
&& \times 
\Bigg\{
x_{\phi_i} x_{H^\pm}^2  
\big(
x_{\phi_i}+x_{\phi_j}
-2 x_{H^\pm}+6 x_W+1 
\big)
\nonumber\\
&&
+
x_{H^\pm} 
\Big\{
x_{\phi_i}^2 
\Big[
x_{\phi_i}
+ 
2 x_W 
-
1
- 
4 x_W \big( x_{\phi_j}-x_t \big)^2
\Big]
\nonumber
\\
&&
\hspace{1.5cm}
-
2 x_W
\Big[
3 x_{\phi_i} x_W
+
2 x_t^2 
\big( x_{\phi_j}-x_t-1 \big)^2
\Big]
\nonumber
\\
&&
\hspace{1.5cm}
+
2 x_{\phi_i} x_W 
\Big[
4 x_{\phi_j} x_t 
\big( x_{\phi_j}+2 x_t-1 \big)
+
4 x_t^2 
\big( x_t+1 \big)
-
3
\Big]
\nonumber
\\
&&
\hspace{1.5cm}
+
x_{\phi_i} x_{\phi_j} 
\big( x_{\phi_j}+2 x_W-1 \big) 
\Big\}
\nonumber \\
&&
-
x_{\phi_i} 
\big( x_{\phi_i} 
+ x_{\phi_j}
-2 x_W-1 \big) 
\Big[
x_{\phi_i} x_{\phi_j}
-
x_W 
\big( x_{\phi_i}
+x_{\phi_j}-x_W-2 \big)
\Big]
\Bigg\},
\nonumber 
\end{eqnarray}
\begin{eqnarray}
\eta_{12,W, H^\pm}^{(2)}
&=&
\dfrac{x_{\phi_i}}{x_W \Big[
(x_{\phi_i}-x_t) (x_{\phi_j}-x_t)
+
x_t
\big]^2}
\times 
\\
&&
\times 
\Big[
x_{\phi_i}^2
+
x_{\phi_j}^2
+
2 
\big(x_t+1 \big)
\big(x_t - x_{\phi_i} 
- x_{\phi_j} \big)
+
1
\Big] 
\times
\nonumber \\
&&\times
\Big[
x_{\phi_i} x_{\phi_j}
-
x_{H^\pm} 
\big(
x_{\phi_i} 
+x_{\phi_j}
-x_{H^\pm}
+2 x_W
\big)
-
x_W 
\big(
x_{\phi_i}
+x_{\phi_j}
-x_W
-2
\big)
\Big],
\nonumber
\end{eqnarray}
\begin{eqnarray}
\eta_{12,W, H^\pm}^{(3)}
&=&
\dfrac{
(x_{\phi_i}-x_t)^2 
}{x_W \Big[
(x_{\phi_i}-x_t) 
(x_{\phi_j}-x_t)
+
x_t
\Big]^2}
\times 
\\
&&
\times 
\Bigg\{
x_{H^\pm} 
\big( 
x_{\phi_i}+x_{\phi_j}
-x_{H^\pm}+2 x_W 
\big)
\Big[
x_{\phi_i} 
\big( x_{\phi_j}-x_t \big)^2
-
x_t 
\big( x_{\phi_j}-x_t-1 \big)^2
\Big] 
\nonumber
\\
&&
+
x_W x_{\phi_i} 
\Big[
x_{\phi_j} 
\big(
x_{\phi_j}^2
-
3 x_{\phi_j} x_t
+ 
3 x_t^2 
+ 
2 x_t 
- 
2
\big)
-
x_t 
\big( x_t^2 + 2 x_t - 1 \big)
\Big]
\nonumber
\\
&&
+
x_{\phi_j} x_t 
x_{\phi_i} 
\big( x_{\phi_j}-x_t-1 \big)^2
-
x_W^2
x_{\phi_i} 
\big( x_{\phi_j}-x_t \big)^2
\nonumber \\
&&
+
\big( x_W-x_{\phi_j} \big)
\Big[
x_t x_W 
\big( x_{\phi_j}-x_t-1 \big)^2 
+
x_{\phi_i}^2 
\big( x_{\phi_j}-x_t \big)^2 
\Big]
\Bigg\}.
\nonumber
\end{eqnarray}
Other terms are presented as
follows:
\begin{eqnarray}
\zeta_{12,W, H^\pm}^{(0)}
&=&
\dfrac{
\hat{s} 
}{x_W 
\Big[
(x_{\phi_i}-x_t)
(x_{\phi_j}-x_t)
+
x_t
\Big]^2}
\times 
\\
&& 
\times
\Bigg\{
x_{\phi_i} x_{H^\pm}
\Big[
2 x_{\phi_i}^2 x_W 
\big( x_{\phi_j}-x_t \big) 
\big( x_{\phi_j}-x_t+1 \big)
\nonumber\\
&&
\hspace{4cm}
-
x_{H^\pm}^2
\big(
x_{\phi_i}+x_{\phi_j}
- x_{H^\pm}+2 x_t+4 x_W
\big)
\Big]
\nonumber\\
&&
\hspace{0cm}
-
2 x_{\phi_i}^3 x_W 
\big( x_{\phi_j}-x_t \big) 
\Big[
2 x_t
+
x_W
\big( x_t-1 \big) 
-
x_{\phi_j} 
\big( x_t+x_W-x_{\phi_j}+1 \big)
\Big]
\nonumber\\
&&
+
2 x_W
x_t^2
\big( x_{\phi_j}-x_t-1 \big)^2 
\Big[
x_W
\big( x_{\phi_j}-x_W \big)
+
x_{H^\pm} 
\big( x_{\phi_j}+2 x_W-2 \big)
\Big]
\nonumber\\
&&
+
x_{\phi_i} x_{\phi_j} 
x_t^2
\big( x_{\phi_i}
- x_{H^\pm} \big)
\nonumber \\
&&
+
x_{H^\pm}^2 
\Big\{
x_{\phi_i}^2 
\Big[
x_{\phi_j}
+
2 x_t
+
x_W
-
2 x_W 
\big( x_{\phi_j}-x_t \big) 
\big( x_{\phi_j}-x_t+1 \big)
\Big]
\nonumber \\
&&
\hspace{1cm}
+
x_{\phi_i} 
\Big[
x_{\phi_j} x_W
\big( 4 x_{\phi_j} x_t+1 \big)
-
2 x_{\phi_j} x_t 
\big( 4 x_t x_W+x_W-1 \big)
\nonumber \\
&&\hspace{1cm}
+
2 x_W 
\big( 2 x_t+1 \big) 
\big( x_t^2+1 \big)
+
x_t^2
+
6 x_W^2
\Big]
-
2 x_t^2 x_W 
\big( x_{\phi_j}-x_t-1 \big)^2
\Big\}
\nonumber\\
&&
+
x_{\phi_i}^2 x_{H^\pm} 
\Big\{
2 x_W 
\big( x_{\phi_j}-x_t \big) 
\Big[
x_t
\big( 2 x_t-3 x_{\phi_j}+3 \big)
+
x_{\phi_j}
\big( x_{\phi_j}-1 \big) 
+
1
\Big]
\nonumber \\
&&\hspace{5cm}
+
x_W^2 
\big( 2 x_{\phi_j}
-2 x_t+1 \big)^2
-
x_t 
\big( 2 x_{\phi_j}
+x_t \big)
\Big\}
\nonumber\\ 
&&
+
x_{\phi_i} x_{H^\pm} 
\Big\{
x_W^2 
\Big[
x_{\phi_j}
-
4 x_W
-
2 x_t
\big(
4 x_{\phi_j}^2 
+
4 x_t^2
+
2 x_t
+
1
\big)
\nonumber \\
&&\hspace{7cm}
+
4 x_{\phi_j} x_t 
\big( 4 x_t+1 \big)
-
4
\Big]
\nonumber \\
&&\hspace{2cm}
+
2 x_t x_W 
\Big[
x_{\phi_j}^2 
\big( 5 x_t
-2 x_{\phi_j}
+5 \big)
-
x_{\phi_j} 
\big( 4 x_t^2
+11x_t
+5 \big)
\nonumber \\
&&\hspace{2cm}
+
x_t 
\big( 
x_t^2+6x_t+6 
\big)
\Big]
\Big\}
\nonumber\\ 
&&
+
x_{\phi_i}^2 
\Big\{
x_W^3 
\Big[
2 x_t
\big(
2 x_{\phi_j}- x_t+1
\big)
- 
2 x_{\phi_j}
\big( x_{\phi_j}+1 \big)
-
1
\Big]
\nonumber \\
&&\hspace{1cm}
+
x_W^2 
\Big[
x_{\phi_j} 
\big( 2 x_{\phi_j}^2
+2 x_{\phi_j}-7 \big)
+
2 x_t^2 
\big( 5 x_{\phi_j}
- 2 x_t-1 \big)
- 8 x_t
\big(
x_{\phi_j}^2-1
\big) \Big]
\nonumber \\
&&\hspace{1cm}
+
x_t x_W 
\Big[
x_t 
\big(
18 x_{\phi_j}
-8 x_{\phi_j}^2-5 \big)
+
2 x_{\phi_j} 
\big( x_{\phi_j}-2 \big) 
\big( 2 x_{\phi_j}-1 \big)
+
4 x_t^2 
\big( x_{\phi_j}-2 \big)
\Big]
\Big\}
\nonumber \\
&&
+
x_{\phi_i} x_W 
\Big\{
2 x_t x_W 
\big( 1-2 x_{\phi_j} \big) 
\big(
x_{\phi_j}^2-x_{\phi_j} x_W-2
\big)
+
2 x_t^4 
\big( x_W-x_{\phi_j}+2 \big)
\nonumber \\
&&\hspace{1cm}
+
4 x_t^3 
\Big[
x_{\phi_j}
\big( x_{\phi_j}-3 \big) 
+
x_W
\big( x_W-2 x_{\phi_j}+1 \big)
+
1
\Big]
\nonumber \\
&&\hspace{1cm}
+
x_t^2 
\Big[
2 x_W^2 
\big( 1-4 x_{\phi_j} \big)
+
2 x_{\phi_j} x_W 
\big( 5 x_{\phi_j}-3 \big)
\nonumber \\
&&\hspace{2cm}
+
x_{\phi_j} 
\big( 8 x_{\phi_j}
-2 x_{\phi_j}^2 -7 \big)
-
5 x_W
+
2
\Big]
+
x_W^2 
\big( x_W-x_{\phi_j}+2 \big)
\Big\}
\Bigg\},
\nonumber
\end{eqnarray}
\begin{eqnarray}
\zeta_{12,W, H^\pm}^{(2)}
&=&
\dfrac{\hat{s}}{x_W
\Big[
(x_{\phi_i}-x_t)
(x_{\phi_j}-x_t)
+
x_t
\Big]^2}
\times
\\
&&\times
\Big[
x_{\phi_i} x_{\phi_j}
-
x_{H^\pm} 
\big( x_{\phi_i}
+x_{\phi_j}
-x_{H^\pm}+2 x_W \big)
-
x_W 
\big( x_{\phi_i}
+x_{\phi_j}-x_W
-2 \big)
\Big] 
\nonumber \\
&&\times
\Bigg\{
x_{\phi_i}
\Big[
x_{H^\pm}^2 
+
\big(
x_t-x_W
\big)^2
\Big]
-
2 x_{H^\pm} 
\Big[
x_{\phi_i}^2 
\big( x_{\phi_j}-x_t \big) 
\big( x_{\phi_j}-x_t+1 \big)
\nonumber \\
&&\hspace{1.0cm}
-
x_{\phi_i} 
x_t 
\big(x_{\phi_j}
- x_t \big)
\big(2 x_{\phi_j}
-2 x_t-1 \big)
+
x_t^2 
\big( x_{\phi_j}
-x_t-1 \big)^2
+
x_{\phi_i} 
x_W
\Big]
\Bigg\},
\nonumber
\end{eqnarray}
\begin{eqnarray}
\zeta_{12,W, H^\pm}^{(4)}
&=&
\dfrac{2 \hat{s} }{x_W \Big[
(x_{\phi_i}-x_t) (x_{\phi_j}-x_t)
+
x_t
\Big]^2}
\times 
\kappa_{12,W, H^\pm}^{(4)},
\end{eqnarray}
with 
\begin{eqnarray}
\kappa_{12,W, H^\pm}^{(4)}
&=&
-
x_{\phi_i}^2
x_{H^\pm}^3
\Big[
x_{\phi_j}
\big( x_{\phi_j}
-2 x_t +1 \big)
-
x_t
\big( 1-x_t \big) 
+
1
\Big] 
-
x_t^2 
x_{H^\pm}^3
\big( x_{\phi_j}-x_t-1 \big)^2
\nonumber \\
&&
-
x_{\phi_i}
x_{H^\pm}^3
\Big[
x_t^2
\big( 4 x_{\phi_j}
-2 x_t-1 \big)
+
x_{\phi_j} x_t
\big( 1-2 x_{\phi_j} \big) 
+
x_t
+
x_{\phi_j}
+
4 x_W
-
x_{H^\pm}
\Big] 
\nonumber\\
&&
+
x_{H^\pm}^2
x_{\phi_i}^3
\big( x_{\phi_j}-x_t \big)
\Big[
x_{\phi_j}
\big(
x_{\phi_j}-2 x_t +1
\big)
+
x_t
\big( x_t-1 \big)
+
1
\Big]
\nonumber \\
&& \hspace{0cm}
+
x_{H^\pm}^2
x_{\phi_i}^2
\Big\{
x_{\phi_j}^2
\Big[
x_{\phi_j}
\big( 1-3 x_t \big)
-
x_t
\big( 1 - 9 x_t \big)
+
x_W
+
1
\Big]
\nonumber \\
&& \hspace{6cm}
+
x_{\phi_j}
\Big[
1
+
x_W
\big( 1-2 x_t \big)
-
x_t^2
\big( 1+9 x_t \big)
\Big]
\nonumber \\
&& \hspace{2cm}
+
x_W
\big( 1-x_t \big)
+
x_t 
\Big[
x_t^2
\big(
1+3 x_t
\big)
+
x_t 
\big(
x_W-1
\big)
+
1
\Big]
\Big\}
\nonumber \\
&&
+
x_{H^\pm}^2
x_{\phi_i}
x_W
\Big[
6 x_W
+
x_t^2
\big( 4 x_{\phi_j}-2 x_t-1 \big) 
+
x_{\phi_j} 
x_t
\big( 1-2 x_{\phi_j} \big) 
+
x_t
+
x_{\phi_j}
+
2
\Big] 
\nonumber \\
&&
+
x_t 
\big( x_{\phi_j}-x_t-1 \big) 
x_{H^\pm}^2
x_{\phi_i}
\Big[
x_{\phi_j}^2
\big( 3 x_t-2 \big) 
-
x_{\phi_j}
\big( 1+6 x_t^2 \big) 
+
x_t 
\big( x_t+1 \big) 
\big( 3 x_t-1 \big)
\Big]
\nonumber \\
&&
+
x_t^2 
x_{H^\pm}^2
\big( x_{\phi_j}-x_t-1 \big)^2 
\Big[
x_t
\big( 1+x_t \big)
+
x_{\phi_j}
\big( 1-x_t \big)
+
x_W
\Big]
\nonumber \\
&&
+
x_{H^\pm}
x_{\phi_i}^3
\big( x_t-x_{\phi_j} \big) 
\times
\nonumber \\
&&\hspace{1cm}
\times
\Big[
x_{\phi_i}
\big( x_{\phi_j}-x_t \big)^2
+
x_{\phi_j}^2
\big(
x_{\phi_j}
-5 x_t
+2 x_W
+1
\big) 
-
2 x_W 
\big( 
x_{\phi_j}+1 \big)
\nonumber \\
&&\hspace{3cm}
+
x_{\phi_j}  
\big(
7 x_t^2
-
4 x_W x_t
+
2 x_t
+
1
\big) 
+
x_t 
\big( x_t+1 \big) 
\big( 2 x_W-3 x_t \big)
\Big]
\nonumber \\
&&
\hspace{0cm}
-
x_{H^\pm}
x_{\phi_i}^2
\Big\{
-
x_W^2
\Big[
1
+
x_{\phi_j}
\big(
x_{\phi_j}-2 x_t +1
\big)
+
x_t 
\big( x_t-1 \big) 
\Big]
+
2 x_W 
\big( x_t-x_{\phi_j} \big) 
\times 
\nonumber \\
&&
\hspace{2cm}
\times 
\Big[
x_t^2
\big( 3 x_t-6 x_{\phi_j}+5 \big) 
+
3 x_t 
\big( x_{\phi_j}-1 \big)^2 
+
x_{\phi_j}
\big( x_{\phi_j}-1 \big)
-
1
\Big] 
\nonumber \\
&&
\hspace{2cm}
+
x_t 
\big( x_{\phi_j}-x_t-1 \big) 
\times 
\nonumber \\
&&
\hspace{2cm}
\times 
\Big[
x_{\phi_j}^2
\big( 9 x_t-3 x_{\phi_j}-2 \big) 
+
\big( x_t+1 \big)
\big( 3 x_t^2-x_{\phi_j} \big)
-
9 x_t^2 x_{\phi_j}
\Big]
\Big\}
\nonumber \\
&&
-
x_{H^\pm}
x_{\phi_i}
\Big\{
x_W^2
\Big[
2 x_t x_{\phi_j}^2
-
x_{\phi_j}
\big( 4 x_t^2+x_t+1 \big) 
+
x_t 
\big( 2 x_t^2+x_t-1 \big)
+
4 \big( x_W+1 \big)
\Big]
\nonumber \\
&&
\hspace{2cm}
+
2 x_t x_W 
\big(
x_{\phi_j}-x_t-1
\big)
\times
\nonumber \\
&&
\hspace{2cm}
\times
\Big[
x_{\phi_j}^2
\big( 3 x_t+2 \big) 
-
3 x_{\phi_j}
\big(
2 x_t^2+2 x_t+1
\big)
+
x_t 
\big(
3 x_t^2 +4 x_t+5
\big)
\Big] 
\nonumber \\
&&
\hspace{2cm}
+
x_t^2 
\big( x_{\phi_j}-x_t-1 \big)^2 
\Big[
x_{\phi_j}
\big( 3 x_{\phi_j}-4 x_t +1 \big)
+
x_t
\big( x_t+1 \big)
\Big]
\Big\}
\nonumber \\ %
&&
-
x_{H^\pm}
x_t^2 
\big( 
x_{\phi_j}-x_t-1 \big)^2 
\times
\nonumber \\
&&
\hspace{2cm}
\times
\Big[
-
x_{\phi_j}^2 
x_t
+
x_{\phi_j}
\big( x_t+1 \big) 
\big( x_t-2 x_W \big) 
+
x_W
\big(
2 x_t^2 + 2 x_t-x_W+4
\big)
\Big]
\nonumber \\
&&
+
x_W^2
\Big[
\big( x_{\phi_i}-x_t \big) 
\big( x_{\phi_j}-x_t \big)
+
x_t
-
x_W
\Big]
\Big\{
x_{\phi_i}^2
\Big[
1
+
\big( x_{\phi_j}-x_t \big)^2
\Big]
-
x_{\phi_i} x_W
\nonumber \\
&&
+
x_{\phi_i}
\Big[
\big( 1-2 x_{\phi_j} x_t \big)
\big( x_{\phi_j}-1 \big) 
+
2 x_t^2 
\big( 2 x_{\phi_j}-x_t-1 \big)
-
1
\Big] 
+
x_t^2 
\big( x_{\phi_j}-x_t-1 \big)^2
\Big\}
\nonumber \\
&&
-
x_W
\Big[
\big( x_{\phi_i}-x_t \big) 
\big( x_{\phi_j}-x_t \big)
+
x_t
-
x_W
\Big]
\Big\{
x_{\phi_i}^2
\Big[
x_{\phi_i}
\big( x_{\phi_j}-x_t \big)^2 
+
2 x_t
\nonumber \\
&&
+
\big( x_{\phi_j}-x_t-1 \big) 
\big(
x_{\phi_j}^2
+
x_{\phi_j}
+
2 x_t^2
-
3 x_t x_{\phi_j}
\big)
\Big]
+
x_t 
\big( x_{\phi_j}-x_t-1 \big) 
\times
\nonumber \\
&&
\hspace{1.5cm}
\times
\Big[
x_{\phi_j} 
x_t
\big(
x_{\phi_j}-x_t-1
\big)
-
x_{\phi_i}
\big( x_{\phi_j}-x_t+1 \big) 
\big( 2 x_{\phi_j}-x_t-2 \big) 
\Big]
\Big\}
\nonumber \\
&&
+
x_{\phi_i} 
x_{\phi_j} 
\Big[
\big( x_{\phi_i}-x_t \big) 
\big( x_{\phi_j}-x_t \big)
+x_t-x_W
\Big]
\Big[
\big( x_{\phi_i}-x_t \big) 
\big( x_{\phi_j}-x_t \big)
+
x_t
\Big]^2
\Bigg\},
\nonumber
\end{eqnarray}
\begin{eqnarray}
\eta_{33,W, H^\pm}^{(0)}
&=&
\dfrac{1}{x_W \Big[
(x_{\phi_i}-x_t) (x_{\phi_j}-x_t)
+
x_t
\Big]^2}
\times 
\\
&& \times 
\Bigg\{
x_W 
\Big[
2 x_{H^\pm} 
\big( x_{\phi_i}
+x_{\phi_j}
- 3 x_{H^\pm}+1 \big)
+
x_{\phi_j}
\big(
x_{\phi_j}
-
3 
\big)
\nonumber \\
&& \hspace{1cm}
-
8 x_t 
\big( x_{\phi_i}
+x_{\phi_j}-x_t-1 \big)
+
x_{\phi_i}
\big(
x_{\phi_i}
+
8 x_{\phi_j}
-
3 
\big)
+
2
\Big]
\nonumber \\
&& \hspace{0cm}
+
2 x_W^2 
\big(
4 x_{H^\pm}
-
x_W
-
1
\big)
\nonumber \\
&&
+
\big( 
2 x_{H^\pm}-x_{\phi_i}-x_{\phi_j}+1 
\big)
\Big[
\big( x_{H^\pm}-x_{\phi_i} \big) 
\big( x_{H^\pm}-x_{\phi_j} \big) 
-
x_W^2
\Big]
\Bigg\},
\nonumber
\end{eqnarray}
\begin{eqnarray}
\eta_{33,W, H^\pm}^{(1)}
&=&
- \dfrac{1}{x_W \Big[
(x_{\phi_i}-x_t) (x_{\phi_j}-x_t)
+x_t
\Big]^2}
\Big[
2 x_{H^\pm}+x_{\phi_i}+x_{\phi_j}-2 x_W-1
\Big] 
\times
\\
&&\times
\Bigg[
x_{\phi_i} x_{\phi_j}
-
x_{H^\pm} 
\big(
x_{\phi_i}+x_{\phi_j}-x_{H^\pm}+2 x_W
\big)
-
x_W 
\big(
x_{\phi_i}+x_{\phi_j}-x_W-2
\big)
\Bigg],
\nonumber\\
\eta_{33,W, H^\pm}^{(2)}
&=&
\dfrac{1}{x_{\phi_i}}
\times
\eta_{12,W, H^\pm}^{(2)},
\\
\eta_{33,W, H^\pm}^{(3)}
&=&
\dfrac{
x_{\phi_i}-x_t
}{x_W \Big[
(x_{\phi_i}-x_t) (x_{\phi_j}-x_t)
+x_t
\Big]^2}
\Big\{
x_{\phi_i} 
\big(
x_{\phi_j} x_t-2 x_{\phi_j} x_W+x_t x_W
\big)
\\
&&
+
x_t \Big[
x_W 
\big(
x_{\phi_j}-2 x_t+x_W
\big)
-
x_{H^\pm} 
\big(
x_{\phi_i}+x_{\phi_j}-x_{H^\pm}+2 x_W
\big)
\Big]
\Big\},
\nonumber
\end{eqnarray}
\begin{eqnarray}
\zeta_{33,W, H^\pm}^{(0)}
&=&
\dfrac{\hat{s} }{x_W \Big[
(x_{\phi_i}-x_t) 
(x_{\phi_j}-x_t)
+
x_t
\Big]^2}
\times
\\
&&
\times
\Bigg\{
x_W^2
x_{H^\pm} 
\Big[
\big( 1-4 x_t \big) 
\big( 
x_{\phi_i}
+x_{\phi_j} 
\big)
+
2 x_t 
\big( 2 x_t-1 \big)
\nonumber\\
&&
+
2
\big( 
3 x_{H^\pm}
+ 2 x_{\phi_i} x_{\phi_j}
- 2
\big)
\Big]
+
x_W^2
\big( x_{\phi_i} x_{\phi_j}
+ x_t^2 \big)
\big( 2 x_{\phi_i}
+2 x_{\phi_j}-7 \big)
\nonumber \\
&&
-
2 x_t x_W^2
\Big[
\big( x_{\phi_i}+x_{\phi_j}-4 \big) 
\big( x_{\phi_i}+x_{\phi_j} \big)
+
2
\Big]
+
\big( x_{H^\pm}-x_{\phi_i} \big) 
\big( x_{H^\pm}-x_{\phi_j} \big) 
\big( x_{H^\pm}-x_t \big)^2
\nonumber \\
&&
+
x_W 
\Big\{
x_{H^\pm}^2 
\Big[
-
4 x_{H^\pm}
-
2 \big(
x_{\phi_i} x_{\phi_j}
+ 
x_t^2
\big)
+
\big( 2 x_t + 1 \big) 
\big( x_{\phi_i}+x_{\phi_j}+2 \big)
\Big]
\nonumber \\
&&
+
2 x_{H^\pm} 
\big( x_{\phi_i}+x_{\phi_j}+1 \big) 
\big( x_{\phi_i}-x_t \big) 
\big( x_{\phi_j}-x_t \big)
\nonumber\\
&&
+
2 x_{\phi_i} x_{\phi_j} 
\Big[
x_{\phi_i} x_{\phi_j}
-
x_t 
\big( 3 x_{\phi_i}+3 x_{\phi_j}-2 \big)
\Big]
-
4 x_t^3 
\big( 2 x_{\phi_i}
+2 x_{\phi_j}-x_t-1 \big)
\nonumber\\
&&
+
x_t^2 
\Big[
2 x_{\phi_i} 
\big( 2 x_{\phi_i}+5 x_{\phi_j} \big)
+
\big( x_{\phi_i} + x_{\phi_j} \big)
\big( 4 x_{\phi_j}-5 \big)
+
2
\Big]
\Big\}
\nonumber \\
&&
-
x_W^3 
\Big[
4 x_{H^\pm}
-
x_W
+
\big( 1-2 x_t \big) 
\big( x_{\phi_i}+x_{\phi_j} \big)
+
2 \big(
x_{\phi_i} x_{\phi_j}
+
x_t^2
-
1
\big)
\Big]
\Bigg\},
\nonumber
\end{eqnarray}
\begin{eqnarray}
\zeta_{33,W, H^\pm}^{(2)}
&=&
\dfrac{\hat{s} }{x_W \Big[
(x_{\phi_i}-x_t) (x_{\phi_j}-x_t)
+
x_t
\Big]^2}
\times
\\
&&\times
\Big[
x_{\phi_i} x_{\phi_j}
-
x_{H^\pm} 
\big(
x_{\phi_i}+x_{\phi_j}-x_{H^\pm}+2 x_W
\big)
-
x_W 
\big(
x_{\phi_i}+x_{\phi_j}-
x_W-2
\big)
\Big] 
\nonumber \\
&&\times
\Big\{
x_{H^\pm}^2-2 x_{H^\pm} 
\Big[
\big(x_{\phi_i}-x_t \big) 
\big(x_{\phi_j}-x_t \big)
+x_W
\Big]
+
\big(x_t-x_W \big)^2
\Big\},
\nonumber
\end{eqnarray}
\begin{eqnarray}
\zeta_{33,W, H^\pm}^{(4)}
&=&
\dfrac{2 \hat{s} }
{x_W
\Big[
(x_{\phi_i}-x_t)
(x_{\phi_j}-x_t)
+
x_t
\Big]^2}
\times
\\
&&\times
\Bigg\{
-
x_W^3 
\Big[
4 x_{H^\pm}
- 
x_W
+
x_{\phi_i} x_{\phi_j}
-
\big( x_t-1 \big) 
\big( x_{\phi_i}
+ x_{\phi_j}-x_t-2 \big)
\Big]
\nonumber\\
&&
+
x_W^2 
\Big\{
x_{H^\pm} 
\Big[
x_{\phi_i} x_{\phi_j}
-
2 \big( 1-3 x_{H^\pm} \big)
-
\big( x_t-1 \big) 
\big( x_{\phi_i}+x_{\phi_j}-x_t-2 \big)
\Big]
\nonumber \\
&&
+
\big( x_{\phi_i}+x_{\phi_j}-4 \big)
\Big[
x_{\phi_i} x_{\phi_j} 
-
x_t 
\big( x_{\phi_i}+x_{\phi_j}-x_t-1 \big)
\Big]
+
x_{\phi_i} x_{\phi_j} 
\Big\}
\nonumber \\
&&
+
x_W 
\Big\{
x_{H^\pm}^2 
\Big[
x_{\phi_i} x_{\phi_j}
-
4 \big( x_{H^\pm}-1 \big)
-
\big( x_t-1 \big) 
\big( x_{\phi_i}
+x_{\phi_j}-x_t-2 \big)
\Big]
\nonumber \\
&&
+
2 x_{H^\pm} 
\Big[
x_{\phi_i} x_{\phi_j} 
\big( x_{\phi_i}
+x_{\phi_j}-1 \big)
-
x_t 
\big( x_{\phi_i}+x_{\phi_j} \big)
\big( x_{\phi_i}+x_{\phi_j}-x_t-1 \big) 
\Big]
\nonumber \\
&&
+
\Big[
\big( x_{\phi_i}-x_t \big) 
\big( x_{\phi_j}-x_t \big)
+
x_t
\Big] 
\Big[
x_{\phi_i} x_{\phi_j}
-
2 x_t 
\big( x_{\phi_i}
+x_{\phi_j}-x_t-1 \big)
\Big]
\Big\}
\nonumber \\
&&
+
x_{H^\pm} 
\big( x_{H^\pm}-x_{\phi_i} \big) 
\big( x_{H^\pm}-x_{\phi_j} \big) 
\Big[
x_{H^\pm}
-
x_t
+
\big( x_{\phi_i}-x_t \big) 
\big( x_t-x_{\phi_j} \big)
\Big]
\Bigg\}.
\nonumber 
\end{eqnarray}
Remaining coefficients
are expressed by the
following relations
as shown,
\begin{eqnarray}
\eta_{ab,W, H^\pm}^{(4)}
&=&
\dfrac{x_{\phi_j}
- x_t}{x_{\phi_i} - x_t}
\times
\eta_{ab,W, H^\pm}^{(3)},
\quad
\eta_{ab,W, H^\pm}^{(5)}
=
\eta_{ab,W, H^\pm}^{(3)}
\big(x_t \leftrightarrow x_u \big),
\\
\eta_{ab,W, H^\pm}^{(6)}
&=&
\dfrac{x_{\phi_j} - x_u}{x_{\phi_i} - x_u}
\times
\eta_{ab,W, H^\pm}^{(5)},
\quad
\zeta_{ab,W, H^\pm}^{(1/3)}
=
\zeta_{ab,W, H^\pm}^{(0/2)}
\big(x_t \leftrightarrow x_u \big).
\end{eqnarray}
\section*{Appendix $C$: 
Lagrangian for Zee-Babu models}   
Deriving all the couplings in Zee-Babu models
are presented in this appendix. 
After the EWSB, the hypercharge field $B_\mu$ 
mixes with the weak isospin field $W_\mu^3$. 
They are decomposed in terms of 
the mass eigenstates as follows:
$B_\mu=c_WA_\mu-s_WZ_\mu$
where $\theta_W$ is the weak mixing angle. 
The kinetic term can be expanded as 
\begin{eqnarray}
\mathcal{L}_K^{ZB}
&=& (D_{\mu}H)^{\dagger}(D^{\mu}H)
+(D_{\mu}K)^{\dagger}(D^{\mu}K) 
\nonumber\\
&{\supset}&
-ig_Yc_WQ_{H}A^{\mu}(H^{\mp}\partial_{\mu}H^{\pm}-H^{\pm}\partial_{\mu}H^{\mp})
+ig_Ys_WQ_{H}Z^{\mu}(H^{\mp}\partial_{\mu}H^{\pm}-H^{\pm}\partial_{\mu}H^{\mp}) 
\notag\\
&& 
+ g_Y^2c_W^2Q_{H}^2A^{\mu}A_{\mu}H^{\pm}H^{\mp}
+ g_Y^2s_W^2Q_{H}^2Z^{\mu}Z_{\mu}H^{\pm}H^{\mp}
- g_Y^2s_{2W}Q_{H}^2A^{\mu}Z_{\mu}H^{\pm}H^{\mp} 
\notag\\
&& 
- ig_Yc_WQ_{K}A^{\mu}(K^{\mp\mp}\partial_{\mu}K^{\pm\pm}
- K^{\pm\pm}\partial_{\mu}K^{\mp\mp})
\\
&&
+ ig_Ys_WQ_{K}Z^{\mu}
(K^{\mp\mp}\partial_{\mu}K^{\pm\pm}
-K^{\pm\pm}\partial_{\mu}K^{\mp\mp})  
+ g_Y^2c_W^2Q_{K}^2A^{\mu}A_{\mu}K^{\pm\pm}K^{\mp\mp}
\nonumber
\\
&&
+ g_Y^2s_W^2Q_{K}^2Z^{\mu}Z_{\mu}K^{\pm\pm}K^{\mp\mp}
- g_Y^2s_{2W}Q_{K}^2A^{\mu}Z_{\mu}K^{\pm\pm}K^{\mp\mp}.
\nonumber
\end{eqnarray}
The scalar potential of $H^{\pm}$ 
and $K^{\pm\pm}$ 
are expressed in the mass basis
\begin{eqnarray}
-\mathcal{V}_{ZB}
&=&
-\mu^2_1H^{\mp}{H^\pm}
-\mu^2_2K^{\mp\mp}{K}^{\pm\pm}
-\lambda_H(H^\mp{H^\pm})^2
-\lambda_K(K^{\mp\mp}{K^{\pm\pm}})^2
\notag\\
&&
-\lambda_{HK}(H^{\mp}{H^\pm})(K^{\mp\mp}{K^{\pm\pm}}) 
-\mu_L({H^{\pm}H^{\pm}K^{\mp\mp}}
+{H^{\mp}H^{\mp}K^{\pm\pm}})
\notag\\
&&
-\lambda_{K\Phi}(K^{\mp\mp}K^{\pm\pm})
[\chi^{\mp}\chi^{\pm}
+\frac{1}{2}(v^2+2vh+hh+\chi_0^2)
] \notag\\
&&
-\lambda_{H\Phi}(H^{\mp\mp}H^{\pm\pm})
[\chi^{\mp}\chi^{\pm}+\frac{1}{2}
(v^2+2vh+hh+\chi_0^2)] 
\\
&\supset& 
-\mu_LH^{\pm}H^{\pm}K^{\mp\mp}
-v\lambda_{H\Phi}hH^{\pm}H^{\mp}
-v\lambda_{K\Phi}hK^{\pm\pm}K^{\mp\mp}
\notag\\
&& 
-\frac{\lambda_{H\Phi}}{2}hhH^{\pm}H^{\mp}
-\frac{\lambda_{K\Phi}}{2}hhK^{\pm\pm}K^{\mp\mp} 
\notag\\
&& 
-\lambda_{HK}H^{\pm}H^{\mp}K^{\pm\pm}K^{\mp\mp}
-\lambda_{H\Phi}H^{\pm}H^{\mp}\chi^{\pm}\chi^{\mp}
-\lambda_{K\Phi}K^{\pm\pm}K^{\mp\mp}\chi^{\pm}\chi^{\mp}.
\end{eqnarray}
\indent The Yukawa of Zee-Babu model is given by
\begin{eqnarray}
    \mathcal{L}_{Y}^{ZB}&=&f_{ij}[\overline{\tilde{L^i}}L^{j}H^\dagger-h.c]+g_{ij}[\overline{(e_R^c)^i}e_R^jK^\dagger+h.c] \notag\\
    &=&f_{ij}[\left(\begin{array}{cc}
        (\overline{e}^c_L)^i & -\overline{\nu}_L^c  \\ 
    \end{array}\right)\left(\begin{array}{c}
         \nu_L  \\
         (e_L)^j 
    \end{array}\right)H^\dagger-h.c]+g_{ij}[\overline{(e_R^c)^i}e_R^jK^\dagger+h.c] \notag\\
    &=&f_{ij}\bigg\{[(\overline{e}^c_L)^i\nu_L-\overline{\nu}_L^ce_L^j]H^\dagger-h.c\bigg\} +g_{ij}[\overline{(e_R^c)^i}e_R^jK^\dagger+h.c]
\end{eqnarray}

\end{document}